\documentclass[camera,letterpaper,nomarginnotes,nonarrowgutter]{jpaper}

\PassOptionsToPackage{hyphens}{url}
\usepackage[bookmarks=true,breaklinks=true,colorlinks,linkcolor=black,citecolor=blue,urlcolor=black]{hyperref}
\usepackage{url}
\usepackage{siunitx}
\usepackage[labelfont=bf]{caption}
\usepackage[normalem]{ulem}
\usepackage[short,nodayofweek,level,12hr]{datetime}
\usepackage{amssymb,amsfonts}
\usepackage{array}
\usepackage{balance}
\usepackage{flushend}
\usepackage{boldline}
\usepackage{booktabs}
\usepackage{cite}
\usepackage{colortbl}
\usepackage{dblfloatfix}
\usepackage{enumitem}
\usepackage{fancyhdr}
\usepackage{float}
\usepackage{graphicx}
\usepackage{hhline}
\usepackage{longfbox}
\usepackage{mathptmx} %
\usepackage{mathtools}
\usepackage{MnSymbol,wasysym}
\usepackage{multirow}
\usepackage{pifont}
\usepackage{setspace}
\usepackage{soul}
\usepackage{ragged2e}
\usepackage{subcaption}
\usepackage{footnotebackref}
\usepackage[firstpage]{draftwatermark}
\usepackage{tablefootnote}
\usepackage{textcomp}
\usepackage{threeparttable}
\usepackage{tikz}
\usepackage{xargs}                    
\usepackage{xcolor}
\usepackage{xspace}
\usepackage[subject={Revision Highlight},author={}]{pdfcomment}
\usepackage[colorinlistoftodos,prependcaption,textsize=tiny,textwidth=1.2cm]{todonotes}
\usepackage[nameinlink]{cleveref}

\SetWatermarkText{
 \hspace*{4.4in}
 \raisebox{10.18in}{
  \href{https://www.acm.org/publications/policies/artifact-review-and-badging-current}{\includegraphics[height=0.7in]{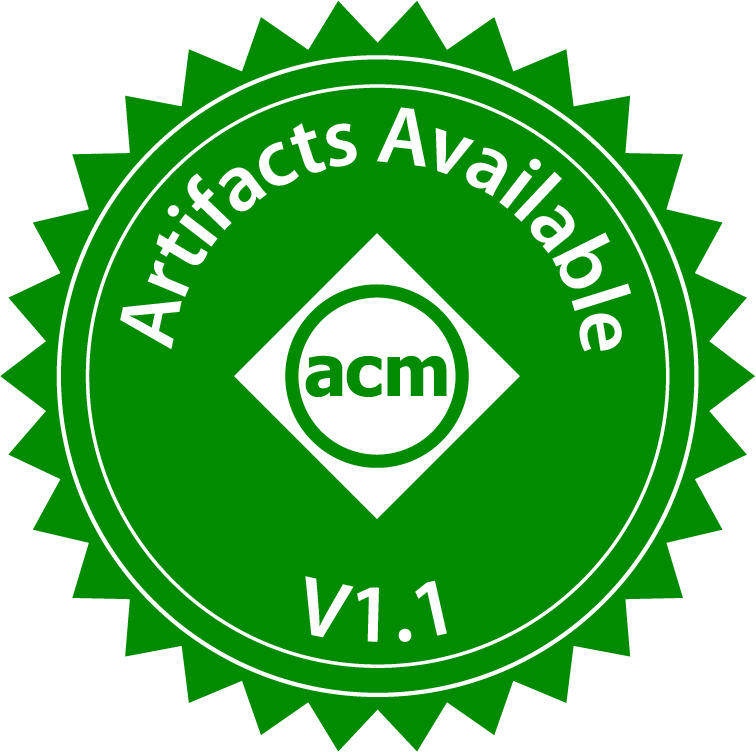}}
  \hskip -1.2cm
  \href{https://www.acm.org/publications/policies/artifact-review-and-badging-current}{\includegraphics[height=0.7in]{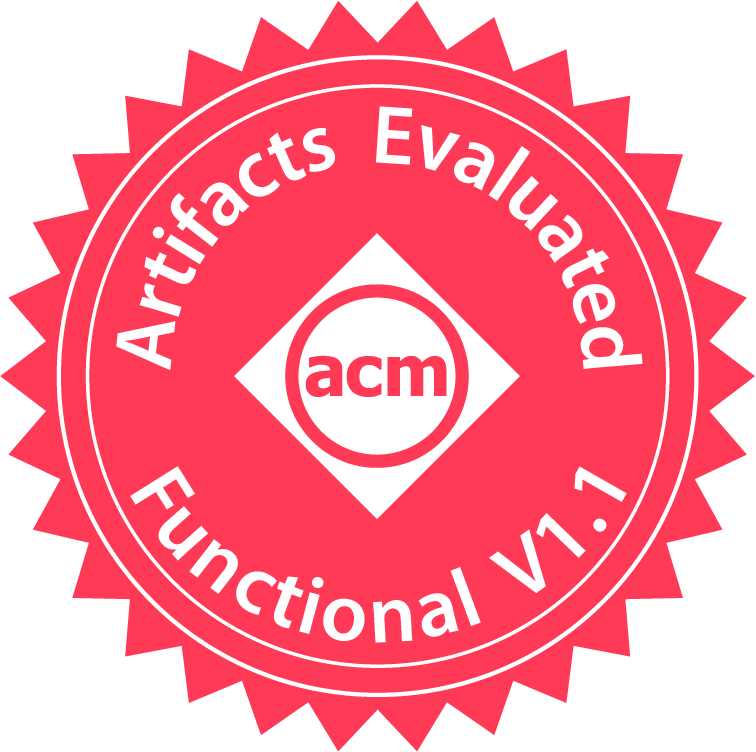}}
  \hskip -1.2cm
  \href{https://www.acm.org/publications/policies/artifact-review-and-badging-current}{\includegraphics[height=0.7in]{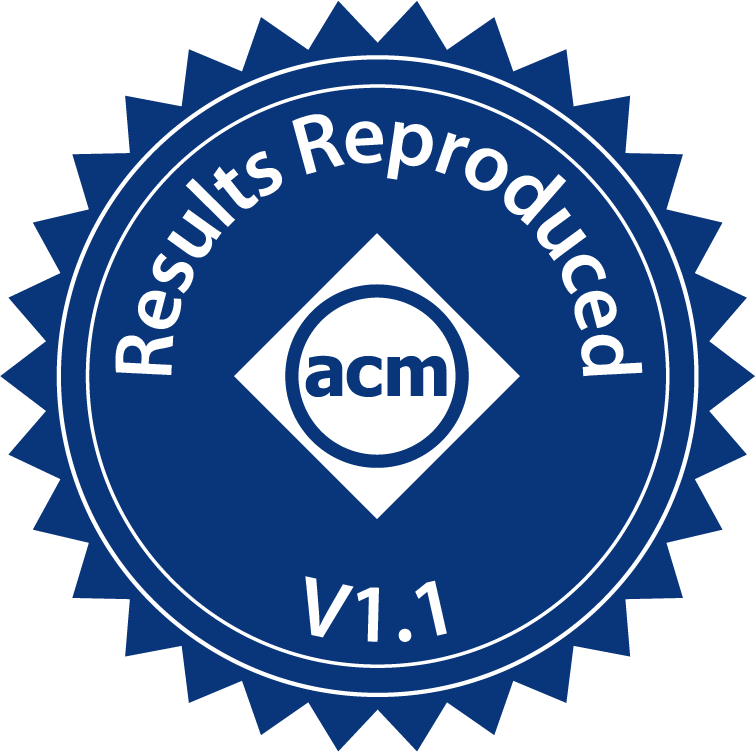}}
 }
}
\SetWatermarkAngle{0}

\makeatletter
\AtEndDocument
  {%
    \immediate\write\@auxout
      {\gdef\noexpand\totaltodos{\the\value{@todonotes@numberoftodonotes}}}%
  }
\newcommand*\iftodos
  {%
    \@ifundefined{totaltodos}
      {}
      {%
        \ifnum\totaltodos>0
          \expandafter\@secondoftwo
        \fi
        \@gobble
      }%
  }
\@ifundefined{totaltodos}{\def\totaltodos{0}}{}
\makeatother

\newif\ifcameraready
\camerareadytrue %

\newif\ifshowtodos
\showtodostrue %

\DeclareRobustCommand{\circledletter}[1]{\tikz[baseline=(char.base)]{ \node[shape=circle,minimum size=11pt,fill=black,inner sep=0pt,outer sep=0,scale=0.85] (char) {\fontfamily{lmss}\selectfont\textbf{\textcolor{white}{#1}}};}}
\DeclareRobustCommand{\circlednumber}[1]{\tikz[baseline=(char.base)]{ \node[shape=circle,minimum size=14pt,fill=black,inner sep=0pt,outer sep=0,scale=0.85] (char) {\fontfamily{lmss}\selectfont\textbf{\textcolor{white}{#1}}};}}

\newcommand{\incircledd}[1]{%
  \tikz[baseline=(char.base)]{
    \node[shape=circle,minimum size=10.5pt,draw,inner sep=0.5pt,fill=white,text=black,font=\itshape\footnotesize] (char) {\fontfamily{lmss}\selectfont#1};}%
}%

\DeclarePairedDelimiter\ceil{\lceil}{\rceil}

\def\genbox#1#2#3#4#5#6{%
\leavevmode\raise#4bp\hbox to#5bp{\vrule height#5bp depth0bp width0bp
\pdfliteral{q .5 w \csname #2COLOR\endcsname\space RG
  \csname #3PDF\endcsname{#5}{#6} S Q
  \ifx1#1 q \csname #2COLOR\endcsname\space rg
    \csname #3PDF\endcsname{#5}{#6} f Q\fi}\hss}}

\newcommand{\head}[1]{{\noindent\textbf{#1.}\xspace}}

\definecolor{airforceblue}{rgb}{0.36, 0.54, 0.66}
\definecolor{aliceblue}{rgb}{0.94, 0.97, 1.0}
\definecolor{alizarin}{rgb}{0.82, 0.1, 0.26}
\definecolor{almond}{rgb}{0.94, 0.87, 0.8}
\definecolor{amaranth}{rgb}{0.9, 0.17, 0.31}
\definecolor{amber}{rgb}{1.0, 0.75, 0.0}
\definecolor{amber(sae/ece)}{rgb}{1.0, 0.49, 0.0}
\definecolor{americanrose}{rgb}{1.0, 0.01, 0.24}
\definecolor{amethyst}{rgb}{0.6, 0.4, 0.8}
\definecolor{anti-flashwhite}{rgb}{0.95, 0.95, 0.96}
\definecolor{antiquebrass}{rgb}{0.8, 0.58, 0.46}
\definecolor{antiquefuchsia}{rgb}{0.57, 0.36, 0.51}
\definecolor{antiquewhite}{rgb}{0.98, 0.92, 0.84}
\definecolor{ao}{rgb}{0.0, 0.0, 1.0}
\definecolor{ao(english)}{rgb}{0.0, 0.5, 0.0}
\definecolor{applegreen}{rgb}{0.55, 0.71, 0.0}
\definecolor{apricot}{rgb}{0.98, 0.81, 0.69}
\definecolor{aqua}{rgb}{0.0, 1.0, 1.0}
\definecolor{aquamarine}{rgb}{0.5, 1.0, 0.83}
\definecolor{armygreen}{rgb}{0.29, 0.33, 0.13}
\definecolor{arsenic}{rgb}{0.23, 0.27, 0.29}
\definecolor{arylideyellow}{rgb}{0.91, 0.84, 0.42}
\definecolor{ashgrey}{rgb}{0.7, 0.75, 0.71}
\definecolor{asparagus}{rgb}{0.53, 0.66, 0.42}
\definecolor{atomictangerine}{rgb}{1.0, 0.6, 0.4}
\definecolor{auburn}{rgb}{0.43, 0.21, 0.1}
\definecolor{aureolin}{rgb}{0.99, 0.93, 0.0}
\definecolor{aurometalsaurus}{rgb}{0.43, 0.5, 0.5}
\definecolor{awesome}{rgb}{1.0, 0.13, 0.32}
\definecolor{azure(colorwheel)}{rgb}{0.0, 0.5, 1.0}
\definecolor{azure(web)(azuremist)}{rgb}{0.94, 1.0, 1.0}
\definecolor{babyblue}{rgb}{0.54, 0.81, 0.94}
\definecolor{babyblueeyes}{rgb}{0.63, 0.79, 0.95}
\definecolor{babypink}{rgb}{0.96, 0.76, 0.76}
\definecolor{ballblue}{rgb}{0.13, 0.67, 0.8}
\definecolor{bananamania}{rgb}{0.98, 0.91, 0.71}
\definecolor{bananayellow}{rgb}{1.0, 0.88, 0.21}
\definecolor{battleshipgrey}{rgb}{0.52, 0.52, 0.51}
\definecolor{bazaar}{rgb}{0.6, 0.47, 0.48}
\definecolor{beaublue}{rgb}{0.74, 0.83, 0.9}
\definecolor{beaver}{rgb}{0.62, 0.51, 0.44}
\definecolor{beige}{rgb}{0.96, 0.96, 0.86}
\definecolor{bisque}{rgb}{1.0, 0.89, 0.77}
\definecolor{bistre}{rgb}{0.24, 0.17, 0.12}
\definecolor{bittersweet}{rgb}{1.0, 0.44, 0.37}
\definecolor{black}{rgb}{0.0, 0.0, 0.0}
\definecolor{blanchedalmond}{rgb}{1.0, 0.92, 0.8}
\definecolor{bleudefrance}{rgb}{0.19, 0.55, 0.91}
\definecolor{blizzardblue}{rgb}{0.67, 0.9, 0.93}
\definecolor{blond}{rgb}{0.98, 0.94, 0.75}
\definecolor{blue}{rgb}{0.0, 0.0, 1.0}
\definecolor{blue(munsell)}{rgb}{0.0, 0.5, 0.69}
\definecolor{blue(ncs)}{rgb}{0.0, 0.53, 0.74}
\definecolor{blue(pigment)}{rgb}{0.2, 0.2, 0.6}
\definecolor{blue(ryb)}{rgb}{0.01, 0.28, 1.0}
\definecolor{bluebell}{rgb}{0.64, 0.64, 0.82}
\definecolor{bluegray}{rgb}{0.4, 0.6, 0.8}
\definecolor{blue-green}{rgb}{0.0, 0.87, 0.87}
\definecolor{blue-violet}{rgb}{0.54, 0.17, 0.89}
\definecolor{blush}{rgb}{0.87, 0.36, 0.51}
\definecolor{bole}{rgb}{0.47, 0.27, 0.23}
\definecolor{bondiblue}{rgb}{0.0, 0.58, 0.71}
\definecolor{bostonuniversityred}{rgb}{0.8, 0.0, 0.0}
\definecolor{brandeisblue}{rgb}{0.0, 0.44, 1.0}
\definecolor{brass}{rgb}{0.71, 0.65, 0.26}
\definecolor{brickred}{rgb}{0.8, 0.25, 0.33}
\definecolor{brightcerulean}{rgb}{0.11, 0.67, 0.84}
\definecolor{brightgreen}{rgb}{0.4, 1.0, 0.0}
\definecolor{brightlavender}{rgb}{0.75, 0.58, 0.89}
\definecolor{brightmaroon}{rgb}{0.76, 0.13, 0.28}
\definecolor{brightpink}{rgb}{1.0, 0.0, 0.5}
\definecolor{brightturquoise}{rgb}{0.03, 0.91, 0.87}
\definecolor{brightube}{rgb}{0.82, 0.62, 0.91}
\definecolor{brilliantlavender}{rgb}{0.96, 0.73, 1.0}
\definecolor{brilliantrose}{rgb}{1.0, 0.33, 0.64}
\definecolor{brinkpink}{rgb}{0.98, 0.38, 0.5}
\definecolor{britishracinggreen}{rgb}{0.0, 0.26, 0.15}
\definecolor{bronze}{rgb}{0.8, 0.5, 0.2}
\definecolor{brown(traditional)}{rgb}{0.59, 0.29, 0.0}
\definecolor{brown(web)}{rgb}{0.65, 0.16, 0.16}
\definecolor{bubblegum}{rgb}{0.99, 0.76, 0.8}
\definecolor{bubbles}{rgb}{0.91, 1.0, 1.0}
\definecolor{buff}{rgb}{0.94, 0.86, 0.51}
\definecolor{bulgarianrose}{rgb}{0.28, 0.02, 0.03}
\definecolor{burgundy}{rgb}{0.5, 0.0, 0.13}
\definecolor{burlywood}{rgb}{0.87, 0.72, 0.53}
\definecolor{burntorange}{rgb}{0.8, 0.33, 0.0}
\definecolor{burntsienna}{rgb}{0.91, 0.45, 0.32}
\definecolor{burntumber}{rgb}{0.54, 0.2, 0.14}
\definecolor{byzantine}{rgb}{0.74, 0.2, 0.64}
\definecolor{byzantium}{rgb}{0.44, 0.16, 0.39}
\definecolor{cadet}{rgb}{0.33, 0.41, 0.47}
\definecolor{cadetblue}{rgb}{0.37, 0.62, 0.63}
\definecolor{cadetgrey}{rgb}{0.57, 0.64, 0.69}
\definecolor{cadmiumgreen}{rgb}{0.0, 0.42, 0.24}
\definecolor{cadmiumorange}{rgb}{0.93, 0.53, 0.18}
\definecolor{cadmiumred}{rgb}{0.89, 0.0, 0.13}
\definecolor{cadmiumyellow}{rgb}{1.0, 0.96, 0.0}
\definecolor{calpolypomonagreen}{rgb}{0.12, 0.3, 0.17}
\definecolor{cambridgeblue}{rgb}{0.64, 0.76, 0.68}
\definecolor{camel}{rgb}{0.76, 0.6, 0.42}
\definecolor{camouflagegreen}{rgb}{0.47, 0.53, 0.42}
\definecolor{canaryyellow}{rgb}{1.0, 0.94, 0.0}
\definecolor{candyapplered}{rgb}{1.0, 0.03, 0.0}
\definecolor{candypink}{rgb}{0.89, 0.44, 0.48}
\definecolor{capri}{rgb}{0.0, 0.75, 1.0}
\definecolor{caputmortuum}{rgb}{0.35, 0.15, 0.13}
\definecolor{cardinal}{rgb}{0.77, 0.12, 0.23}
\definecolor{caribbeangreen}{rgb}{0.0, 0.8, 0.6}
\definecolor{carmine}{rgb}{0.59, 0.0, 0.09}
\definecolor{carminepink}{rgb}{0.92, 0.3, 0.26}
\definecolor{carminered}{rgb}{1.0, 0.0, 0.22}
\definecolor{carnationpink}{rgb}{1.0, 0.65, 0.79}
\definecolor{carnelian}{rgb}{0.7, 0.11, 0.11}
\definecolor{carolinablue}{rgb}{0.6, 0.73, 0.89}
\definecolor{carrotorange}{rgb}{0.93, 0.57, 0.13}
\definecolor{ceil}{rgb}{0.57, 0.63, 0.81}
\definecolor{celadon}{rgb}{0.67, 0.88, 0.69}
\definecolor{celestialblue}{rgb}{0.29, 0.59, 0.82}
\definecolor{cerise}{rgb}{0.87, 0.19, 0.39}
\definecolor{cerisepink}{rgb}{0.93, 0.23, 0.51}
\definecolor{cerulean}{rgb}{0.0, 0.48, 0.65}
\definecolor{ceruleanblue}{rgb}{0.16, 0.32, 0.75}
\definecolor{chamoisee}{rgb}{0.63, 0.47, 0.35}
\definecolor{champagne}{rgb}{0.97, 0.91, 0.81}
\definecolor{charcoal}{rgb}{0.21, 0.27, 0.31}
\definecolor{chartreuse(traditional)}{rgb}{0.87, 1.0, 0.0}
\definecolor{chartreuse(web)}{rgb}{0.5, 1.0, 0.0}
\definecolor{cherryblossompink}{rgb}{1.0, 0.72, 0.77}
\definecolor{chestnut}{rgb}{0.8, 0.36, 0.36}
\definecolor{chocolate(traditional)}{rgb}{0.48, 0.25, 0.0}
\definecolor{chocolate(web)}{rgb}{0.82, 0.41, 0.12}
\definecolor{chromeyellow}{rgb}{1.0, 0.65, 0.0}
\definecolor{cinereous}{rgb}{0.6, 0.51, 0.48}
\definecolor{cinnabar}{rgb}{0.89, 0.26, 0.2}
\definecolor{cinnamon}{rgb}{0.82, 0.41, 0.12}
\definecolor{citrine}{rgb}{0.89, 0.82, 0.04}
\definecolor{classicrose}{rgb}{0.98, 0.8, 0.91}
\definecolor{cobalt}{rgb}{0.0, 0.28, 0.67}
\definecolor{cocoabrown}{rgb}{0.82, 0.41, 0.12}
\definecolor{columbiablue}{rgb}{0.61, 0.87, 1.0}
\definecolor{coolblack}{rgb}{0.0, 0.18, 0.39}
\definecolor{coolgrey}{rgb}{0.55, 0.57, 0.67}
\definecolor{copper}{rgb}{0.72, 0.45, 0.2}
\definecolor{copperrose}{rgb}{0.6, 0.4, 0.4}
\definecolor{coquelicot}{rgb}{1.0, 0.22, 0.0}
\definecolor{coral}{rgb}{1.0, 0.5, 0.31}
\definecolor{coralpink}{rgb}{0.97, 0.51, 0.47}
\definecolor{coralred}{rgb}{1.0, 0.25, 0.25}
\definecolor{cordovan}{rgb}{0.54, 0.25, 0.27}
\definecolor{corn}{rgb}{0.98, 0.93, 0.36}
\definecolor{cornellred}{rgb}{0.7, 0.11, 0.11}
\definecolor{cornflowerblue}{rgb}{0.39, 0.58, 0.93}
\definecolor{cornsilk}{rgb}{1.0, 0.97, 0.86}
\definecolor{cosmiclatte}{rgb}{1.0, 0.97, 0.91}
\definecolor{cottoncandy}{rgb}{1.0, 0.74, 0.85}
\definecolor{cream}{rgb}{1.0, 0.99, 0.82}
\definecolor{crimson}{rgb}{0.86, 0.08, 0.24}
\definecolor{crimsonglory}{rgb}{0.75, 0.0, 0.2}
\definecolor{cyan}{rgb}{0.0, 1.0, 1.0}
\definecolor{cyan(process)}{rgb}{0.0, 0.72, 0.92}
\definecolor{daffodil}{rgb}{1.0, 1.0, 0.19}
\definecolor{dandelion}{rgb}{0.94, 0.88, 0.19}
\definecolor{darkblue}{rgb}{0.0, 0.0, 0.55}
\definecolor{darkbrown}{rgb}{0.4, 0.26, 0.13}
\definecolor{darkbyzantium}{rgb}{0.36, 0.22, 0.33}
\definecolor{darkcandyapplered}{rgb}{0.64, 0.0, 0.0}
\definecolor{darkcerulean}{rgb}{0.03, 0.27, 0.49}
\definecolor{darkchampagne}{rgb}{0.76, 0.7, 0.5}
\definecolor{darkchestnut}{rgb}{0.6, 0.41, 0.38}
\definecolor{darkcoral}{rgb}{0.8, 0.36, 0.27}
\definecolor{darkcyan}{rgb}{0.0, 0.55, 0.55}
\definecolor{darkelectricblue}{rgb}{0.33, 0.41, 0.47}
\definecolor{darkgoldenrod}{rgb}{0.72, 0.53, 0.04}
\definecolor{darkgray}{rgb}{0.66, 0.66, 0.66}
\definecolor{darkgreen}{rgb}{0.0, 0.2, 0.13}
\definecolor{darkjunglegreen}{rgb}{0.1, 0.14, 0.13}
\definecolor{darkkhaki}{rgb}{0.74, 0.72, 0.42}
\definecolor{darklava}{rgb}{0.28, 0.24, 0.2}
\definecolor{darklavender}{rgb}{0.45, 0.31, 0.59}
\definecolor{darkmagenta}{rgb}{0.55, 0.0, 0.55}
\definecolor{darkmidnightblue}{rgb}{0.0, 0.2, 0.4}
\definecolor{darkolivegreen}{rgb}{0.33, 0.42, 0.18}
\definecolor{darkorange}{rgb}{1.0, 0.55, 0.0}
\definecolor{darkorchid}{rgb}{0.6, 0.2, 0.8}
\definecolor{darkpastelblue}{rgb}{0.47, 0.62, 0.8}
\definecolor{darkpastelgreen}{rgb}{0.01, 0.75, 0.24}
\definecolor{darkpastelpurple}{rgb}{0.59, 0.44, 0.84}
\definecolor{darkpastelred}{rgb}{0.76, 0.23, 0.13}
\definecolor{darkpink}{rgb}{0.91, 0.33, 0.5}
\definecolor{darkpowderblue}{rgb}{0.0, 0.2, 0.6}
\definecolor{darkraspberry}{rgb}{0.53, 0.15, 0.34}
\definecolor{darkred}{rgb}{0.55, 0.0, 0.0}
\definecolor{darksalmon}{rgb}{0.91, 0.59, 0.48}
\definecolor{darkscarlet}{rgb}{0.34, 0.01, 0.1}
\definecolor{darkseagreen}{rgb}{0.56, 0.74, 0.56}
\definecolor{darksienna}{rgb}{0.24, 0.08, 0.08}
\definecolor{darkslateblue}{rgb}{0.28, 0.24, 0.55}
\definecolor{darkslategray}{rgb}{0.18, 0.31, 0.31}
\definecolor{darkspringgreen}{rgb}{0.09, 0.45, 0.27}
\definecolor{darktan}{rgb}{0.57, 0.51, 0.32}
\definecolor{darktangerine}{rgb}{1.0, 0.66, 0.07}
\definecolor{darktaupe}{rgb}{0.28, 0.24, 0.2}
\definecolor{darkterracotta}{rgb}{0.8, 0.31, 0.36}
\definecolor{darkturquoise}{rgb}{0.0, 0.81, 0.82}
\definecolor{darkviolet}{rgb}{0.58, 0.0, 0.83}
\definecolor{dartmouthgreen}{rgb}{0.05, 0.5, 0.06}
\definecolor{davy\'sgrey}{rgb}{0.33, 0.33, 0.33}
\definecolor{debianred}{rgb}{0.84, 0.04, 0.33}
\definecolor{deepcarmine}{rgb}{0.66, 0.13, 0.24}
\definecolor{deepcarminepink}{rgb}{0.94, 0.19, 0.22}
\definecolor{deepcarrotorange}{rgb}{0.91, 0.41, 0.17}
\definecolor{deepcerise}{rgb}{0.85, 0.2, 0.53}
\definecolor{deepchampagne}{rgb}{0.98, 0.84, 0.65}
\definecolor{deepchestnut}{rgb}{0.73, 0.31, 0.28}
\definecolor{deepfuchsia}{rgb}{0.76, 0.33, 0.76}
\definecolor{deepjunglegreen}{rgb}{0.0, 0.29, 0.29}
\definecolor{deeplilac}{rgb}{0.6, 0.33, 0.73}
\definecolor{deepmagenta}{rgb}{0.8, 0.0, 0.8}
\definecolor{deeppeach}{rgb}{1.0, 0.8, 0.64}
\definecolor{deeppink}{rgb}{1.0, 0.08, 0.58}
\definecolor{deepsaffron}{rgb}{1.0, 0.6, 0.2}
\definecolor{deepskyblue}{rgb}{0.0, 0.75, 1.0}
\definecolor{denim}{rgb}{0.08, 0.38, 0.74}
\definecolor{desert}{rgb}{0.76, 0.6, 0.42}
\definecolor{desertsand}{rgb}{0.93, 0.79, 0.69}
\definecolor{dimgray}{rgb}{0.41, 0.41, 0.41}
\definecolor{dodgerblue}{rgb}{0.12, 0.56, 1.0}
\definecolor{dogwoodrose}{rgb}{0.84, 0.09, 0.41}
\definecolor{dollarbill}{rgb}{0.52, 0.73, 0.4}
\definecolor{drab}{rgb}{0.59, 0.44, 0.09}
\definecolor{dukeblue}{rgb}{0.0, 0.0, 0.61}
\definecolor{earthyellow}{rgb}{0.88, 0.66, 0.37}
\definecolor{ecru}{rgb}{0.76, 0.7, 0.5}
\definecolor{eggplant}{rgb}{0.38, 0.25, 0.32}
\definecolor{eggshell}{rgb}{0.94, 0.92, 0.84}
\definecolor{egyptianblue}{rgb}{0.06, 0.2, 0.65}
\definecolor{electricblue}{rgb}{0.49, 0.98, 1.0}
\definecolor{electriccrimson}{rgb}{1.0, 0.0, 0.25}
\definecolor{electriccyan}{rgb}{0.0, 1.0, 1.0}
\definecolor{electricgreen}{rgb}{0.0, 1.0, 0.0}
\definecolor{electricindigo}{rgb}{0.44, 0.0, 1.0}
\definecolor{electriclavender}{rgb}{0.96, 0.73, 1.0}
\definecolor{electriclime}{rgb}{0.8, 1.0, 0.0}
\definecolor{electricpurple}{rgb}{0.75, 0.0, 1.0}
\definecolor{electricultramarine}{rgb}{0.25, 0.0, 1.0}
\definecolor{electricviolet}{rgb}{0.56, 0.0, 1.0}
\definecolor{electricyellow}{rgb}{1.0, 1.0, 0.0}
\definecolor{emerald}{rgb}{0.31, 0.78, 0.47}
\definecolor{etonblue}{rgb}{0.59, 0.78, 0.64}
\definecolor{fallow}{rgb}{0.76, 0.6, 0.42}
\definecolor{falured}{rgb}{0.5, 0.09, 0.09}
\definecolor{fandango}{rgb}{0.71, 0.2, 0.54}
\definecolor{fashionfuchsia}{rgb}{0.96, 0.0, 0.63}
\definecolor{fawn}{rgb}{0.9, 0.67, 0.44}
\definecolor{feldgrau}{rgb}{0.3, 0.36, 0.33}
\definecolor{ferngreen}{rgb}{0.31, 0.47, 0.26}
\definecolor{ferrarired}{rgb}{1.0, 0.11, 0.0}
\definecolor{fielddrab}{rgb}{0.42, 0.33, 0.12}
\definecolor{firebrick}{rgb}{0.7, 0.13, 0.13}
\definecolor{fireenginered}{rgb}{0.81, 0.09, 0.13}
\definecolor{flame}{rgb}{0.89, 0.35, 0.13}
\definecolor{flamingopink}{rgb}{0.99, 0.56, 0.67}
\definecolor{flavescent}{rgb}{0.97, 0.91, 0.56}
\definecolor{flax}{rgb}{0.93, 0.86, 0.51}
\definecolor{floralwhite}{rgb}{1.0, 0.98, 0.94}
\definecolor{fluorescentorange}{rgb}{1.0, 0.75, 0.0}
\definecolor{fluorescentpink}{rgb}{1.0, 0.08, 0.58}
\definecolor{fluorescentyellow}{rgb}{0.8, 1.0, 0.0}
\definecolor{folly}{rgb}{1.0, 0.0, 0.31}
\definecolor{forestgreen(traditional)}{rgb}{0.0, 0.27, 0.13}
\definecolor{forestgreen(web)}{rgb}{0.13, 0.55, 0.13}
\definecolor{frenchbeige}{rgb}{0.65, 0.48, 0.36}
\definecolor{frenchblue}{rgb}{0.0, 0.45, 0.73}
\definecolor{frenchlilac}{rgb}{0.53, 0.38, 0.56}
\definecolor{frenchrose}{rgb}{0.96, 0.29, 0.54}
\definecolor{fuchsia}{rgb}{1.0, 0.0, 1.0}
\definecolor{fuchsiapink}{rgb}{1.0, 0.47, 1.0}
\definecolor{fulvous}{rgb}{0.86, 0.52, 0.0}
\definecolor{fuzzywuzzy}{rgb}{0.8, 0.4, 0.4}
\definecolor{gainsboro}{rgb}{0.86, 0.86, 0.86}
\definecolor{gamboge}{rgb}{0.89, 0.61, 0.06}
\definecolor{ghostwhite}{rgb}{0.97, 0.97, 1.0}
\definecolor{ginger}{rgb}{0.69, 0.4, 0.0}
\definecolor{glaucous}{rgb}{0.38, 0.51, 0.71}
\definecolor{gold(metallic)}{rgb}{0.83, 0.69, 0.22}
\definecolor{gold(web)(golden)}{rgb}{1.0, 0.84, 0.0}
\definecolor{goldenbrown}{rgb}{0.6, 0.4, 0.08}
\definecolor{goldenpoppy}{rgb}{0.99, 0.76, 0.0}
\definecolor{goldenyellow}{rgb}{1.0, 0.87, 0.0}
\definecolor{goldenrod}{rgb}{0.85, 0.65, 0.13}
\definecolor{grannysmithapple}{rgb}{0.66, 0.89, 0.63}
\definecolor{gray}{rgb}{0.5, 0.5, 0.5}
\definecolor{gray(html/cssgray)}{rgb}{0.5, 0.5, 0.5}
\definecolor{gray(x11gray)}{rgb}{0.75, 0.75, 0.75}
\definecolor{gray-asparagus}{rgb}{0.27, 0.35, 0.27}
\definecolor{green(colorwheel)(x11green)}{rgb}{0.0, 1.0, 0.0}
\definecolor{green(html/cssgreen)}{rgb}{0.0, 0.5, 0.0}
\definecolor{green(munsell)}{rgb}{0.0, 0.66, 0.47}
\definecolor{green(ncs)}{rgb}{0.0, 0.62, 0.42}
\definecolor{green(pigment)}{rgb}{0.0, 0.65, 0.31}
\definecolor{green(ryb)}{rgb}{0.4, 0.69, 0.2}
\definecolor{green-yellow}{rgb}{0.68, 1.0, 0.18}
\definecolor{grullo}{rgb}{0.66, 0.6, 0.53}
\definecolor{guppiegreen}{rgb}{0.0, 1.0, 0.5}
\definecolor{halayaube}{rgb}{0.4, 0.22, 0.33}
\definecolor{hanblue}{rgb}{0.27, 0.42, 0.81}
\definecolor{hanpurple}{rgb}{0.32, 0.09, 0.98}
\definecolor{hansayellow}{rgb}{0.91, 0.84, 0.42}
\definecolor{harlequin}{rgb}{0.25, 1.0, 0.0}
\definecolor{harvardcrimson}{rgb}{0.79, 0.0, 0.09}
\definecolor{harvestgold}{rgb}{0.85, 0.57, 0.0}
\definecolor{heartgold}{rgb}{0.5, 0.5, 0.0}
\definecolor{heliotrope}{rgb}{0.87, 0.45, 1.0}
\definecolor{hollywoodcerise}{rgb}{0.96, 0.0, 0.63}
\definecolor{honeydew}{rgb}{0.94, 1.0, 0.94}
\definecolor{hooker\'sgreen}{rgb}{0.0, 0.44, 0.0}
\definecolor{hotmagenta}{rgb}{1.0, 0.11, 0.81}
\definecolor{hotpink}{rgb}{1.0, 0.41, 0.71}
\definecolor{huntergreen}{rgb}{0.21, 0.37, 0.23}
\definecolor{iceberg}{rgb}{0.44, 0.65, 0.82}
\definecolor{icterine}{rgb}{0.99, 0.97, 0.37}
\definecolor{inchworm}{rgb}{0.7, 0.93, 0.36}
\definecolor{indiagreen}{rgb}{0.07, 0.53, 0.03}
\definecolor{indianred}{rgb}{0.8, 0.36, 0.36}
\definecolor{indianyellow}{rgb}{0.89, 0.66, 0.34}
\definecolor{indigo(dye)}{rgb}{0.0, 0.25, 0.42}
\definecolor{indigo(web)}{rgb}{0.29, 0.0, 0.51}
\definecolor{internationalkleinblue}{rgb}{0.0, 0.18, 0.65}
\definecolor{internationalorange}{rgb}{1.0, 0.31, 0.0}
\definecolor{iris}{rgb}{0.35, 0.31, 0.81}
\definecolor{isabelline}{rgb}{0.96, 0.94, 0.93}
\definecolor{islamicgreen}{rgb}{0.0, 0.56, 0.0}
\definecolor{ivory}{rgb}{1.0, 1.0, 0.94}
\definecolor{jade}{rgb}{0.0, 0.66, 0.42}
\definecolor{jasper}{rgb}{0.84, 0.23, 0.24}
\definecolor{jazzberryjam}{rgb}{0.65, 0.04, 0.37}
\definecolor{jonquil}{rgb}{0.98, 0.85, 0.37}
\definecolor{junebud}{rgb}{0.74, 0.85, 0.34}
\definecolor{junglegreen}{rgb}{0.16, 0.67, 0.53}
\definecolor{kellygreen}{rgb}{0.3, 0.73, 0.09}
\definecolor{khaki(html/css)(khaki)}{rgb}{0.76, 0.69, 0.57}
\definecolor{khaki(x11)(lightkhaki)}{rgb}{0.94, 0.9, 0.55}
\definecolor{lasallegreen}{rgb}{0.03, 0.47, 0.19}
\definecolor{languidlavender}{rgb}{0.84, 0.79, 0.87}
\definecolor{lapislazuli}{rgb}{0.15, 0.38, 0.61}
\definecolor{laserlemon}{rgb}{1.0, 1.0, 0.13}
\definecolor{lava}{rgb}{0.81, 0.06, 0.13}
\definecolor{lavender(floral)}{rgb}{0.71, 0.49, 0.86}
\definecolor{lavender(web)}{rgb}{0.9, 0.9, 0.98}
\definecolor{lavenderblue}{rgb}{0.8, 0.8, 1.0}
\definecolor{lavenderblush}{rgb}{1.0, 0.94, 0.96}
\definecolor{lavendergray}{rgb}{0.77, 0.76, 0.82}
\definecolor{lavenderindigo}{rgb}{0.58, 0.34, 0.92}
\definecolor{lavendermagenta}{rgb}{0.93, 0.51, 0.93}
\definecolor{lavendermist}{rgb}{0.9, 0.9, 0.98}
\definecolor{lavenderpink}{rgb}{0.98, 0.68, 0.82}
\definecolor{lavenderpurple}{rgb}{0.59, 0.48, 0.71}
\definecolor{lavenderrose}{rgb}{0.98, 0.63, 0.89}
\definecolor{lawngreen}{rgb}{0.49, 0.99, 0.0}
\definecolor{lemon}{rgb}{1.0, 0.97, 0.0}
\definecolor{lemonchiffon}{rgb}{1.0, 0.98, 0.8}
\definecolor{lightapricot}{rgb}{0.99, 0.84, 0.69}
\definecolor{lightblue}{rgb}{0.68, 0.85, 0.9}
\definecolor{lightbrown}{rgb}{0.71, 0.4, 0.11}
\definecolor{lightcarminepink}{rgb}{0.9, 0.4, 0.38}
\definecolor{lightcoral}{rgb}{0.94, 0.5, 0.5}
\definecolor{lightcornflowerblue}{rgb}{0.6, 0.81, 0.93}
\definecolor{lightcyan}{rgb}{0.88, 1.0, 1.0}
\definecolor{lightfuchsiapink}{rgb}{0.98, 0.52, 0.9}
\definecolor{lightgoldenrodyellow}{rgb}{0.98, 0.98, 0.82}
\definecolor{lightgray}{rgb}{0.83, 0.83, 0.83}
\definecolor{lightgreen}{rgb}{0.56, 0.93, 0.56}
\definecolor{lightkhaki}{rgb}{0.94, 0.9, 0.55}
\definecolor{lightmauve}{rgb}{0.86, 0.82, 1.0}
\definecolor{lightpastelpurple}{rgb}{0.69, 0.61, 0.85}
\definecolor{lightpink}{rgb}{1.0, 0.71, 0.76}
\definecolor{lightsalmon}{rgb}{1.0, 0.63, 0.48}
\definecolor{lightsalmonpink}{rgb}{1.0, 0.6, 0.6}
\definecolor{lightseagreen}{rgb}{0.13, 0.7, 0.67}
\definecolor{lightskyblue}{rgb}{0.53, 0.81, 0.98}
\definecolor{lightslategray}{rgb}{0.47, 0.53, 0.6}
\definecolor{lighttaupe}{rgb}{0.7, 0.55, 0.43}
\definecolor{lightthulianpink}{rgb}{0.9, 0.56, 0.67}
\definecolor{lightyellow}{rgb}{1.0, 1.0, 0.88}
\definecolor{lilac}{rgb}{0.78, 0.64, 0.78}
\definecolor{lime(colorwheel)}{rgb}{0.75, 1.0, 0.0}
\definecolor{lime(web)(x11green)}{rgb}{0.0, 1.0, 0.0}
\definecolor{limegreen}{rgb}{0.2, 0.8, 0.2}
\definecolor{lincolngreen}{rgb}{0.11, 0.35, 0.02}
\definecolor{linen}{rgb}{0.98, 0.94, 0.9}
\definecolor{liver}{rgb}{0.33, 0.29, 0.31}
\definecolor{lust}{rgb}{0.9, 0.13, 0.13}
\definecolor{macaroniandcheese}{rgb}{1.0, 0.74, 0.53}
\definecolor{magenta}{rgb}{1.0, 0.0, 1.0}
\definecolor{magenta(dye)}{rgb}{0.79, 0.08, 0.48}
\definecolor{magenta(process)}{rgb}{1.0, 0.0, 0.56}
\definecolor{magicmint}{rgb}{0.67, 0.94, 0.82}
\definecolor{magnolia}{rgb}{0.97, 0.96, 1.0}
\definecolor{mahogany}{rgb}{0.75, 0.25, 0.0}
\definecolor{maize}{rgb}{0.98, 0.93, 0.37}
\definecolor{majorelleblue}{rgb}{0.38, 0.31, 0.86}
\definecolor{malachite}{rgb}{0.04, 0.85, 0.32}
\definecolor{manatee}{rgb}{0.59, 0.6, 0.67}
\definecolor{mangotango}{rgb}{1.0, 0.51, 0.26}
\definecolor{maroon(html/css)}{rgb}{0.5, 0.0, 0.0}
\definecolor{maroon(x11)}{rgb}{0.69, 0.19, 0.38}
\definecolor{mauve}{rgb}{0.88, 0.69, 1.0}
\definecolor{mauvetaupe}{rgb}{0.57, 0.37, 0.43}
\definecolor{mauvelous}{rgb}{0.94, 0.6, 0.67}
\definecolor{mayablue}{rgb}{0.45, 0.76, 0.98}
\definecolor{meatbrown}{rgb}{0.9, 0.72, 0.23}
\definecolor{mediumaquamarine}{rgb}{0.4, 0.8, 0.67}
\definecolor{mediumblue}{rgb}{0.0, 0.0, 0.8}
\definecolor{mediumcandyapplered}{rgb}{0.89, 0.02, 0.17}
\definecolor{mediumcarmine}{rgb}{0.69, 0.25, 0.21}
\definecolor{mediumchampagne}{rgb}{0.95, 0.9, 0.67}
\definecolor{mediumelectricblue}{rgb}{0.01, 0.31, 0.59}
\definecolor{mediumjunglegreen}{rgb}{0.11, 0.21, 0.18}
\definecolor{mediumlavendermagenta}{rgb}{0.8, 0.6, 0.8}
\definecolor{mediumorchid}{rgb}{0.73, 0.33, 0.83}
\definecolor{mediumpersianblue}{rgb}{0.0, 0.4, 0.65}
\definecolor{mediumpurple}{rgb}{0.58, 0.44, 0.86}
\definecolor{mediumred-violet}{rgb}{0.73, 0.2, 0.52}
\definecolor{mediumseagreen}{rgb}{0.24, 0.7, 0.44}
\definecolor{mediumslateblue}{rgb}{0.48, 0.41, 0.93}
\definecolor{mediumspringbud}{rgb}{0.79, 0.86, 0.54}
\definecolor{mediumspringgreen}{rgb}{0.0, 0.98, 0.6}
\definecolor{mediumtaupe}{rgb}{0.4, 0.3, 0.28}
\definecolor{mediumtealblue}{rgb}{0.0, 0.33, 0.71}
\definecolor{mediumturquoise}{rgb}{0.28, 0.82, 0.8}
\definecolor{mediumviolet-red}{rgb}{0.78, 0.08, 0.52}
\definecolor{melon}{rgb}{0.99, 0.74, 0.71}
\definecolor{midnightblue}{rgb}{0.1, 0.1, 0.44}
\definecolor{midnightgreen(eaglegreen)}{rgb}{0.0, 0.29, 0.33}
\definecolor{mikadoyellow}{rgb}{1.0, 0.77, 0.05}
\definecolor{mint}{rgb}{0.24, 0.71, 0.54}
\definecolor{mintcream}{rgb}{0.96, 1.0, 0.98}
\definecolor{mintgreen}{rgb}{0.6, 1.0, 0.6}
\definecolor{mistyrose}{rgb}{1.0, 0.89, 0.88}
\definecolor{moccasin}{rgb}{0.98, 0.92, 0.84}
\definecolor{modebeige}{rgb}{0.59, 0.44, 0.09}
\definecolor{moonstoneblue}{rgb}{0.45, 0.66, 0.76}
\definecolor{mordantred19}{rgb}{0.68, 0.05, 0.0}
\definecolor{mossgreen}{rgb}{0.68, 0.87, 0.68}
\definecolor{mountainmeadow}{rgb}{0.19, 0.73, 0.56}
\definecolor{mountbattenpink}{rgb}{0.6, 0.48, 0.55}
\definecolor{mulberry}{rgb}{0.77, 0.29, 0.55}
\definecolor{mustard}{rgb}{1.0, 0.86, 0.35}
\definecolor{myrtle}{rgb}{0.13, 0.26, 0.12}
\definecolor{msugreen}{rgb}{0.09, 0.27, 0.23}
\definecolor{nadeshikopink}{rgb}{0.96, 0.68, 0.78}
\definecolor{napiergreen}{rgb}{0.16, 0.5, 0.0}
\definecolor{naplesyellow}{rgb}{0.98, 0.85, 0.37}
\definecolor{navajowhite}{rgb}{1.0, 0.87, 0.68}
\definecolor{navyblue}{rgb}{0.0, 0.0, 0.5}
\definecolor{neoncarrot}{rgb}{1.0, 0.64, 0.26}
\definecolor{neonfuchsia}{rgb}{1.0, 0.25, 0.39}
\definecolor{neongreen}{rgb}{0.22, 0.88, 0.08}
\definecolor{non-photoblue}{rgb}{0.64, 0.87, 0.93}
\definecolor{oceanboatblue}{rgb}{0.0, 0.47, 0.75}
\definecolor{ochre}{rgb}{0.8, 0.47, 0.13}
\definecolor{officegreen}{rgb}{0.0, 0.5, 0.0}
\definecolor{oldgold}{rgb}{0.81, 0.71, 0.23}
\definecolor{oldlace}{rgb}{0.99, 0.96, 0.9}
\definecolor{oldlavender}{rgb}{0.47, 0.41, 0.47}
\definecolor{oldmauve}{rgb}{0.4, 0.19, 0.28}
\definecolor{oldrose}{rgb}{0.75, 0.5, 0.51}
\definecolor{olive}{rgb}{0.5, 0.5, 0.0}
\definecolor{olivedrab(web)(olivedrab3)}{rgb}{0.42, 0.56, 0.14}
\definecolor{olivedrab7}{rgb}{0.24, 0.2, 0.12}
\definecolor{olivine}{rgb}{0.6, 0.73, 0.45}
\definecolor{onyx}{rgb}{0.06, 0.06, 0.06}
\definecolor{operamauve}{rgb}{0.72, 0.52, 0.65}
\definecolor{orange(colorwheel)}{rgb}{1.0, 0.5, 0.0}
\definecolor{orange(ryb)}{rgb}{0.98, 0.6, 0.01}
\definecolor{orange(webcolor)}{rgb}{1.0, 0.65, 0.0}
\definecolor{orangepeel}{rgb}{1.0, 0.62, 0.0}
\definecolor{orange-red}{rgb}{1.0, 0.27, 0.0}
\definecolor{orchid}{rgb}{0.85, 0.44, 0.84}
\definecolor{otterbrown}{rgb}{0.4, 0.26, 0.13}
\definecolor{outerspace}{rgb}{0.25, 0.29, 0.3}
\definecolor{outrageousorange}{rgb}{1.0, 0.43, 0.29}
\definecolor{oxfordblue}{rgb}{0.0, 0.13, 0.28}
\definecolor{oucrimsonred}{rgb}{0.6, 0.0, 0.0}
\definecolor{pakistangreen}{rgb}{0.0, 0.4, 0.0}
\definecolor{palatinateblue}{rgb}{0.15, 0.23, 0.89}
\definecolor{palatinatepurple}{rgb}{0.41, 0.16, 0.38}
\definecolor{paleaqua}{rgb}{0.74, 0.83, 0.9}
\definecolor{paleblue}{rgb}{0.69, 0.93, 0.93}
\definecolor{palebrown}{rgb}{0.6, 0.46, 0.33}
\definecolor{palecarmine}{rgb}{0.69, 0.25, 0.21}
\definecolor{palecerulean}{rgb}{0.61, 0.77, 0.89}
\definecolor{palechestnut}{rgb}{0.87, 0.68, 0.69}
\definecolor{palecopper}{rgb}{0.85, 0.54, 0.4}
\definecolor{palecornflowerblue}{rgb}{0.67, 0.8, 0.94}
\definecolor{palegold}{rgb}{0.9, 0.75, 0.54}
\definecolor{palegoldenrod}{rgb}{0.93, 0.91, 0.67}
\definecolor{palegreen}{rgb}{0.6, 0.98, 0.6}
\definecolor{palemagenta}{rgb}{0.98, 0.52, 0.9}
\definecolor{palepink}{rgb}{0.98, 0.85, 0.87}
\definecolor{paleplum}{rgb}{0.8, 0.6, 0.8}
\definecolor{palered-violet}{rgb}{0.86, 0.44, 0.58}
\definecolor{palerobineggblue}{rgb}{0.59, 0.87, 0.82}
\definecolor{palesilver}{rgb}{0.79, 0.75, 0.73}
\definecolor{palespringbud}{rgb}{0.93, 0.92, 0.74}
\definecolor{paletaupe}{rgb}{0.74, 0.6, 0.49}
\definecolor{paleviolet-red}{rgb}{0.86, 0.44, 0.58}
\definecolor{pansypurple}{rgb}{0.47, 0.09, 0.29}
\definecolor{papayawhip}{rgb}{1.0, 0.94, 0.84}
\definecolor{parisgreen}{rgb}{0.31, 0.78, 0.47}
\definecolor{pastelblue}{rgb}{0.68, 0.78, 0.81}
\definecolor{pastelbrown}{rgb}{0.51, 0.41, 0.33}
\definecolor{pastelgray}{rgb}{0.81, 0.81, 0.77}
\definecolor{pastelgreen}{rgb}{0.47, 0.87, 0.47}
\definecolor{pastelmagenta}{rgb}{0.96, 0.6, 0.76}
\definecolor{pastelorange}{rgb}{1.0, 0.7, 0.28}
\definecolor{pastelpink}{rgb}{1.0, 0.82, 0.86}
\definecolor{pastelpurple}{rgb}{0.7, 0.62, 0.71}
\definecolor{pastelred}{rgb}{1.0, 0.41, 0.38}
\definecolor{pastelviolet}{rgb}{0.8, 0.6, 0.79}
\definecolor{pastelyellow}{rgb}{0.99, 0.99, 0.59}
\definecolor{patriarch}{rgb}{0.5, 0.0, 0.5}
\definecolor{payne\'sgrey}{rgb}{0.25, 0.25, 0.28}
\definecolor{peach}{rgb}{1.0, 0.9, 0.71}
\definecolor{peach-orange}{rgb}{1.0, 0.8, 0.6}
\definecolor{peachpuff}{rgb}{1.0, 0.85, 0.73}
\definecolor{peach-yellow}{rgb}{0.98, 0.87, 0.68}
\definecolor{pear}{rgb}{0.82, 0.89, 0.19}
\definecolor{pearl}{rgb}{0.94, 0.92, 0.84}
\definecolor{peridot}{rgb}{0.9, 0.89, 0.0}
\definecolor{periwinkle}{rgb}{0.8, 0.8, 1.0}
\definecolor{persianblue}{rgb}{0.11, 0.22, 0.73}
\definecolor{persiangreen}{rgb}{0.0, 0.65, 0.58}
\definecolor{persianindigo}{rgb}{0.2, 0.07, 0.48}
\definecolor{persianorange}{rgb}{0.85, 0.56, 0.35}
\definecolor{peru}{rgb}{0.8, 0.52, 0.25}
\definecolor{persianpink}{rgb}{0.97, 0.5, 0.75}
\definecolor{persianplum}{rgb}{0.44, 0.11, 0.11}
\definecolor{persianred}{rgb}{0.8, 0.2, 0.2}
\definecolor{persianrose}{rgb}{1.0, 0.16, 0.64}
\definecolor{persimmon}{rgb}{0.93, 0.35, 0.0}
\definecolor{phlox}{rgb}{0.87, 0.0, 1.0}
\definecolor{phthaloblue}{rgb}{0.0, 0.06, 0.54}
\definecolor{phthalogreen}{rgb}{0.07, 0.21, 0.14}
\definecolor{piggypink}{rgb}{0.99, 0.87, 0.9}
\definecolor{pinegreen}{rgb}{0.0, 0.47, 0.44}
\definecolor{pink}{rgb}{1.0, 0.75, 0.8}
\definecolor{pink-orange}{rgb}{1.0, 0.6, 0.4}
\definecolor{pinkpearl}{rgb}{0.91, 0.67, 0.81}
\definecolor{pinksherbet}{rgb}{0.97, 0.56, 0.65}
\definecolor{pistachio}{rgb}{0.58, 0.77, 0.45}
\definecolor{platinum}{rgb}{0.9, 0.89, 0.89}
\definecolor{plum(traditional)}{rgb}{0.56, 0.27, 0.52}
\definecolor{plum(web)}{rgb}{0.8, 0.6, 0.8}
\definecolor{portlandorange}{rgb}{1.0, 0.35, 0.21}
\definecolor{powderblue(web)}{rgb}{0.69, 0.88, 0.9}
\definecolor{princetonorange}{rgb}{1.0, 0.56, 0.0}
\definecolor{prune}{rgb}{0.44, 0.11, 0.11}
\definecolor{prussianblue}{rgb}{0.0, 0.19, 0.33}
\definecolor{psychedelicpurple}{rgb}{0.87, 0.0, 1.0}
\definecolor{puce}{rgb}{0.8, 0.53, 0.6}
\definecolor{pumpkin}{rgb}{1.0, 0.46, 0.09}
\definecolor{purple(html/css)}{rgb}{0.5, 0.0, 0.5}
\definecolor{purple(munsell)}{rgb}{0.62, 0.0, 0.77}
\definecolor{purple(x11)}{rgb}{0.63, 0.36, 0.94}
\definecolor{purpleheart}{rgb}{0.41, 0.21, 0.61}
\definecolor{purplemountainmajesty}{rgb}{0.59, 0.47, 0.71}
\definecolor{purplepizzazz}{rgb}{1.0, 0.31, 0.85}
\definecolor{purpletaupe}{rgb}{0.31, 0.25, 0.3}
\definecolor{radicalred}{rgb}{1.0, 0.21, 0.37}
\definecolor{raspberry}{rgb}{0.89, 0.04, 0.36}
\definecolor{raspberryglace}{rgb}{0.57, 0.37, 0.43}
\definecolor{raspberrypink}{rgb}{0.89, 0.31, 0.61}
\definecolor{raspberryrose}{rgb}{0.7, 0.27, 0.42}
\definecolor{rawumber}{rgb}{0.51, 0.4, 0.27}
\definecolor{razzledazzlerose}{rgb}{1.0, 0.2, 0.8}
\definecolor{razzmatazz}{rgb}{0.89, 0.15, 0.42}
\definecolor{red}{rgb}{1.0, 0.0, 0.0}
\definecolor{red(munsell)}{rgb}{0.95, 0.0, 0.24}
\definecolor{red(ncs)}{rgb}{0.77, 0.01, 0.2}
\definecolor{red(pigment)}{rgb}{0.93, 0.11, 0.14}
\definecolor{red(ryb)}{rgb}{1.0, 0.15, 0.07}
\definecolor{red-brown}{rgb}{0.65, 0.16, 0.16}
\definecolor{red-violet}{rgb}{0.78, 0.08, 0.52}
\definecolor{redwood}{rgb}{0.67, 0.31, 0.32}
\definecolor{regalia}{rgb}{0.32, 0.18, 0.5}
\definecolor{richblack}{rgb}{0.0, 0.25, 0.25}
\definecolor{richbrilliantlavender}{rgb}{0.95, 0.65, 1.0}
\definecolor{richcarmine}{rgb}{0.84, 0.0, 0.25}
\definecolor{richelectricblue}{rgb}{0.03, 0.57, 0.82}
\definecolor{richlavender}{rgb}{0.67, 0.38, 0.8}
\definecolor{richlilac}{rgb}{0.71, 0.4, 0.82}
\definecolor{richmaroon}{rgb}{0.69, 0.19, 0.38}
\definecolor{riflegreen}{rgb}{0.25, 0.28, 0.2}
\definecolor{robineggblue}{rgb}{0.0, 0.8, 0.8}
\definecolor{rose}{rgb}{1.0, 0.0, 0.5}
\definecolor{rosebonbon}{rgb}{0.98, 0.26, 0.62}
\definecolor{roseebony}{rgb}{0.4, 0.3, 0.28}
\definecolor{rosegold}{rgb}{0.72, 0.43, 0.47}
\definecolor{rosemadder}{rgb}{0.89, 0.15, 0.21}
\definecolor{rosepink}{rgb}{1.0, 0.4, 0.8}
\definecolor{rosequartz}{rgb}{0.67, 0.6, 0.66}
\definecolor{rosetaupe}{rgb}{0.56, 0.36, 0.36}
\definecolor{rosevale}{rgb}{0.67, 0.31, 0.32}
\definecolor{rosewood}{rgb}{0.4, 0.0, 0.04}
\definecolor{rossocorsa}{rgb}{0.83, 0.0, 0.0}
\definecolor{rosybrown}{rgb}{0.74, 0.56, 0.56}
\definecolor{royalazure}{rgb}{0.0, 0.22, 0.66}
\definecolor{royalblue(traditional)}{rgb}{0.0, 0.14, 0.4}
\definecolor{royalblue(web)}{rgb}{0.25, 0.41, 0.88}
\definecolor{royalfuchsia}{rgb}{0.79, 0.17, 0.57}
\definecolor{royalpurple}{rgb}{0.47, 0.32, 0.66}
\definecolor{ruby}{rgb}{0.88, 0.07, 0.37}
\definecolor{ruddy}{rgb}{1.0, 0.0, 0.16}
\definecolor{ruddybrown}{rgb}{0.73, 0.4, 0.16}
\definecolor{ruddypink}{rgb}{0.88, 0.56, 0.59}
\definecolor{rufous}{rgb}{0.66, 0.11, 0.03}
\definecolor{russet}{rgb}{0.5, 0.27, 0.11}
\definecolor{rust}{rgb}{0.72, 0.25, 0.05}
\definecolor{sacramentostategreen}{rgb}{0.0, 0.34, 0.25}
\definecolor{saddlebrown}{rgb}{0.55, 0.27, 0.07}
\definecolor{safetyorange(blazeorange)}{rgb}{1.0, 0.4, 0.0}
\definecolor{saffron}{rgb}{0.96, 0.77, 0.19}
\definecolor{st.patrick\'sblue}{rgb}{0.14, 0.16, 0.48}
\definecolor{salmon}{rgb}{1.0, 0.55, 0.41}
\definecolor{salmonpink}{rgb}{1.0, 0.57, 0.64}
\definecolor{sand}{rgb}{0.76, 0.7, 0.5}
\definecolor{sanddune}{rgb}{0.59, 0.44, 0.09}
\definecolor{sandstorm}{rgb}{0.93, 0.84, 0.25}
\definecolor{sandybrown}{rgb}{0.96, 0.64, 0.38}
\definecolor{sandytaupe}{rgb}{0.59, 0.44, 0.09}
\definecolor{sangria}{rgb}{0.57, 0.0, 0.04}
\definecolor{sapgreen}{rgb}{0.31, 0.49, 0.16}
\definecolor{sapphire}{rgb}{0.03, 0.15, 0.4}
\definecolor{satinsheengold}{rgb}{0.8, 0.63, 0.21}
\definecolor{scarlet}{rgb}{1.0, 0.13, 0.0}
\definecolor{schoolbusyellow}{rgb}{1.0, 0.85, 0.0}
\definecolor{screamin\'green}{rgb}{0.46, 1.0, 0.44}
\definecolor{seagreen}{rgb}{0.18, 0.55, 0.34}
\definecolor{sealbrown}{rgb}{0.2, 0.08, 0.08}
\definecolor{seashell}{rgb}{1.0, 0.96, 0.93}
\definecolor{selectiveyellow}{rgb}{1.0, 0.73, 0.0}
\definecolor{sepia}{rgb}{0.44, 0.26, 0.08}
\definecolor{shadow}{rgb}{0.54, 0.47, 0.36}
\definecolor{shamrockgreen}{rgb}{0.0, 0.62, 0.38}
\definecolor{shockingpink}{rgb}{0.99, 0.06, 0.75}
\definecolor{sienna}{rgb}{0.53, 0.18, 0.09}
\definecolor{silver}{rgb}{0.75, 0.75, 0.75}
\definecolor{sinopia}{rgb}{0.8, 0.25, 0.04}
\definecolor{skobeloff}{rgb}{0.0, 0.48, 0.45}
\definecolor{skyblue}{rgb}{0.53, 0.81, 0.92}
\definecolor{skymagenta}{rgb}{0.81, 0.44, 0.69}
\definecolor{slateblue}{rgb}{0.42, 0.35, 0.8}
\definecolor{slategray}{rgb}{0.44, 0.5, 0.56}
\definecolor{smalt(darkpowderblue)}{rgb}{0.0, 0.2, 0.6}
\definecolor{smokeytopaz}{rgb}{0.58, 0.25, 0.03}
\definecolor{smokyblack}{rgb}{0.06, 0.05, 0.03}
\definecolor{snow}{rgb}{1.0, 0.98, 0.98}
\definecolor{spirodiscoball}{rgb}{0.06, 0.75, 0.99}
\definecolor{splashedwhite}{rgb}{1.0, 0.99, 1.0}
\definecolor{springbud}{rgb}{0.65, 0.99, 0.0}
\definecolor{springgreen}{rgb}{0.0, 1.0, 0.5}
\definecolor{steelblue}{rgb}{0.27, 0.51, 0.71}
\definecolor{stildegrainyellow}{rgb}{0.98, 0.85, 0.37}
\definecolor{straw}{rgb}{0.89, 0.85, 0.44}
\definecolor{sunglow}{rgb}{1.0, 0.8, 0.2}
\definecolor{sunset}{rgb}{0.98, 0.84, 0.65}
\definecolor{tan}{rgb}{0.82, 0.71, 0.55}
\definecolor{tangelo}{rgb}{0.98, 0.3, 0.0}
\definecolor{tangerine}{rgb}{0.95, 0.52, 0.0}
\definecolor{tangerineyellow}{rgb}{1.0, 0.8, 0.0}
\definecolor{taupe}{rgb}{0.28, 0.24, 0.2}
\definecolor{taupegray}{rgb}{0.55, 0.52, 0.54}
\definecolor{teagreen}{rgb}{0.82, 0.94, 0.75}
\definecolor{tearose(orange)}{rgb}{0.97, 0.51, 0.47}
\definecolor{tearose(rose)}{rgb}{0.96, 0.76, 0.76}
\definecolor{teal}{rgb}{0.0, 0.5, 0.5}
\definecolor{tealblue}{rgb}{0.21, 0.46, 0.53}
\definecolor{tealgreen}{rgb}{0.0, 0.51, 0.5}
\definecolor{tenné(tawny)}{rgb}{0.8, 0.34, 0.0}
\definecolor{terracotta}{rgb}{0.89, 0.45, 0.36}
\definecolor{thistle}{rgb}{0.85, 0.75, 0.85}
\definecolor{thulianpink}{rgb}{0.87, 0.44, 0.63}
\definecolor{ticklemepink}{rgb}{0.99, 0.54, 0.67}
\definecolor{tiffanyblue}{rgb}{0.04, 0.73, 0.71}
\definecolor{tiger\'seye}{rgb}{0.88, 0.55, 0.24}
\definecolor{timberwolf}{rgb}{0.86, 0.84, 0.82}
\definecolor{titaniumyellow}{rgb}{0.93, 0.9, 0.0}
\definecolor{tomato}{rgb}{1.0, 0.39, 0.28}
\definecolor{toolbox}{rgb}{0.45, 0.42, 0.75}
\definecolor{tractorred}{rgb}{0.99, 0.05, 0.21}
\definecolor{trolleygrey}{rgb}{0.5, 0.5, 0.5}
\definecolor{tropicalrainforest}{rgb}{0.0, 0.46, 0.37}
\definecolor{trueblue}{rgb}{0.0, 0.45, 0.81}
\definecolor{tuftsblue}{rgb}{0.28, 0.57, 0.81}
\definecolor{tumbleweed}{rgb}{0.87, 0.67, 0.53}
\definecolor{turkishrose}{rgb}{0.71, 0.45, 0.51}
\definecolor{turquoise}{rgb}{0.19, 0.84, 0.78}
\definecolor{turquoiseblue}{rgb}{0.0, 1.0, 0.94}
\definecolor{turquoisegreen}{rgb}{0.63, 0.84, 0.71}
\definecolor{tuscanred}{rgb}{0.51, 0.21, 0.21}
\definecolor{twilightlavender}{rgb}{0.54, 0.29, 0.42}
\definecolor{tyrianpurple}{rgb}{0.4, 0.01, 0.24}
\definecolor{uablue}{rgb}{0.0, 0.2, 0.67}
\definecolor{uared}{rgb}{0.85, 0.0, 0.3}
\definecolor{ube}{rgb}{0.53, 0.47, 0.76}
\definecolor{uclablue}{rgb}{0.33, 0.41, 0.58}
\definecolor{uclagold}{rgb}{1.0, 0.7, 0.0}
\definecolor{ufogreen}{rgb}{0.24, 0.82, 0.44}
\definecolor{ultramarine}{rgb}{0.07, 0.04, 0.56}
\definecolor{ultramarineblue}{rgb}{0.25, 0.4, 0.96}
\definecolor{ultrapink}{rgb}{1.0, 0.44, 1.0}
\definecolor{umber}{rgb}{0.39, 0.32, 0.28}
\definecolor{unitednationsblue}{rgb}{0.36, 0.57, 0.9}
\definecolor{unmellowyellow}{rgb}{1.0, 1.0, 0.4}
\definecolor{upforestgreen}{rgb}{0.0, 0.27, 0.13}
\definecolor{upmaroon}{rgb}{0.48, 0.07, 0.07}
\definecolor{upsdellred}{rgb}{0.68, 0.09, 0.13}
\definecolor{urobilin}{rgb}{0.88, 0.68, 0.13}
\definecolor{usccardinal}{rgb}{0.6, 0.0, 0.0}
\definecolor{uscgold}{rgb}{1.0, 0.8, 0.0}
\definecolor{utahcrimson}{rgb}{0.83, 0.0, 0.25}
\definecolor{vanilla}{rgb}{0.95, 0.9, 0.67}
\definecolor{vegasgold}{rgb}{0.77, 0.7, 0.35}
\definecolor{venetianred}{rgb}{0.78, 0.03, 0.08}
\definecolor{verdigris}{rgb}{0.26, 0.7, 0.68}
\definecolor{vermilion}{rgb}{0.89, 0.26, 0.2}
\definecolor{veronica}{rgb}{0.63, 0.36, 0.94}
\definecolor{violet}{rgb}{0.56, 0.0, 1.0}
\definecolor{violet(colorwheel)}{rgb}{0.5, 0.0, 1.0}
\definecolor{violet(ryb)}{rgb}{0.53, 0.0, 0.69}
\definecolor{violet(web)}{rgb}{0.93, 0.51, 0.93}
\definecolor{viridian}{rgb}{0.25, 0.51, 0.43}
\definecolor{vividauburn}{rgb}{0.58, 0.15, 0.14}
\definecolor{vividburgundy}{rgb}{0.62, 0.11, 0.21}
\definecolor{vividcerise}{rgb}{0.85, 0.11, 0.51}
\definecolor{vividtangerine}{rgb}{1.0, 0.63, 0.54}
\definecolor{vividviolet}{rgb}{0.62, 0.0, 1.0}
\definecolor{warmblack}{rgb}{0.0, 0.26, 0.26}
\definecolor{wenge}{rgb}{0.39, 0.33, 0.32}
\definecolor{wheat}{rgb}{0.96, 0.87, 0.7}
\definecolor{white}{rgb}{1.0, 1.0, 1.0}
\definecolor{whitesmoke}{rgb}{0.96, 0.96, 0.96}
\definecolor{wildblueyonder}{rgb}{0.64, 0.68, 0.82}
\definecolor{wildstrawberry}{rgb}{1.0, 0.26, 0.64}
\definecolor{wildwatermelon}{rgb}{0.99, 0.42, 0.52}
\definecolor{wisteria}{rgb}{0.79, 0.63, 0.86}
\definecolor{xanadu}{rgb}{0.45, 0.53, 0.47}
\definecolor{yaleblue}{rgb}{0.06, 0.3, 0.57}
\definecolor{yellow}{rgb}{1.0, 1.0, 0.0}
\definecolor{yellow(munsell)}{rgb}{0.94, 0.8, 0.0}
\definecolor{yellow(ncs)}{rgb}{1.0, 0.83, 0.0}
\definecolor{yellow(process)}{rgb}{1.0, 0.94, 0.0}
\definecolor{yellow(ryb)}{rgb}{1.0, 1.0, 0.2}
\definecolor{yellow-green}{rgb}{0.6, 0.8, 0.2}
\definecolor{zaffre}{rgb}{0.0, 0.08, 0.66}
\definecolor{zinnwalditebrown}{rgb}{0.17, 0.09, 0.03}

\usepackage{glossaries}

\newcommand{\mechanism}{{pLUTo}\xspace}
\newcommand{\mechanismit}{\emph{pLUTo}\xspace}
\newcommand{\mechanismTDS}{{pLUTo-3DS}\xspace}

\newcommand{\mechanismAFull}{{pLUTo-GSA \emph{(\underline{G}ated \underline{S}ense \underline{A}mplifier)}}\xspace}
\newcommand{\mechanismBFull}{{pLUTo-BSA \emph{(\underline{B}uffered \underline{S}ense \underline{A}mplifier)}}\xspace}
\newcommand{\mechanismCFull}{{pLUTo-GMC \emph{ (\underline{G}ated \underline{M}emory \underline{C}ell)}}\xspace}
\newcommand{\mechanismA}{{pLUTo-GSA}\xspace}
\newcommand{\mechanismB}{{pLUTo-BSA}\xspace}
\newcommand{\mechanismC}{{pLUTo-GMC}\xspace}

\newcommand{\enabledmechanism}{{pLUTo-enabled}\xspace}
\newcommand{\enabledmechanismtitle}{{pLUTo-Enabled}\xspace}

\newcommand{\lutquery}{{pLUTo LUT Query}\xspace}
\newcommand{\lutqueries}{{pLUTo LUT Queries}\xspace}
\newcommand{\lutqueryit}{\emph{pLUTo LUT Query}\xspace}

\newcommand{\plutoop}{{\texttt{pluto\_op}}\xspace}
\newcommand{\rowsweep}{{\mechanism Row Sweep}\xspace}

\newcommand{\rowsweepit}{\emph{\mechanism Row Sweep}\xspace}
\newcommand{\matchlogic}{{\mechanism Match Logic}\xspace}
\newcommand{\matchlogicit}{\emph{\mechanism Match Logic}\xspace}
\newcommand{\mechcompiler}{\mechanism Compiler\xspace}
\newcommand{\mechcontroller}{\mechanism Controller\xspace}
\newcommand{\allocationroutines}{Memory Allocation\xspace}
\newcommand{\mechregisters}{pLUTo Registers\xspace}
\newcommand{\mechrowregister}{\mechanism Row Register\xspace}
\newcommand{\mechrowregisters}{\mechanism Row Registers\xspace}
\newcommand{\mechsubarrayregister}{\mechanism Subarray Register\xspace}

\newcommand{\trcd}{\texttt{{tRCD}}\xspace}

\newcommand{\trp}{\texttt{{tRP}}\xspace}

\newcommand{\tfaw}{\texttt{{tFAW}}\xspace}

\newcommand{\erp}{\texttt{E\textsubscript{RP}}\xspace}
\newcommand{\ercd}{\texttt{E\textsubscript{RCD}}\xspace}
\newcommand{\elisa}{\texttt{E\textsubscript{$LISA_{RBM}$}}\xspace}

\newcommand{\cmdact}{\texttt{{ACT}}\xspace}

\newcommand{\cmdread}{\texttt{{RD}}\xspace}
\newcommand{\cmdwrite}{\texttt{{WR}}\xspace}
\newcommand{\cmdprech}{\texttt{{PRE}}\xspace}

\newcommand{\speedupavgmechA}{$357\times$\xspace}
\newcommand{\speedupavgmechB}{$713\times$\xspace}
\newcommand{\speedupavgmechC}{$1413\times$\xspace}
\newcommand{\speedupavgmechAovergpu}{$0.6\times$\xspace}
\newcommand{\speedupavgmechBovergpu}{$1.2\times$\xspace}
\newcommand{\speedupavgmechCovergpu}{$2.3\times$\xspace}
\newcommand{\speedupavgmechAoverpnm}{$9.2\times$\xspace}
\newcommand{\speedupavgmechBoverpnm}{$18.3\times$\xspace}
\newcommand{\speedupavgmechCoverpnm}{$36.2\times$\xspace}
\newcommand{\speedupavgmechATDS}{$496\times$\xspace}
\newcommand{\speedupavgmechBTDS}{$990\times$\xspace}
\newcommand{\speedupavgmechCTDS}{$1962\times$\xspace}
\newcommand{\speedupavgmechATDSovergpu}{$0.8\times$\xspace}
\newcommand{\speedupavgmechBTDSovergpu}{$1.6\times$\xspace}
\newcommand{\speedupavgmechCTDSovergpu}{$3.2\times$\xspace}
\newcommand{\speedupavgmechATDSoverpnm}{$12.7\times$\xspace}
\newcommand{\speedupavgmechBTDSoverpnm}{$25.4\times$\xspace}
\newcommand{\speedupavgmechCTDSoverpnm}{$50.3\times$\xspace}

\newcommand{\energyavgmechA}{$1361.7\times$\xspace}
\newcommand{\energyavgmechB}{$1855\times$\xspace}
\newcommand{\energyavgmechC}{$3071.4\times$\xspace}
\newcommand{\energyavgmechAovergpu}{$29\times$\xspace}
\newcommand{\energyavgmechBovergpu}{$39.5\times$\xspace}
\newcommand{\energyavgmechCovergpu}{$65.3\times$\xspace}
\newcommand{\energyavgmechATDS}{$154.3\times$\xspace}
\newcommand{\energyavgmechBTDS}{$235.8\times$\xspace}
\newcommand{\energyavgmechCTDS}{$430.8\times$\xspace}
\newcommand{\energyavgmechATDSovergpu}{$3.3\times$\xspace}
\newcommand{\energyavgmechBTDSovergpu}{$5\times$\xspace}
\newcommand{\energyavgmechCTDSovergpu}{$9.2\times$\xspace}

\newcommand{\areaoverheadA}{{10.2\%}\xspace}
\newcommand{\areaoverheadB}{{16.7\%}\xspace}
\newcommand{\areaoverheadC}{{23.1\%}\xspace}

\newcommand{\versionnum}[0]{3.0}

\ifcameraready
\else    
    \fancypagestyle{firstpage}
    {
        \fancyhead{}
        \fancyhead[C]{\textcolor{MidnightBlue}{\emph{Version \versionnum~---~\today, \ampmtime}}}
    }
\fi

\makeatletter
\def\bstctlcite{\@ifnextchar[{\@bstctlcite}{\@bstctlcite[@auxout]}}
\def\@bstctlcite[#1]#2{\@bsphack
  \@for\@citeb:=#2\do{%
    \edef\@citeb{\expandafter\@firstofone\@citeb}%
    \if@filesw\immediate\write\csname #1\endcsname{\string\citation{\@citeb}}\fi}%
  \@esphack}
\makeatother

\begin{document}
\bstctlcite{IEEEexample:BSTcontrol}

\ifcameraready
  \renewcommand{\todo}[1]{}
\fi

\title{\fontfamily{bch}\selectfont
  \mechanism: Enabling Massively Parallel Computation \\ in DRAM via Lookup Tables\vspace{-6pt}}

\newcommand{\affilETH}[0]{\textsuperscript{\S}}
\newcommand{\affilUC}[0]{\textsuperscript{$\dagger$}}
\newcommand{\affilDelft}[0]{\textsuperscript{$\ddagger$}}
\newcommand{\affilIntel}[0]{\textsuperscript{$\star$}}
\newcommand{\affilCESGA}[0]{\textsuperscript{$\nabla$}}

\author{
  \scalebox{0.93}{\href{https://joaof.eu/}{João~Dinis~Ferreira\affilETH}}\qquad~~~%
  \scalebox{0.93}{Gabriel~Falcao\affilUC}\qquad~~~%
  \scalebox{0.93}{Juan~Gómez-Luna\affilETH}\qquad~~~%
  \vspace{1pt}
  \scalebox{0.93}{Mohammed~Alser\affilETH}\\
  \scalebox{0.93}{Lois~Orosa\affilETH\affilCESGA}\qquad~~~%
  \scalebox{0.93}{Mohammad~Sadrosadati\affilETH}\qquad~~~%
  \scalebox{0.93}{Jeremie~S.~Kim\affilETH}\qquad~~~%
  \vspace{1pt}
  \scalebox{0.93}{Geraldo~F.~Oliveira\affilETH}\\
  \scalebox{0.93}{Taha~Shahroodi\affilDelft}\qquad~~~%
  \scalebox{0.93}{Anant~Nori\affilIntel}\qquad~~~%
  \vspace{3pt}
  \scalebox{0.93}{Onur~Mutlu\affilETH}\\%
  \scalebox{0.93}{\emph{{\affilETH ETH Z{\"u}rich}~~~~~\affilUC IT, University of Coimbra~~~~~\affilCESGA Galicia Supercomputing Center~~~~~\affilDelft TU Delft~~~~~\affilIntel Intel}}%
  \vspace{-10pt}
}%

\renewcommand{\citepunct}{,\penalty\citepunctpenalty\,}
\renewcommand{\citedash}{--}%

\newcommand{\prtag}[1]{\lfbox[padding=1pt, border-color=blue, background-color=blue!40]{\textbf{\small #1}}}

\maketitle

\ifcameraready
  \thispagestyle{plain}
\else
  \thispagestyle{firstpage}
\fi
\pagestyle{plain}

\begin{abstract}
        Data movement between the main memory and the processor is a key contributor to execution time and energy consumption in memory-intensive applications.
        This \emph{data movement bottleneck} can be alleviated using Processing-in-Memory (PiM).
        One category of PiM is Processing-using-Memory (PuM),
        in which computation takes place \emph{inside} the memory array by exploiting intrinsic analog properties of the memory device.
        PuM yields high {performance and energy efficiency}, but {existing PuM techniques support} a limited range of operations.
        As a result, {current} PuM architectures cannot efficiently perform some complex operations (e.g., multiplication, division, exponentiation) without {large} increases in chip area and design complexity.

        To overcome {these limitations of existing} PuM architectures, we introduce \textbf{\mechanismit} \textbf{(\underline{p}rocessing-using-memory with lookup table (\underline{LUT}) \underline{o}perations)}, a DRAM-based PuM architecture that leverages the {high storage density} of DRAM to enable the massively parallel storing and querying of lookup tables (LUTs).
                {The key idea of \mechanism is to replace complex operations with low-cost, bulk memory reads (i.e., LUT queries) instead of relying on complex extra logic.}

                {We evaluate \mechanism across 11 real-world workloads that showcase the limitations of prior PuM approaches and show that our solution outperforms optimized CPU and GPU baselines by an average {of
                                        \speedupavgmechB and \speedupavgmechBovergpu, respectively, while simultaneously reducing energy consumption by an average of \energyavgmechB and \energyavgmechBovergpu.
                                        Across these workloads, \mechanism outperforms state-of-the-art PiM architectures by an average of \speedupavgmechBoverpnm.}}
                {We also show that different versions of \mechanism provide different levels of flexibility and performance at different {additional DRAM area overheads} (between \areaoverheadA and \areaoverheadC).}
                {\mechanism's source code and all {scripts} required to reproduce the results of this paper are openly and fully available at \url{https://github.com/CMU-SAFARI/pLUTo}}.
        \vspace{-5pt}
\end{abstract}

\section{Introduction}
\label{sec:introduction}

\emph{Processing-in-Memory} (PiM) is a promising paradigm that augments a system's memory with compute capability~\cite{ghose2019pim,mutlu2019processing,mutlu2019enabling,ghose2019processing,siegl2016data} to alleviate the \emph{data movement bottleneck} between %
processing %
and %
memory {units}%
~{\cite{pandiyan2014,
  kanev2015profiling,
  mutlu2019processing,
  paul2015harmonia,
  ware2010architecting,
  lefurgy2003energy,
  boroumand2018google,
  vogelsang2010understanding,
  wulf1995hitting,
  BillDally2015,
  oliveira2021damov,
  mutlu2013memory,
  mutlu2015research}}.
PiM architectures can be classified into two {categories~\cite{ghose2019pim,mutlu2021modern}}:
1) \textbf{Processing-near-Memory (PnM)}, where computation takes place in {dedicated} processing elements {(e.g., accelerators~{\cite{nai2017graphpim,boroumand2018google,kim2018grim,PEI,gao2017tetris,kim2016neurocube,cali2020genasm,akin2016data,hsieh2016accelerating,babarinsa2015jafar,boroumand2021mitigating,boroumand2021google,boroumand2022polynesia,boroumand2021polynesia,amiraliphd,ghiasi2022genstore,fernandez2020natsa,singh2020nero,lee2022isscc,kwon202125,lee2021hardware,niu2022184qps,rosenfeld2014performance,oliveira2022accelerating,cho2020mcdram,shin2018mcdram}},
processing cores~{\cite{ahn2016scalable,boroumand2018google,boroumand2016lazypim,zhang2014top,drumond2017mondrian,boroumand2019conda,hsieh2016transparent,pattnaik2016scheduling,devaux2019,boroumand2022polynesia,boroumand2021polynesia,amiraliphd,gomez2021benchmarkingcut,gomez2021benchmarkingarxiv,gomez2021benchmarking,besta2021sisa,syncron,singh2019napel,giannoula2022sparsep,oliveira2022accelerating}}, reconfigurable logic~{\cite{gao2016hrl,santos2017operand,NIM,farmahini2015nda,ke2021near}}) placed} {near
the} memory array {(e.g.,~\cite{lee201425,pawlowski2011hybrid,loh2008isca,hmc_spec,nai2017graphpim,kim2018grim,PEI,gao2017tetris,kim2016neurocube,cali2020genasm,akin2016data,hsieh2016accelerating,babarinsa2015jafar,boroumand2021mitigating,boroumand2021google,ghiasi2022genstore,fernandez2020natsa,singh2020nero,lee2022isscc,kwon202125,lee2021hardware,niu2022184qps,rosenfeld2014performance,ahn2016scalable,boroumand2018google,boroumand2016lazypim,zhang2014top,drumond2017mondrian,boroumand2019conda,hsieh2016transparent,pattnaik2016scheduling,devaux2019,boroumand2022polynesia,boroumand2021polynesia,amiraliphd,gomez2021benchmarkingcut,gomez2021benchmarkingarxiv,gomez2021benchmarking,besta2021sisa,syncron,singh2019napel,giannoula2022sparsep,gao2016hrl,santos2017operand,NIM,farmahini2015nda,ke2021near,oliveira2022accelerating})}, and
2) \textbf{Processing-using-Memory (PuM)}, where computation takes place \emph{inside} the memory array by exploiting intrinsic analog {operational} properties of the memory device {(e.g.,~{\cite{aga2017compute,besta2021sisa,chi2016prime,deng2018dracc,eckert2018neural,flashcosmos,fujiki2019duality,gao2019computedram,hajinazar2020simdram,he2020sparse,imani2019floatpim,li2016pinatubo,li2017drisa,oliveira2022accelerating,seshadri.arxiv16,seshadri2013rowclone,seshadri2015fast,seshadri2016processing,seshadri2017ambit,seshadri2017simple,seshadri2018rowclone,seshadri2019dram,shafiee2016isaac,song2017pipelayer,song2018graphr,xin2020elp2im}})}.

In {DRAM-based} PnM, data is transferred from the DRAM {array} to nearby processors or specialized accelerators, which {could be}
1) part of the DRAM chip, but separate from the DRAM array~{\cite{devaux2019,kwon202125,gao2017tetris,lee2022isscc,lee2021hardware,gomez2021benchmarkingcut,gomez2021benchmarkingarxiv,gomez2021benchmarking,syncron,giannoula2022sparsep,cho2020mcdram,shin2018mcdram}}, e.g., near the DRAM {banks,
    2)} integrated into the logic layer of 3D-stacked memories~{\cite{hmc_spec,loh2008isca,kim2018grim,PEI,kim2016neurocube,cali2020genasm,akin2016data,boroumand2021mitigating,boroumand2021google,fernandez2020natsa,niu2022184qps,ahn2016scalable,boroumand2018google,boroumand2016lazypim,zhang2014top,drumond2017mondrian,boroumand2019conda,hsieh2016transparent,pattnaik2016scheduling,boroumand2022polynesia,boroumand2021polynesia,amiraliphd,besta2021sisa,singh2019napel,gao2016hrl,santos2017operand}}, or
{3) inside the memory controller~{\cite{seshadri2015gather,hashemi2016accelerating,singh2020nero}}.}
PnM enables the design of flexible substrates that support a diverse range of operations.
However, {the performance, efficiency, and scalability of near-bank PnM architectures~{\cite{devaux2019,kwon202125,gao2017tetris,lee2022isscc,lee2021hardware,gomez2021benchmarkingcut,gomez2021benchmarkingarxiv,gomez2021benchmarking,syncron,giannoula2022sparsep,cho2020mcdram,shin2018mcdram}} can be limited by %
design and fabrication challenges, such as
1) the difficulty in designing complex logic due to the limited number of DRAM metal layers~\cite{weber2005current,peng2015design}, and
2) the inefficiency of the DRAM process for the implementation of digital logic due to its {heavy} optimization for memory %
density~\cite{devaux2019,gomez2021benchmarking}.}
In 3D-stacked memories, the logic layer's limited area and thermal {budgets} impose {additional constraints.}
All these design and fabrication issues lead to {generally very simple} PnM execution engines%
, which {are unable to} exploit the entire DRAM bandwidth~\cite{drumond2017mondrian,oliveira2021damov,gomez2021benchmarking}.

In contrast, PuM architectures enable computation \emph{within} the memory array.
The key benefit of PuM architectures is that \emph{data does not leave the memory {array} during computation.}
As a result, PuM architectures can provide high compute throughput by performing operations in a bulk parallel manner, often at the granularity of memory rows.
Prior PuM works~{\cite{seshadri2017ambit,seshadri2015fast,gao2019computedram,li2017drisa,deng2018dracc,deng2019lacc,sutradhar2020ppim,hajinazar2020simdram, flashcosmos}} propose mechanisms for the execution of {bulk} bitwise operations {(e.g., bitwise \texttt{MAJority,AND,OR,NOT})}~{\cite{seshadri2017ambit,seshadri2015fast,gao2019computedram,flashcosmos,seshadri2019dram,seshadri2016processing,seshadri2017simple,seshadri.arxiv16,li2016pinatubo,xin2020elp2im,angizi2019redram}} and {bulk} arithmetic operations{~\cite{li2017drisa,deng2018dracc,deng2019lacc,sutradhar2020ppim,hajinazar2020simdram}}. {However, these proposals have two important limitations:
{1) the execution of some complex operations (e.g., multiplication, division) incurs high latency and energy consumption~\cite{hajinazar2020simdram}, and}
2) other complex operations (e.g., exponentiation, trigonometric functions) are not {even} supported.}

{We aim to overcome these two limitations of prior PuM architectures in this work.
To this end, we employ \emph{LUT-based computing}, i.e., the use of memory read operations \emph{(LUT queries)} to retrieve the results of complex operations from lookup tables that hold precomputed values.}
Concretely, a \emph{LUT query} is a \emph{memory read operation}
{that, for a given input value $x$, returns $f(x)$, i.e., the result {of applying} some function $f$ to the input $x$.}
Many PuM architectures~\cite{deng2019lacc,gao2016draf,sutradhar2020ppim} exploit LUT-based computing to improve the performance of a few complex operations.
However, {
    \emph{no} prior work supports the} general-purpose execution of LUT-based complex operations.

Our \textbf{goal} in this work is to \emph{extend the functionality of DRAM-based PuM {systems} to provide support for {general-purpose execution of} complex {operations.}}
\textbf{To this end}, we propose \textit{\mechanismit: \underline{p}rocessing-using-memory with lookup table (\underline{LUT}) \underline{o}perations,} a DRAM-based PuM architecture that leverages LUT-based computing {via bulk querying of LUTs} to perform complex operations beyond the scope of prior {DRAM-based} PuM proposals.
\mechanism introduces a novel LUT-querying mechanism, the \lutqueryit, which enables the simultaneous querying of multiple LUTs stored in a single %
  {DRAM subarray}.
{In \mechanism, the number of elements stored in each LUT may be as {large}} as {the number of} rows in each {DRAM} subarray ({e.g., 512{--}1024 rows~\cite{kim2012case,kim2018solar,lee2017design}}).
\mechanism requires the following two {modest} modifications to DRAM hardware:
1) \textbf{\emph{row sweeping logic}}, which enables the \emph{sweeping} of {DRAM} rows, i.e., the successive activation of consecutive %
rows in a {DRAM} {subarray};
2) \textbf{\emph{match logic}}, which identifies \emph{matches} between the \emph{elements} in the input row and the \emph{index} of the currently active row in the subarray {that} holds {multiple copies of one or more LUTs}.
We describe three \mechanism designs{: \mechanismBFull, \mechanismAFull, and \mechanismCFull.}
These designs achieve different %
performance, energy efficiency, and area overhead {trade-offs}.

{To} enable {the} {seamless} integration of \mechanism with the system, we {methodically} describe {the changes} that {allow programmers} to offload their applications to {\mechanism.}
  {These changes comprise
    1)
    \mechanism ISA {instructions} that enable support for each of the DRAM operations required for \mechanism's operation,
    2)
    {the {\mechanism Library, an API library that includes}
        routines that programmers can use to conveniently express \mechanism operations at a high level of abstraction,
      }
    3)
    the \mechanism Compiler, which analyzes {an} application's data dependency graph to {plan the in-memory placement {and alignment} of data}, and
    4)
    the \mechanism Controller, a modified memory controller {that} supports the execution of \mechanism ISA {instructions}.
  }

  {We evaluate \mechanism on a diverse range of real-world arithmetic, bitwise logic, cryptographic, image processing, and neural network workloads that {demonstrate} the limitations of existing PuM architectures and how \mechanism is able to overcome them.
    These workloads include
      {1)~}bitwise (\texttt{AND/OR/XOR}), arithmetic (addition, multiplication), and nonlinear operations (substitution tables, image binarization, and color grading){;} and
      {2)~}a quantized neural network.}
We compare \mechanism to state-of-the-art processor-centric architectures (CPU~\cite{intel_gold_cpu}, GPU~\cite{nvidia_3080ti}, FPGA~\cite{xilinxzcu102}) and PiM architectures (PnM~\cite{hmc_spec}, PuM~\cite{seshadri2017ambit,li2017drisa,hajinazar2020simdram}).
Our evaluations show that \mechanism consistently {and considerably} outperforms these {five} baselines {in
    performance and energy consumption}.

{We} make the following \textbf{key contributions}:

\begin{itemize}[leftmargin=3mm,itemsep=0mm,parsep=0mm,topsep=0mm]
  \item We introduce \mechanism, a {new DRAM-based PuM %
            architecture that introduces support for general-purpose operations through the use of bulk {lookup table (LUT) queries.}}

  \item We propose three different {and new} \mechanism designs with varying trade-offs in {performance,} energy efficiency, and {DRAM} {area overhead}.

  \item We describe the end-to-end system integration of \mechanism, {including
            1) ISA {instructions},
            2) an API library,
            3) a compiler, and
            4) modifications to the {memory controller}.}

  \item {We experimentally demonstrate that \mechanism significantly outperforms CPU~\cite{intel_gold_cpu}, GPU~\cite{nvidia_3080ti}, FPGA~\cite{xilinxzcu102}, and PiM~\cite{hmc_spec,seshadri2017ambit,li2017drisa,hajinazar2020simdram} baselines across a wide variety of real-world {bitwise logic, arithmetic,} cryptographic, image processing, and neural network workloads.}

  \item {We open-source
        {
        \mechanism's source code and all {scripts} required to reproduce the results presented in this paper on}
        \mbox{\url{https://github.com/CMU-SAFARI/pLUTo}}.}
\end{itemize}

\section{Background}

This section describes the hierarchical organization of DRAM and provides an overview of relevant prior work {we build on}.
{We refer the reader to prior work~\cite{kim2012case,lee2013tiered,lee2015decoupled,chang2016lisa,seshadri2017ambit,li2017drisa} for more detailed descriptions of DRAM operation.}

\subsection{DRAM Background}
\label{dram-background}

\begin{figure}[ht]
  \centering
  \includegraphics[width=\linewidth]{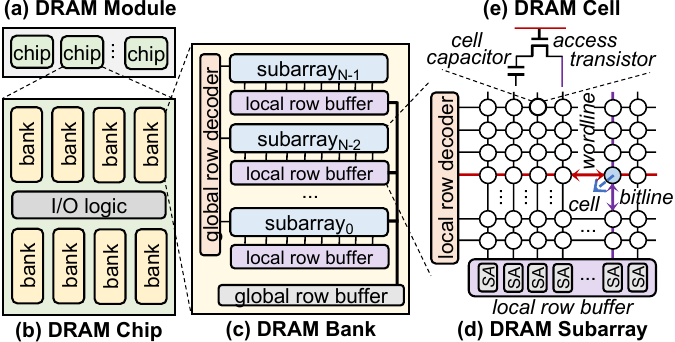}
  \caption{{Internal} organization of {a DRAM {module}.}}
  \label{fig:dram_organization}
\end{figure}

{DRAM is organized {hierarchically: each} DRAM module consists of multiple chips, banks, and subarrays, as {\Cref{fig:dram_organization} shows}.} {A DRAM \emph{module} {(\Cref{fig:dram_organization}a)} consists of multiple DRAM \emph{chips} {(\Cref{fig:dram_organization}b)}, each of which contains} %
multiple {DRAM} \emph{banks} ({e.g.,} 8 for DDR3{~\cite{jedec2012sdram}}, 16 for DDR4{~\cite{jedec2017sdram}}) and I/O {logic.} %
{Each DRAM} bank {(\Cref{fig:dram_organization}c)} {is divided} into {DRAM} \emph{subarrays}~\cite{kim2012case} {(\Cref{fig:dram_organization}d)}, which are two-dimensional arrays of {DRAM} \emph{cells} {(\Cref{fig:dram_organization}e)}.
DRAM subarrays in a bank share peripheral circuitry (e.g., {a global} row decoder {and a global row buffer}).
Each {DRAM} cell contains one {cell} capacitor and one access transistor.
The {cell} capacitor encodes a single bit as stored electrical charge.
The {access} transistor connects the {cell} capacitor to the \emph{bitline} wire.
Each bitline is shared by all {DRAM} cells in a column and {is connected} to a \emph{sense amplifier} {(\emph{SA} in \Cref{fig:dram_organization}d)}.
The set of sense amplifiers in a subarray makes up the \textit{local row buffer}.

Reading and writing data in DRAM occurs over three phases{.}%

\noindent
{\textit{\textit{1) Row Activation.}}}
{The memory controller initiates a memory request by issuing an {activate (\cmdact)} command {together} with a DRAM row address to the target DRAM bank.
  Once the DRAM {chip} receives the \cmdact command, it asserts the corresponding wordline to activate the DRAM row.
  The row activation process happens in three main steps.
  First, the wordline of the accessed row is driven {with} high {voltage}, turning on the row's access transistors and creating a path for \emph{charge sharing} between each {DRAM} cell and its bitline.
  This process induces a voltage fluctuation{,~$\pm\delta$,} {that affects the voltage level of the \emph{precharged} (i.e., set to $V_{DD}/2$) bitline.}
  If the cell is charged, the bitline voltage becomes $V_{DD}/2+\delta$.
  Otherwise, it becomes $V_{DD}/2-\delta$.
    {Second, {each sense amplifier} in the local row buffer {\emph{amplifies} its} {{bitline's voltage} fluctuation ($\pm\delta$)
          until the bitline voltage reaches either $V_{DD}$ or $0$.}
        {Third, %
          the original voltage {level} of
            {each cell} capacitor {in the activated row is}
          \emph{restored}, since {each access transistor allows} charge to flow between {each bitline and {each} corresponding DRAM cell in the activated row}.}
        {Once the row activation process {is complete}, the local row buffer contains the values originally stored {in} the cells along the asserted wordline.}
    }}

\noindent
{\textit{\textit{2) Reading/Writing.}}}
{
  {The} memory controller issues {read (\cmdread) or write (\cmdwrite)} commands {together} with a DRAM {\emph{column address}} to {read or write} chunks of the data latched {in} the local row buffer.
  In the case of a read, data {in the corresponding {columns}} is sent to the CPU through the DRAM's I/O circuitry and the memory bus.
  In the case of a write, data received from the CPU modifies the corresponding {columns'} bitline {(and thus cell)} voltages.}

\noindent
{\textit{\textit{3) Precharging.}}} {{The memory controller issues a {precharge (\cmdprech)} command to} {free} up the {local} row buffer and allow the activation of other {DRAM} rows.
  To achieve this, {the wordline is de-asserted, which turns off} the access transistors {along a DRAM row}, and the subarray's bitlines are precharged (i.e., restored to $V_{DD}/2$).}

\subsection{{Enhanced DRAM Architectures}}
\label{sec:dram_extensions}

\mechanism optimizes key operations by incorporating the following previous proposals for enhanced DRAM architectures.

\head{Intra-Subarray Data Copy}~{The \emph{RowClone-FPM}  (\underline{F}ast \underline{P}arallel \underline{M}ode)~\cite{seshadri2013rowclone} operation enables data to be copied between two DRAM rows belonging to the same {DRAM} subarray.
  {This is achieved {via} two consecutive activations: first to the source row, {then} to the destination row.
    Doing so asserts the} destination row's wordline while the contents of the source row {are already in} the subarray's row buffer, which causes the {entire} row buffer's contents to be written %
to the destination row.
}

\head{Inter-Subarray Data Copy}~The \textit{LISA-RBM} (\underline{R}ow \underline{B}uffer \underline{M}ovement) operation~\cite{chang2016lisa} copies the contents of one {local} row buffer to another {local} row buffer in a different subarray {in the same bank} without relying on the external memory channel.
This is achieved by linking neighboring subarrays with isolation transistors.

\head{Subarray-Level Parallelism}~\textit{MASA}~\cite{kim2012case} is a mechanism that enables subarray-level parallelism by overlapping the latency of memory accesses directed to different {subarrays' bitlines}.
{MASA enables multiple rows in different subarrays to be activated and to be accessed (read or written to) in parallel.}

\head{{Bulk} Bitwise Operations}~\textit{Ambit}~\cite{seshadri2017ambit} {is a PuM architecture that} introduces {native} support for bulk bitwise logic operations {(\texttt{MAJority,AND,OR,NOT})} between rows in a DRAM subarray.
  {Ambit uses the \emph{triple row activation} primitive {(which concurrently asserts {three} wordlines, leading to the execution of the {majority} function between the contents of {three DRAM} rows)}
    and the copy operation enabled by RowClone~\cite{seshadri2013rowclone} to enable these simple, row-granularity bitwise operations.}

\head{Shifting}~\textit{DRISA}~\cite{li2017drisa} {is a PuM architecture that features} support for intra-row shifting in DRAM.
Using this mechanism, the contents of a {DRAM} row can be shifted by 1 or 8 bits at a time, {at the} cost of \emph{one} {\texttt{ACT}-\texttt{ACT}-\texttt{PRE}}~\cite{seshadri2017ambit} command sequence.

\section{Motivation}
\label{sec:motivation}

{Our \textbf{goal} in this work is to \emph{extend the functionality of {Processing-using-Memory (PuM) architectures to provide support for {the} general-purpose execution of complex operations.}}}
In \mbox{particular}, \mechanism is motivated by the following two key observations.
First, state-of-the-art PuM architectures~{\cite{aga2017compute,besta2021sisa,chi2016prime,deng2018dracc,eckert2018neural,flashcosmos,fujiki2019duality,gao2019computedram,hajinazar2020simdram,he2020sparse,imani2019floatpim,li2016pinatubo,li2017drisa,oliveira2022accelerating,seshadri.arxiv16,seshadri2013rowclone,seshadri2015fast,seshadri2016processing,seshadri2017ambit,seshadri2017simple,seshadri2018rowclone,seshadri2019dram,shafiee2016isaac,song2017pipelayer,song2018graphr,xin2020elp2im,deng2019lacc}}
provide very high performance and energy efficiency by mitigating data movement, but {they} {\emph{only}} support a limited range of operations.
For example,
{prior DRAM-based PuM accelerators}
only support the execution of basic operations (e.g., {bitwise logic, addition})~{\cite{li2017drisa,deng2018dracc,deng2019lacc,sutradhar2020ppim,hajinazar2020simdram,seshadri2017ambit,seshadri2015fast,gao2019computedram}} or {require} long sequences of DRAM commands {to support} more complex operations (e.g., multiplication, division)~\cite{hajinazar2020simdram}.
Second, {lookup tables (LUTs)} {enable the replacement of complex computations with cheaper LUT {query} operations} (i.e., memory reads).
\mechanism improves prior PuM works by leveraging their best features (i.e., high parallelism, reduced data movement) and addressing their main drawbacks (i.e., reduced range of supported operations and low {performance for} complex {operations).
        We achieve this via the introduction of the \lutqueryit operation {(described in \Cref{sec:lut_query_operation})},} which enables the {bulk} querying of all the values in a given input {DRAM} {row.}

\section{An Overview of \mechanism}
\label{sec:mechanism_overview}

{The key contribution of \mechanism is the \lutqueryit, {an operation} that enables the bulk execution of a large number of LUT queries \emph{inside} {a DRAM subarray}.
  Each individual \emph{LUT query} is defined as a \emph{memory read operation} that, given an \emph{input value} $x$, {returns} as its \emph{output value} the result of applying some arbitrary function $f$ to $x$, i.e., $f(x)$.}
{Building on the \lutquery, \mechanism employs \emph{LUT-based computing} (i.e., \emph{the replacement of complex operations with equivalent LUT queries}) to perform computation under the Processing-using-Memory paradigm.
  LUT-based computing requires that each complex operation to be replaced with a LUT query be \emph{deterministic}; in other words, the behavior of the function $f$ being replaced with a LUT query \emph{should {only depend} on its input value $x$.}
  The construction of a LUT requires a one-time effort of computing all its values, i.e., all \emph{LUT elements}.}

\begin{figure}[ht]
  \vspace{2mm}
  \centering
  \includegraphics[width=\linewidth]{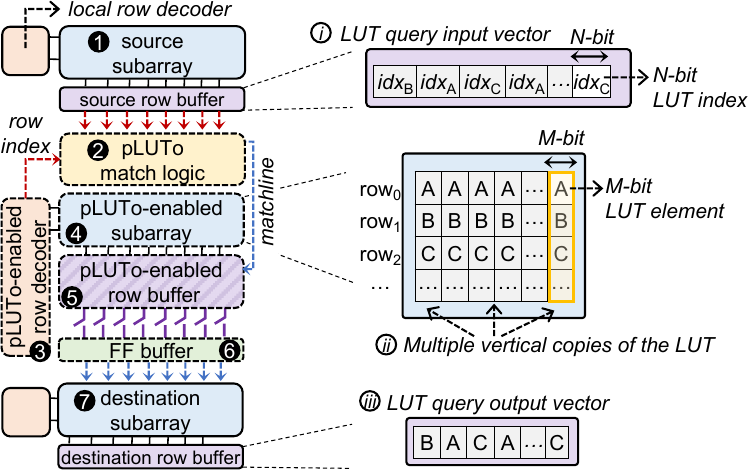}
  \caption{{
        Main components of \mechanism.}}
  \label{fig:subarray_layout}
  \vspace{2mm}
\end{figure}

{\Cref{fig:subarray_layout} shows an overview of the DRAM structures required to {perform a} {\lutqueryit.%
      \footnote{{\Cref{fig:subarray_layout} assumes that \mechanism has been implemented with the \mechanismB design (described in \Cref{sec:mechArch}).
            However, the key ideas of the \lutquery described in this section {apply to} all three \mechanism designs described in \Cref{sec:mechArch,sec:pluto_a,sec:pluto_c}.}} }
  First, the \emph{source subarray} (\circlednumber{1} in \Cref{fig:subarray_layout}) stores the \emph{LUT query input vector} (\incircledd{i} in \Cref{fig:subarray_layout}), which consists of {a set of} $N$-bit LUT indices associated with LUT elements.
  Second, the \emph{\mechanism Match Logic} (\circlednumber{2}) comprises a set of comparators that identify matches between 1) the row index of the currently activated row in the \mechanism-enabled subarray, and 2) each LUT index in the {LUT query input vector (i.e., the source subarray's row buffer)}.
  Third, the \emph{\mechanism-enabled row decoder} (\circlednumber{3}) enables the {successive} activation {of} consecutive DRAM rows in the \mechanism-enabled subarray with a single DRAM command.
  It also outputs the row index of the currently activated row {as input} to the \mechanism Match Logic.
  Fourth, the \emph{\mechanism-enabled subarray} (\circlednumber{4}) stores multiple vertical copies of a given LUT (\incircledd{ii}), which consists of $M$-bit LUT elements.
  Fifth, the \emph{\mechanism-enabled row buffer} (\circlednumber{5}) allows {the reading of} individual LUT elements from the activated row in the \emph{\mechanism-enabled subarray}.
  This is possible by extending the DRAM sense amplifier design of the \mechanism-enabled row buffer with switches controlled by the \mechanism Match Logic (using the \emph{matchline} signal).
  Sixth, the \emph{flip-flop (FF) buffer} (\circlednumber{6}) enables {\mechanism to temporarily store select LUT elements by copying them from the \enabledmechanism row buffer, conditioned on the output of the \matchlogic following each row activation.}
  Seventh, {a LISA-RBM operation copies the entire contents of the FF buffer (i.e., the LUT query output vector, \incircledd{iii}) into the destination row buffer, i.e., the row buffer of the \emph{destination subarray} (\circlednumber{7})}.
}

{In contrast to the bit-serial paradigm employed by prior PuM architectures (e.g., SIMDRAM~\cite{hajinazar2020simdram}), \mechanism operates in a \emph{bit-parallel} manner; in other words, the bits that make up {each LUT element (e.g., \texttt{A}) are stored \emph{horizontally} (i.e., in adjacent bitlines), and {all the} copies of each LUT element (i.e., \texttt{\{A,A,...,A\}}) take up \emph{one {whole} row} in the depicted \enabledmechanism subarray {(\incircledd{ii})}.}}\footnote{{{Throughout} this example, we assume that all \texttt{\{A,B,C,...\}} values are 8 bits wide, although this bit width is a parameter of each \lutquery and may vary {(see \Cref{sec:mech_library})}{.}}}

\begin{figure*}[t]
  \centering
  \includegraphics[width=0.925\textwidth]{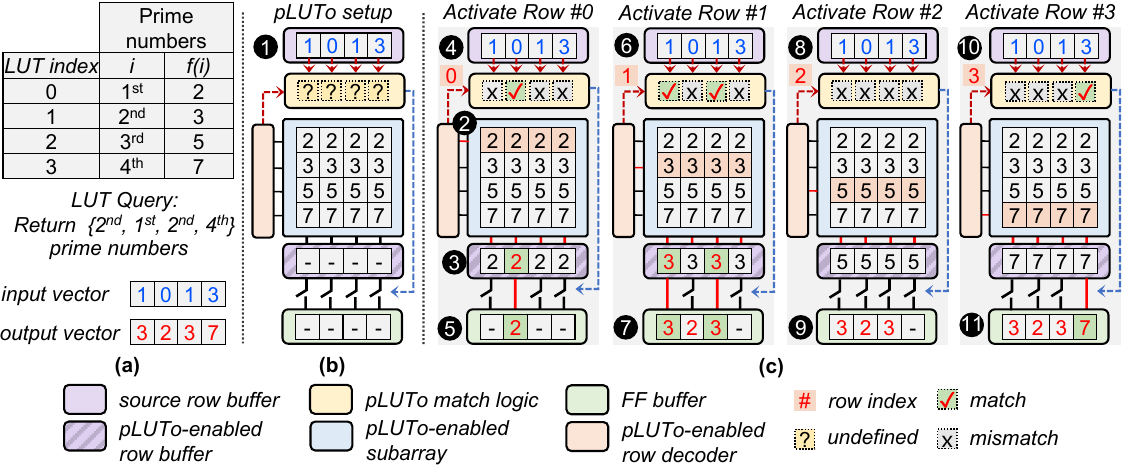}
  \caption{
    A \lutquery:
    (a) a LUT containing the first four prime numbers {and an example user-specified LUT query,}
    (b) {setup of \mechanism's main components prior to the execution of the \lutquery}, and
    (c) steps of the \lutquery.
    This \lutquery returns {into the destination row buffer {(not depicted)}} the i-th prime number for each {LUT index} in the {source row buffer}.
  }
  \label{fig:pluto_single_bitcount}
\end{figure*}

{
  \subsection{The \lutquery}
  \label{sec:lut_query_operation}
}
\vspace{1mm}
The \lutquery enables all the elements stored in a source row buffer to {simultaneously be used to} query a LUT.
We illustrate the \lutquery using a {simple} example that employs a small LUT {to store the first} four prime numbers (i.e., \texttt{\{2,3,5,7\}}) at {LUT} indices \texttt{\{0,1,2,3\}}, as shown in \Cref{fig:pluto_single_bitcount}a.
{In our example, the user-defined LUT query will return the \{$2^{nd},1^{st},2^{nd},4^{th}$\} prime numbers, which corresponds to a LUT query input vector \textcolor{blue}{\texttt{[1,0,1,3]}} and an expected LUT query output vector \textcolor{red}{\texttt{[3,2,3,7]}} (i.e., \{$2^{nd},1^{st},2^{nd},4^{th}$\} prime numbers $\Rightarrow$ {\texttt{\{LUT[\textcolor{blue}{1}],LUT[\textcolor{blue}{0}],LUT[\textcolor{blue}{1}],LUT[\textcolor{blue}{3}]\}}\ $=$\ \textcolor{red}{\texttt{\{3,2,3,7\}}}}).} {Note that, in this example, \emph{four} {individual} lookup operations are performed
  {by} a \emph{single} \lutquery.}

Four copies of this LUT are stored in a \enabledmechanism subarray, as shown in \Cref{fig:pluto_single_bitcount}b: each row $i$ contains repeated copies of the element corresponding to the entry at the $i$-th index of the LUT.
\mechanism performs {the} \lutquery {operation in five steps.}
First, the memory controller loads the {LUT query input vector} from the source subarray (not shown) into the source row buffer (\circlednumber{1} in \Cref{fig:pluto_single_bitcount}).
Second, the memory controller issues a \rowsweepit operation (\Cref{sec:rowdecoder}) to {consecutively activate all four rows {in the \mechanism-enabled subarray} that hold LUT {elements}, in order (i.e., the row indices to be activated are \texttt{\{\#0,\#1,\#2,\#3\}}).
      {{After each row activation during a \rowsweep operation, the \matchlogic identifies matches between 1)
            the row index of the currently activated row in the \enabledmechanism subarray, and
            2) {each element} of the LUT query input vector (i.e., the source row buffer).}
        The aim of this procedure is to \emph{allow} for
          {consecutive} LUT elements, in turn, to be copied to the FF buffer, \emph{if {they are part of the final output row}} for the ongoing \lutquery {operation.}}
      {Consider the activation of row \texttt{\#0} (\circlednumber{2}), which {creates} four copies of \texttt{LUT[0]} (i.e., {the LUT element with the value} \textcolor{red}{2}) {in} the \mechanism-enabled row buffer (\circlednumber{3}).
          {
            Concurrently with this row activation, the \matchlogic
            1)~identifies a match between the index of the currently activated row \texttt{(\#0)} and the \emph{second} LUT index in the LUT query input vector (\circlednumber{4}), and
            2)~asserts the matchlines corresponding to {the} \emph{second element} in the \mechanism-enabled row buffer.}
        As a result, the switch at the \emph{second LUT element} in the \mechanism-enabled row buffer is closed, enabling the LUT element to be copied to the \emph{second position in the FF buffer} (\circlednumber{5}).
      }

      {Third, the activation of row index \texttt{\#1} {illustrates how} multiple LUT indices may be matched at once: in this case, the LUT element of \texttt{LUT[1]} is required by \emph{both the first and the third} positions of the LUT query input vector (\circlednumber{6}).
        Therefore, the \matchlogic asserts the {matchlines} corresponding to \emph{both the first and third} LUT elements, which copies both LUT elements into the FF buffer (\circlednumber{7}).
        Fourth, the \lutquery operation progresses by activating row \texttt{\#2}, which produces \emph{no} matches with LUT indices stored in the {LUT Query input vector} (\circlednumber{8}).
        As a result, \emph{no} LUT {elements are} copied into the FF buffer (\circlednumber{9}).
        Fifth, when activating row \texttt{\#3}, the \matchlogic identifies a {match} between row index \texttt{\#1} and the \emph{fourth} element of the LUT query input vector (\circlednumber{10}), which {leads to the copy of the LUT element \texttt{LUT[3]} (i.e., \textcolor{red}{7})} into the \emph{fourth} position in the FF buffer (\circlednumber{11}).}

    At this point, the \rowsweep operation {has been completed, and the FF buffer holds the results of the \lutquery.
        The contents of the FF buffer are then copied} to the \emph{{destination row buffer}} (not shown in \Cref{fig:pluto_single_bitcount}) using a LISA-RBM command ({see Section~\ref{sec:dram_extensions}).
      }}

{
\section{{\mechanism's Hardware Design}}
}

{This section describes the hardware design of \mechanism that enables the \lutquery operation.
First, we} propose three different \mechanism architectures.
  {Each of these architectures provides a different trade-off between} performance, energy efficiency, and area overhead:
{
1)~\mechanismBFull {incurs} moderate hardware overhead and provides intermediate performance {and energy efficiency gains;}
2)~\mechanismAFull {incurs} the lowest hardware overhead {but} provides the lowest performance {and energy efficiency;}
3)~\mechanismCFull {incurs} the highest hardware overhead {but} provides the highest performance {and energy efficiency}.}
Second, we {describe} the synergistic integration of the novel components of \mechanism with prior PuM-based operations~\cite{seshadri2017ambit,li2017drisa,chang2016lisa} and subarray-level parallelism~\cite{kim2012case}.

\subsection{{\mechanismBFull Design}}
\label{sec:mechArch}

To enhance a DRAM subarray with {support for {the execution of} \lutqueries, we modify {the DRAM subarray's row decoder and local row buffer to implement the \matchlogicit}.}
  {Uniquely, \mechanismB employs the \emph{key idea} of relying on a secondary {row} buffer {(the \emph{FF buffer}, see \Cref{sec:bsa_row_buffer})}
    to store {matching LUT elements
        during a \rowsweep.}}

\subsubsection{\enabledmechanismtitle Row Decoder}
\label{sec:rowdecoder}

The {pLUTo-enabled} row decoder enhances the DRAM row decoder by introducing support for the \rowsweepit operation.
The \rowsweep extends the self-refresh operation ({already present in} commodity DRAM~\cite{lazar2004selfrefresh,patel1994dram,kim2005self})
to activate consecutive rows quickly.
  {With support for the \rowsweep operation, \mechanism activates all the rows in the \mechanism-enabled subarray that {store} LUT elements {during a \lutquery operation \emph{{via} a single {new} DRAM command.}}}
The latency of the \rowsweep is equal to $(\trcd + \trp) \times LUT_{\# Elems}$,
{where $\trcd$
    {($\approx$ \SI{12.5}{\nano\second}
        in DDR4~\cite{jedec2017sdram})} is the time that must elapse to ensure that the sense amplifiers can reliably amplify the voltage perturbation on the bitline,
    $\trp$ {($\approx$ \SI{12.5}{\nano\second} in DDR4~\cite{jedec2017sdram})} is the time that must elapse between a \cmdprech command and the next \cmdact command, and {$LUT_{\# Elems}$} is the total number of rows swept.}

\subsubsection{\matchlogic}
\label{sec:match_logic}

{{
      As shown in \Cref{fig:subarray_layout}a, we implement
        {the} \matchlogicit between the source subarray and the \enabledmechanism subarray.
        {This logic comprises a set of \emph{comparators}; there are as many comparators in the \matchlogic as there are \emph{elements} in the source row {buffer.
              Every $i$-th comparator in the \matchlogic} receives as input the following two $N$-bit values, where $N$ is \emph{the bit width of each LUT element:}
          1)~the row index of the currently activated row in the \enabledmechanism subarray, and
          2)~the $i$-th element in the source subarray's row buffer.
          Each comparator outputs an $N$-bit value (the \emph{matchlines} in \mechanism's design) that depends on the result of the comparison between its two $N$-bit inputs: if the two inputs exactly match, all $N$ matchlines at the output are driven high; otherwise, all matchlines are driven low. %
        }
    }}

\subsubsection{\enabledmechanismtitle Row Buffer}
\label{sec:bsa_row_buffer}

\begin{figure}[ht]
  \centering
  \includegraphics[width=\linewidth]{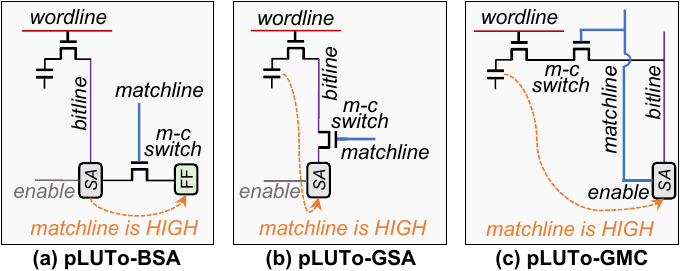}
  \caption{
    {{The three} \mechanism designs.
        \emph{m-c switch} stands for matchline-controlled switch.
        Orange-dashed lines show how charge flows in case the matchline signal is asserted.}}
  \label{fig:pluto_designs_combined}
\end{figure}

{{Performing} the \lutquery {operation} as described in \Cref{sec:lut_query_operation} requires a mechanism to perform many fine-grained (i.e., {LUT}-element-wise) {operations throughout a \rowsweep, to both}
  {1)~}read {data} from the row buffer of the \enabledmechanism subarray, and
  {2)~}write the result of the \lutquery operation to some output buffer.
To realize this functionality, which commodity DRAM does {\emph{not}} support, we connect one flip-flop (FF) to every sense amplifier in the {\mechanism-enabled} row buffer using a {\emph{matchline-controlled switch}} %
(\emph{m-c switch,} shown in {\Cref{fig:pluto_designs_combined}{a}).
      {Each m-c switch is closed only if there is a} match between the row index of the currently activated row in the \mechanism-enabled subarray and {the corresponding LUT index in the LUT Query input vector}}}.
The complete {row} of FFs constitutes {an} \emph{FF buffer}{, which gives \mechanismB \emph{(Buffered Sense Amplifier)} its name.}
In this design, when a sense amplifier reads a {DRAM} cell's value, this value {is also {immediately} written into} the corresponding FF, {but} {only} if the {corresponding} \emph{matchline} signal is high (i.e., if the {m-c} switch {connected to the FF} is closed).

\subsubsection{{Analysis of \mechanismB}}
\label{sec:bsa_analysis}
{\mechanismB's design entails the following {\lutquery throughput, \lutquery energy consumption, and area {overhead} {values:}}}
\begin{itemize}
  \item {\textit{Throughput:} \mechanismB's maximum \lutquery throughput (in \emph{number of LUT queries per second}, LUTs/s) for a single \mechanism-enabled subarray depends on the number of LUT indices that fit in the source subarray
        ($\frac{RowSize_{bytes}}{LUT_{ElementSize}}$)
        and the latency of a \rowsweep
        ($(\trcd + \trp) \times LUT_{\# Elems}$).
        Thus, \mechanismB's maximum throughput  {is}}

        \vspace{-\baselineskip}
        {\footnotesize
          \begin{align*}
            BSA_{Throughput} ={} \frac{RowSize_{bits}/input_{bit\ width}}{(\trcd + \trp) \times LUT_{\# Elems}}\ LUTs/s.
          \end{align*}}%

  \item {\textit{Energy Consumption:}
        The energy \mechanismB consumes during a \rowsweep {depends on the energy consumed by each DRAM row activation/precharge and $LUT_{\# Elems}$, the total number of rows swept in the \mechanism-enabled subarray:}}

        \vspace{-\baselineskip}
        {\footnotesize
          \begin{align*}
            BSA_{Energy} ={} (\ercd + \erp) \times LUT_{\# Elems}.
          \end{align*}}%

  \item \textit{Area Overhead:} The area overhead of \mbox{\mechanismB} includes the area of the \matchlogic ($RowSize_{bytes} \times Area_{ByteComp}$), matchline-controlled switches ($RowSize_{bits} \times Area_{m-c\ switch}$), and FF buffer ($RowSize_{bits} \times Area_{FF}$).
        Thus, \mechanismB's area overhead is

        \vspace{-\baselineskip}
        {\footnotesize
          \begin{align*}
            BSA_{Area} ={} & RowSize_{bytes} \times Area_{ByteComp} +   \\
                           & RowSize_{bits} \times Area_{m-c\ switch} + \\
                           & RowSize_{bits} \times Area_{FF}.
          \end{align*}}%
\end{itemize}

\subsection{{\mechanismAFull Design} %
}
\label{sec:pluto_a}

{
\textit{\mechanismA} differs from \mechanismB in its {\mechanism-enabled} row buffer design and in its implementation of the \rowsweep {operation}.
{\mechanismA's \emph{key idea} is to use the sense amplifier as a buffer that stores {\emph{only} the} LUT elements indicated as a match by the \matchlogic.
{By doing this, \mechanismA eliminates the need for a secondary buffer (such as} the FF buffer in \mechanismB) {to store} LUT elements during a \rowsweep {operation.}}
\mechanismA provides reduced area overhead over \mechanismB, at the expense of reduced {throughput} and energy {efficiency.}
}

{
\subsubsection{\mechanismA Row Buffer}
Each sense amplifier in \mechanismA's row buffer is \emph{gated} from its bitline by a matchline-controlled switch (\Cref{fig:pluto_designs_combined}b).
  {This switch {(i.e., isolation transistor)} incurs} a lower area overhead than the FF buffer used by \mechanismB, resulting in \mechanismA's {higher area efficiency}.
During a \lutquery, if the matchline signal is high (i.e., the switch is closed), the sense amplifier is electrically connected to its bitline, thus amplifying bitline voltage perturbations normally.
However, if the matchline signal is low (i.e., the switch is open), the sense amplifier {does} \emph{not} respond to the perturbation induced by its corresponding {DRAM} cell, which {leads} to the loss of the cell's contents during row activation {(i.e., implementing \mechanismA's row buffer leads to destructive reads)}.
This represents a potential for LUT data loss, which means that \emph{{a} LUT must be loaded into {the} \enabledmechanism subarray before every \lutquery in \mechanismA}, {leading to performance overheads compared to \mechanismB.}
}

\subsubsection{The \rowsweep}
{
  The latency of the \rowsweep operation is lower in \mechanismA than in \mechanismB.
  This is because
  1) row activations in \mechanismA's \rowsweep do \emph{not} require the full-fledged activation process required to read DRAM data, but \emph{only the triggering} of the charge sharing process, and
  2) the \rowsweep does \emph{not} require a {precharge (\cmdprech)} command to be issued after every {activate (\cmdact)} command, since {each unmatched bitline {remains} in a}
  precharged state {({i.e.,} the voltage in each bitline remains at $V_{DD}/2${)} \emph{until} the \matchlogic registers a match.}
  Instead, a single \cmdprech command may be issued at the end of the \rowsweep.
  However, row activations in \mechanismA's \rowsweep lead to LUT data loss for unmatched elements: when an \emph{unmatched} DRAM cell capacitor is discharged, its charge level will \emph{not} be restored since the matchline-controlled switch leaves the path between the bitline and the sense amplifier open.}

The total time required to perform a \rowsweep in \mechanismA {is equal} to {$\trcd \times N + \trp$}, where {$\trcd$ {($\approx$ \SI{12.5}{\nano\second} in DDR4~\cite{jedec2017sdram})}} is the time that must elapse between {the DRAM chip receiving {an} \cmdact command and the sense amplifier {finishing} sensing the voltage perturbation in the bitline,}
$\trp$ {($\approx$ \SI{12.5}{\nano\second} in DDR4~\cite{jedec2017sdram})} is the precharge time, and {$LUT_{\# Elems}$} is the total number of rows swept.
This is about half the time a \rowsweep requires in \mechanismB.\footnote{\label{fn:gsa_vs_bsa}{The ratio between {the latencies of the \rowsweep in \mechanismB and \mechanismA} is given by $\frac{(\trcd + \trp) \times N}{\trcd \times N + \trp}$.
    Under the assumption that $\trcd \approx \trp$, this simplifies to
    $\frac{2 \times N}{1+N}$, which {approaches} 2 for large values of $N$}.}
{However, due to the destruction of row contents during the \rowsweep, the calculation of the average} latency of a \lutquery in \mechanismA must {also factor in the latency of loading data into a \enabledmechanism subarray before {\emph{every}} \rowsweep $(LISA_{RBM} \times LUT_{\# Elems})$.}
As a result, while the latency of the \rowsweep operation is lower in \mechanismA than in \mechanismB, the latency of the \emph{total} \lutquery is higher in \mechanismA than in \mechanismB.

\subsubsection{{Analysis of \mechanismA}}
\label{sec:gsa_analysis}
{\mechanismA's design entails the following {\lutquery throughput, \lutquery energy consumption, and area {overhead} {values:}}}
\begin{itemize}
  \item {\textit{Throughput:} \mechanismA's maximum throughput (in \emph{number of LUT queries per second}, LUTs/s) for a single \mechanism-enabled subarray depends on the number of LUT indices that fit in the source subarray ($\frac{RowSize_{bytes}}{LUT_{ElementSize}}$), the latency of a \rowsweep operation ($\trcd \times LUT_{\# Elems} + \trp$), and the latency of loading LUT elements into the \mechanism-enabled subarray{, since LUT elements are destroyed after each \rowsweep $(LISA_{RBM} \times LUT_{\# Elems})$.}
        Thus, the maximum throughput achievable with \mechanismA is}

        \vspace{-\baselineskip}
        {\footnotesize
          \begin{align*}
            GSA_{Throughput} = \frac{RowSize_{bits}/input_{bit\ width}}{LISA_{RBM} \times LUT_{\# Elems} + (\trcd \times LUT_{\# Elems} + \trp)}\ LUTs/s.
          \end{align*}}%

  \item {\textit{Energy Consumption:} The energy \mechanismA consumes during a \rowsweep operation depends on the number of elements in the LUT and the energy consumed by a DRAM row activation and precharge operation:}

        \vspace{-\baselineskip}
        {\footnotesize
          \begin{align*}
            GSA_{Energy} = \elisa \times LUT_{\# Elems} + \ercd \times LUT_{\# Elems} + \erp.
          \end{align*}}%

  \item {\textit{Area {Overhead:}}
        The area overhead of \mechanismA's design includes the area of the \matchlogic ($RowSize_{bytes} \times Area_{ByteComp}$), and matchline-controlled switches ($RowSize_{bits} \times Area_{m-c\ switch}$).
        Thus, the total area {overhead} of \mechanismA is}

        \vspace{-\baselineskip}
        {\footnotesize
          \begin{align*}
            GSA_{Area} ={} & RowSize_{bytes} \times Area_{ByteComp} +  \\
                           & RowSize_{bits} \times Area_{m-c\ switch}.
          \end{align*}}%
\end{itemize}

\subsection{{\mechanismCFull Design}}
\label{sec:pluto_c}

{
\textit{\mechanismC} provides {higher throughput} and energy efficiency over \mechanismB, at the expense of increased area overhead.
\mechanismC differs from \mechanismB in its DRAM cell design, {\mechanism-enabled} row buffer design, and \rowsweep implementation.
  {Similarly to \mechanismA, \textit{\mechanismC}'s \emph{key idea} is to use the sense amplifier in the \mechanism-enabled subarray as a buffer to store {the} matched LUT elements during a \lutquery (instead of adding a new {buffer, as} \mechanismB does).
    However, in contrast to \mechanismA, {row activations during the \rowsweep in \mechanismC are \emph{not} destructive, since charge is allowed to flow from the DRAM cell to the bitline} {\emph{only if}} {there is} a match between the row index and {the} LUT indices in the source subarray.
    To do so, \mechanismC adds an extra transistor in \emph{each DRAM cell} of the \mechanism-enabled subarray.
    As such, \mechanismC is the most intrusive to the subarray design (because it changes the DRAM cell itself).}}

{
\subsubsection{\mechanismC DRAM Cell}
\mechanismC implements {a} 2T1C {DRAM} cell instead of the conventional 1T1C design {{(described in Section~\ref{dram-background}).
        An} additional transistor connects the DRAM cell's access transistor to the bitline}.
{The output of the \matchlogic (i.e., the matchline)} controls the additional transistor in each  cell, as shown {in \Cref{fig:pluto_designs_combined}c.}
The matchline signal thus {controls whether a cell} in an activated row {shares} charge with the bitline.
This significantly reduces the overall movement of charge, {since charge only flows between the DRAM cell and the bitline if the \matchlogic outputs a match during the \lutquery operation,} which reduces the overall energy consumption of \mechanismC {during a \rowsweep {compared} to both \mechanismB and \mechanismA}.}

\subsubsection{\mechanismC Row Buffer}
{
  In \mechanismC, additional matchline-controlled switches exist between each sense amplifier and its enable signal.
  The role of these switches is to ensure that, when a row is activated, the sense amplifier connected to a given {DRAM} cell only senses the bitline voltage if \emph{both the wordline and the matchline signals are high}.
  Without this safeguard, the sense amplifiers would be activated when cells in the active row are not connected to the bitline (i.e., when the wordline signal is high for a given row, but the matchline signals are low for one or more cells in that row), which would lead to undefined behavior.
  In addition, these matchline-controlled switches enable \mechanismC to perform back-to-back activations \emph{without needing to {precharge the subarray.}}
  This happens because, when a matchline signal is low, the corresponding bitline behaves as if it had remained inactive, which keeps it in its precharged state.}
{Conversely, \emph{only when the matchline is high} does the sense amplifier become enabled.}

\subsubsection{The \rowsweep}
{\mechanismC optimizes the \rowsweep operation by introducing the ability to perform back-to-back activations \emph{without} the need to precharge the bitlines.
  Leveraging this optimization, \mechanismC outperforms \mechanismB in the  \rowsweep by almost $2\times$.\footnote{{See \Cref{fn:gsa_vs_bsa}.
        The latency of the \rowsweep in \mechanismA and \mechanismC is the same.}}
  To achieve this optimization, \mechanismC adopts the following two key design features.
  First, a sense amplifier is only enabled when there is a match in the corresponding \matchlogic.
  This means that {an activation \emph{only} perturbs a bitline if the associated matchline signal is high, and that the voltage in the bitlines is \emph{kept} at $V_{DD}/2$ (i.e., in the precharged state) if the matchline signal is low.
      Second, since each source row element necessarily only has one match in a LUT, the sense amplifier is only enabled for a single row activation during an entire \lutquery.
      Therefore, we can guarantee that back-to-back row activations will \emph{not} open the gating transistors of any two cells sharing the same bitline, and thus will not destroy the data in the cell.
      As in \mechanismA, the total time required to perform a \rowsweep in \mechanismC is {$\trcd \times LUT_{\# Elems} + \trp$}.
      In addition, due to matchline-controlled switches (\Cref{sec:bsa_row_buffer}), \mechanismC does \emph{not} destroy the data in the LUTs; this translates into significant performance gains, as there is no need to repeatedly load LUT data into the subarray.}}

\subsubsection{{Analysis of \mechanismC}}
\label{sec:gmc_analysis}
{\mechanismC's design entails the following {\lutquery throughput, \lutquery energy consumption, and area {overhead} {values}:}}
\begin{itemize}
  \item {\textit{Throughput:} \mechanismC's maximum throughput (in \emph{number of LUT queries per second}, LUTs/s) for a single \mechanism-enabled subarray depends on the number of LUT indices that fit in the source subarray ($RowSize_{bits}/input_{bit\ width}$) and the latency of a \rowsweep operation ($\trcd + \trp \times LUT_{\# Elems}$).
        Thus, the maximum throughput achievable with \mechanismC is}

        \vspace{-\baselineskip}
        {\footnotesize
          \begin{align*}
            GMC_{Throughput} = \frac{RowSize_{bits}/input_{bit\ width}}{\trcd \times LUT_{\# Elems} + \trp}\ LUTs/s.
          \end{align*}}%

  \item {\textit{Energy Consumption}: The energy \mechanismC consumes during a \lutquery operation depends on the number of elements in the LUT and the energy consumed by a DRAM row activation and precharge operation:}

        \vspace{-\baselineskip}
        {\footnotesize
          \begin{align*}
            GMC_{Energy} ={} \ercd \times LUT_{\# Elems} + \erp.
          \end{align*}}%

  \item {\textit{Area {Overhead}:} The area overhead of \mechanismC's design includes the area of the \matchlogic ($RowSize_{bytes} \times Area_{ByteComp}$) and matchline-controlled switches ($\#Rows \times RowSize_{bits} \times Area_{m-c\ switch}$).
        Thus, the total area {overhead} of \mechanismC is}

        \vspace{-\baselineskip}
        {\footnotesize
          \begin{align*}
            GMC_{Area} ={} & RowSize_{bytes} \times Area_{ByteComp} +                \\
                           & \#Rows \times RowSize_{bits} \times Area_{m-c\ switch}.
          \end{align*}}%

\end{itemize}

\subsection{{Summary of} \mechanism Architectures}
\label{sec:alternative_architectures}

{While the \mechanismB design provides a balanced trade-off between performance, energy efficiency{,} and area overhead,} {the system designer could} %
prefer to optimize for one of these three metrics in isolation.
  {To} provide this added flexibility, we {described \mechanismA and \mechanismC, {two additional \mechanism designs with different trade-offs in performance, energy efficiency, and area overhead.}}
  {\Cref{tab:pluto-modes} summarizes the trade-offs of each of the proposed \mechanism designs.}

\begin{table}[ht]
  \centering
  \caption{Comparison of pLUTo designs' core attributes.
      {Bold cells represent key benefits of a \mechanism design compared to others.} {$N$ corresponds to LUT elements (i.e., $LUT_{\#Elems}$).}}
  \begin{small}
    \renewcommand{\arraystretch}{1.2}
    \resizebox{\columnwidth}{!}{%
      \begin{tabular}{c|c|c|c|}
        \cline{2-4}
                                                           & \textbf{pLUTo-BSA}        & \textbf{pLUTo-GSA}                              & \textbf{pLUTo-GMC}        \\ \hhline{-|=|=|=|}
        \multicolumn{1}{|c||}{\textbf{Area Efficiency}}    & Medium                    & \textbf{High}                                   & Low                       \\ \hline
        \multicolumn{1}{|c||}{\textbf{Throughput}}         & Medium                    & Low                                             & \textbf{High}             \\ \hline
        \multicolumn{1}{|c||}{\textbf{Energy Efficiency}}  & Medium                    & Low                                             & \textbf{High}             \\ \hline
        \multicolumn{1}{|c||}{\textbf{Destructive Reads}}  & \textbf{No}               & Yes                                             & \textbf{No}               \\ \hline
        \multicolumn{1}{|c||}{\textbf{{LUT} Data Loading}} & \textbf{Once}             & After every use                                 & \textbf{Once}             \\ \hline
        \multicolumn{1}{|c||}{\textbf{Query Latency}}      & $(\trcd + \trp) \times N$ & {$LISA_{RBM} \times N + \trcd \times N + \trp$} & {$\trcd \times N + \trp$} \\ \hline
        \multicolumn{1}{|c||}{\textbf{Query Energy}}       & $(\ercd + \erp) \times N$ & {$\elisa \times N + \ercd \times N + \erp$}     & {$\ercd \times N + \erp$} \\ \hline
      \end{tabular}
    }
  \end{small}
  \label{tab:pluto-modes}
\end{table}

{
\Cref{tab:pluto-modes} and the expressions for throughput, energy consumption and area overhead derived in \Cref{sec:bsa_analysis,sec:gsa_analysis,sec:gmc_analysis} enable three key observations.
First, \mechanismC provides the highest throughput of the three designs ($GMC_{Throughput} > BSA_{Throughput} > GSA_{Throughput}$).
\mechanismC achieves this by 1) eliminating the need to issue \cmdprech commands following every row activation in the \enabledmechanism subarray (as in \mechanismB), and 2) eliminating the need to load all LUT elements before each \lutquery operation (as required by \mechanismA, due to the destructive {row activations required by its \rowsweep operation)}.
Second, GMC provides the highest energy efficiency of the three designs ($GMC_{Energy} < BSA_{Energy} < GSA_{Energy}$).
\mechanismC achieves this by
1) eliminating the energy overhead associated with the issuance of \cmdprech commands following every row activation in the \enabledmechanism subarray (as in \mechanismB), and
2) eliminating the energy overhead associated with {loading the LUTs} before each \lutquery (as in \mechanismA).
Third, GSA incurs the smallest area overhead of the three designs ($GSA_{Area} < BSA_{Area} < GMC_{Area}$).
\mechanismA's area overhead is minimized by making as few modifications to the DRAM array as possible.
In particular, \mechanismA does \emph{not} employ the logic components required for the operation of \mechanismB (i.e., the FF buffer) or \mechanismC (i.e., per-cell {matchline-controlled} switches).
}

We conclude that \mechanismA is the most well-suited design to minimize {area overhead}, \mechanismC is the most well-suited {design to} maximize either performance or energy efficiency, and \mechanismB provides a trade-off point that offers intermediate throughput, energy efficiency, and area overhead metrics.

\subsection{Subarray-Level Parallelism}
\label{sec:salp}

{Since the lookup operations of different input values are independent {of} one another, {many} \lutqueries {can be} executed simultaneously across {\emph{multiple}} subarrays {by} exploiting \emph{subarray-level parallelism} (SALP)~\cite{kim2012case}, as described in \Cref{sec:dram_extensions}.}
{Two important use cases benefit from the distribution of LUT queries across multiple subarrays:}
1) {LUT query input vectors} with a very large number of {LUT indices can} be partitioned across multiple source subarrays to be queried simultaneously{;} and
2) independent \lutqueries (possibly belonging to different threads or applications) {can} be executed concurrently.

  {The achievable degree of subarray-level parallelism is limited by the \tfaw DRAM timing {constant}~\cite{jedec2012sdram, jedec2017sdram}, which corresponds to the duration of the time window during which \emph{at most four} \cmdact commands {can} be issued, per DRAM rank.}
This constraint protects against
the deterioration of the DRAM reference voltage, although DRAM manufacturers have been able to mitigate it {substantially in commodity DRAM chips in recent years~\cite{micron2013tfaw}, as well as to perform a targeted reduction of this parameter specifically for PiM architectures where it becomes a performance bottleneck~\cite{he2020newton}.
    These advances suggest that this parameter may not limit \mechanism's scalability severely.}

\subsection{Limitations of \mechanism's Subarray Design}

{
For a single-subarray \lutquery, the number of LUT {elements} can scale up to \emph{the number of rows in the subarray.}
To query LUTs with a greater number of {elements}, it is possible to partition a \lutquery across subarrays.
Note that partitioning the query \emph{does not increase} latency (since multiple subarrays operate simultaneously), but \emph{does increase} energy consumption $N$-fold, for a \lutquery distributed across $N$ subarrays.
For this reason, the design of \mechanism is not well suited for executing large-bit-width lookup queries.
We {leave} the potential exploration of alternative designs that address this limitation for future work.}

{
\subsection{The Role of \mechanism in the PiM Landscape}
\label{sec:integration_prior_pum}
As discussed in \Cref{sec:introduction}, PnM and PuM are complementary approaches: the former enables flexible substrates that support a diverse range of operations, while the latter yields maximal {performance and energy efficiency} benefits. %
{\mechanism does {\emph{not}} aim to replace prior PuM proposals.
Instead, it addresses an important gap in the literature and enables PuM to support {more complex} operations, as many applications require.
Ideally, a real-world PiM system would combine the strengths of different proposals: for example, relying on SIMDRAM~\cite{hajinazar2020simdram} for addition, \mechanism for trigonometric functions {and bit counting operations}, and near-memory general-purpose cores~\cite{hmc_spec} for serial reduction {and other irregular tasks}.
{Mapping} application segments to their most suitable {PiM} substrates is {a rich area for} future work.
}
}

\section{System Integration}
\label{sec:system_integration}

\begin{figure}[ht]
  \centering
  \includegraphics[width=\linewidth]{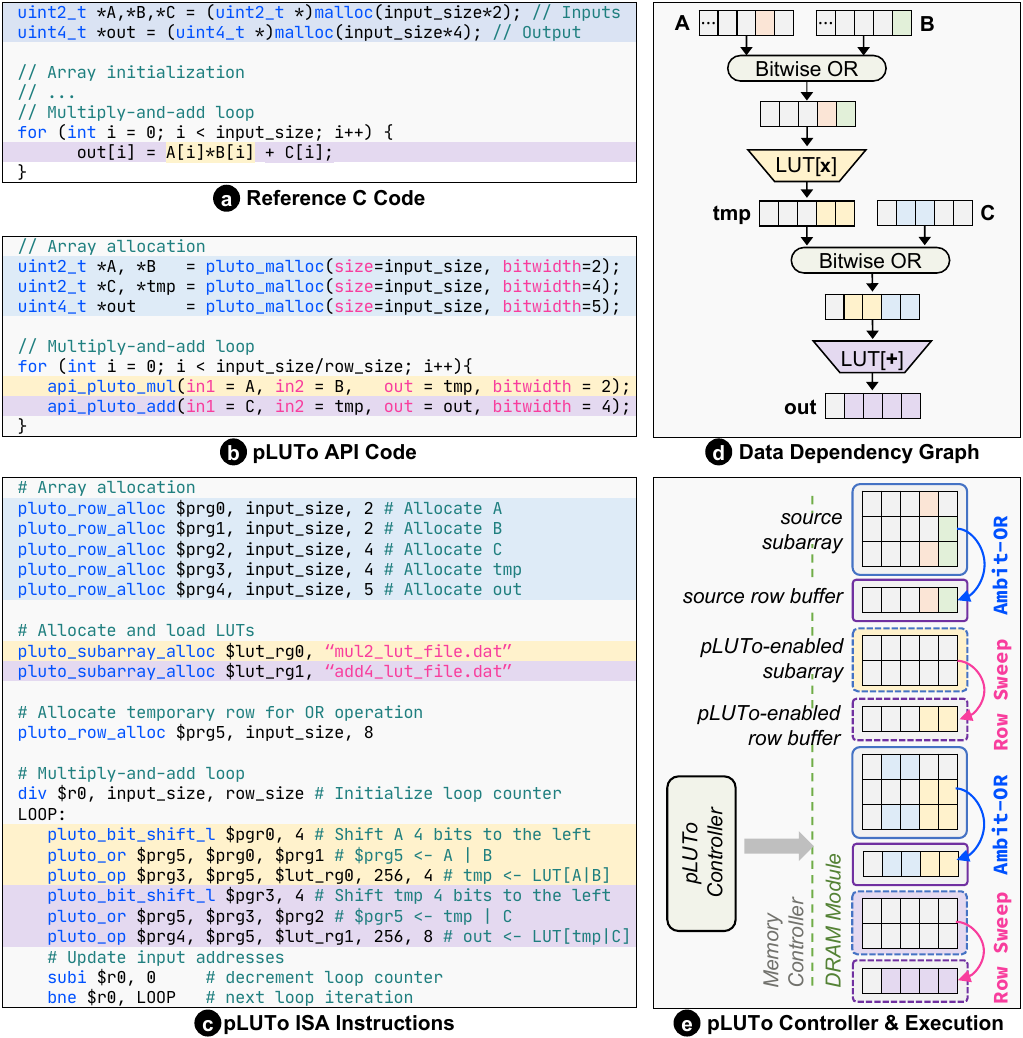}
  \caption{\mechanism's system integration stack.
    An example is shown for the C code displayed in \circledletter{a}.
    Subsequent steps are shown in top-down, left-to-right order: \circledletter{b} implementation using \mechanism's API Library, {\circledletter{c} the {transformation of the \mechanism API code performed by the {\mechcompiler}},} {\circledletter{d} data dependency graph analysis,} \circledletter{e} the role of the {\mechcontroller} and in-memory execution.
  }
  \label{fig:pluto_framework}
\end{figure}

This section describes the system integration stack that enables \mechanism to {operate} seamlessly with the host system.
  {
    There are four key components in this stack:
    1) {the \emph{\mechanism ISA}} (\Cref{sec:mech_isa_extensions}), a set of instructions that express
    i) \rowsweep operations (\plutoop),
    ii) bitwise logic operations~\cite{seshadri2017ambit},
    iii) bit- and byte-level shifting operations~\cite{li2017drisa}, and
    iv) data movement operations {involving multiple {DRAM} rows~\cite{chang2016lisa};}
    2) the \emph{\mechanism library} (\Cref{sec:mech_library}), a set of routines
    that implement complex operations (i.e., operations that involve several \mechanism ISA {instructions});
    3) the \emph{{\mechcompiler}} (\Cref{sec:mech_compiler}), which {analyzes data {dependences} (necessary for data allocation and alignment) and} translates \mechanism library routines to \mechanism ISA {instructions};
    4) the \emph{{\mechcontroller}} (\Cref{sec:mech_controller}), a modified DRAM controller that supports the execution of \mechanism ISA {instructions} (i.e., given a \plutoop, the {\mechcontroller} carries out the corresponding \rowsweep operation through a series of {\cmdact} and {\cmdprech}  DRAM commands).
    The description of the entire system integration stack is {depicted} by~\Cref{fig:pluto_framework}, which {shows} an end-to-end example of \mechanism's operation, from {reference C code (\circledletter{a}) to in-memory computation (\circledletter{e}).
        This example describes} the implementation of the multiply-and-add
    ($\texttt{A} \odot \texttt{B} + \texttt{C}$)
    operation between three vectors{: \texttt{Row A} (2-bit elements), \texttt{Row B} (2-bit elements), and \texttt{Row C} (4-bit elements).}
    \Cref{sec:create_load_store_luts} describes the methods of creating the LUTs used in \lutqueries.
    Finally, \Cref{sec:mech_stack_limitations} discusses the limitations of our proposed system integration of \mechanism and how we can mitigate these limitations.
  }

\subsection{{The \mechanism ISA}}
\label{sec:mech_isa_extensions}

{We propose ISA {extension instructions} that express the operations required by \mechanism to perform in-memory computation.
    {The instructions in the \mechanism ISA} {manipulate {special-purpose} \emph{\mechanism registers} that keep track of the currently allocated \mechanism data structures}.
  We describe ISA {instructions for} {
      1) allocating memory,
      2) querying LUTs (the \plutoop),
      and
      3) manipulating data} (in-memory bitwise logic~\cite{seshadri2017ambit}, bit shifting~\cite{li2017drisa}, and data copy~\cite{chang2016lisa} {operations), as summarized in \Cref{tab:isa_extensions}.}}
{
  {\Cref{fig:pluto_framework}~\circledletter{c}}
    {shows the translation of}
  instructions from the reference program (\circledletter{a}) {into a sequence of} {\mechanism ISA instructions.}
}

\begin{table}[h]
  \centering
  \caption{Summary of \mechanism ISA extension instructions.}
  \renewcommand{\arraystretch}{0.93}
  \resizebox{1.0\columnwidth}{!}{%
    \begin{tabular}{@{}clc@{}}
      \toprule
      \textbf{Operation}                                                                       & \multicolumn{1}{c}{\textbf{Instruction}}                       & \textbf{Proposed in}     \\ \midrule
      \multirow{2}{*}{\begin{tabular}[c]{@{}c@{}}pLUTo \\ Register Allocation\end{tabular}}    & \texttt{pluto\_row\_alloc dst, size, bitwidth}                 & This work                \\
                                                                                               & \texttt{pluto\_subarray\_alloc dst, num\_rows, lut\_file}      & This work                \\ \midrule
      \begin{tabular}[c]{@{}c@{}}pLUTo \\ Row Sweep\end{tabular}                               & \texttt{pluto\_op dst, src, lut\_subarr, lut\_size, lut\_bitw} & This work                \\ \midrule
      \begin{tabular}[c]{@{}c@{}}Bitwise \\ Logic Operations\end{tabular}                      & \texttt{pluto\_\{not, and, or\} dst, src1, scr2}               & \cite{seshadri2017ambit} \\ \midrule
      \multirow{2}{*}{\begin{tabular}[c]{@{}c@{}}Bit- and Byte-\\ Level Shifting\end{tabular}} & \texttt{pluto\_bit\_\{shift\_l, shift\_r\} src, \#N}           & \cite{li2017drisa}       \\
                                                                                               & \texttt{pluto\_byte\_\{shift\_l, shift\_r\} src, \#N}          & \cite{li2017drisa}       \\ \midrule
      \begin{tabular}[c]{@{}c@{}}In-DRAM \\ Data Movement\end{tabular}                         & \texttt{pluto\_move dst, src}                                  & \cite{chang2016lisa}     \\ \bottomrule
    \end{tabular}
  }

  \label{tab:isa_extensions}
\end{table}

\head{\mechregisters}
\phantomsection
\label{sec:mech_registers}
{\mechanism's instructions operate at the granularity of \textit{{contiguously allocated} {DRAM} rows} ({for both LUT query input and output vectors}%
) and \textit{{contiguously allocated} {DRAM} subarrays} (\enabledmechanism subarrays that hold LUTs).
{{To} guarantee that the physical memory {addresses} of all DRAM rows involved in a \lutquery operation are contiguously allocated in the DRAM array,} we define two data structures (\emph{row registers} and \emph{subarray registers}) that {separately} capture these {two} abstractions.
  {Each \emph{\mechrowregister} is a special-purpose architectural register that identifies a DRAM \emph{row} to be used either as the input or the output of a \lutquery.}
  {Each \emph{\mechsubarrayregister} is a special-purpose architectural register that identifies a LUT-holding DRAM \emph{subarray} to be used in a \lutquery.}
These two types of registers {are} used as arguments {of \mechanism's ISA {instructions}} where appropriate.
  {The allocation of \mechanism Registers is performed} by the operating system {via a} call to a \mechanism allocation routine (see ``\allocationroutines'', \Cref{sec:allocation_routines}).
The allocation of both register types is recorded in an in-memory allocation table{,} which the {\mechcontroller} (\Cref{sec:mech_controller}) {accesses} to derive the \emph{physical memory addresses} required {to issue DRAM commands during the execution of \lutqueries.}}

\head{\allocationroutines}
\phantomsection
\label{sec:allocation_routines}
{
  We {introduce
      two instructions to} enable the compiler or the programmer to allocate {\mechanism Registers.}
  The {first instruction,}
  \texttt{pluto\_row\_alloc}, allocates the memory space (as a \emph{whole number} of memory rows) to be used by a source or destination row involved in the execution of \mechanism ISA {instructions}.
}
{{This instruction {has two inputs (\texttt{size} and \texttt{bitwidth}) and one output (\texttt{dst}). It} sets \texttt{dst} to a valid
      \mechanism Row Register that is used to reference the allocated \texttt{size}-byte memory row(s) whose elements are \texttt{bitwidth}-bits wide.}
  \texttt{bitwidth} is only a meaningful parameter for data structures used as {\emph{inputs}}, and is equal to $\log_{2}(\texttt{lut\_size})$, where \texttt{lut\_size} is number of elements in the LUT to query.}

{
  {
      The {second instruction,} \texttt{pluto\_subarray\_alloc}, allocates memory space corresponding to consecutive rows belonging to a single subarray, in which the LUT required by {a} \plutoop will be stored.}}
{{This instruction {has two inputs (\texttt{num\_rows} and \texttt{lut\_file}) and one output (\texttt{dst}). It} sets \texttt{dst} to a valid pointer that {references} the allocated {DRAM} subarray.}
    {\texttt{num\_rows} is the number of rows to be reserved, i.e., the number of elements in the associated LUT, and \texttt{lut\_file} is a {memory location} that holds the {LUT data to be stored in} the allocated subarray.}}

\head{{LUT Querying}}
{
  {A \lutquery} (\Cref{sec:lut_query_operation}) uniquely {maps} to a \plutoop instruction{{, which has three inputs (\texttt{src}, \texttt{lut\_subarr}, \texttt{lut\_size}, and \texttt{lut\_bitw}) and one output (\texttt{dst}). }}%
}
{Here,} {\texttt{dst} and \texttt{src} are the \mechrowregisters of the destination and source rows.}
\texttt{lut\_subarr} is the
physical address of the \enabledmechanism subarray where the LUT to query is stored.
  {
    \texttt{lut\_size} is the number of
    LUT elements,
    i.e., the number of rows to sweep.
    \texttt{lut\_size} must be a power of two: more specifically, \texttt{lut\_size}~$:=2^N$, where $N$ is the bit width of each source row value;
    for example, a 4-bit-input LUT contains $2^4=16$ elements, and thus requires the sweeping of 16 rows.}
  {\texttt{lut\_bitw} specifies \emph{the bit width of the LUT elements},\footnote{{\texttt{lut\_bitw} can only be greater than or equal to $N$.
          If $\texttt{lut\_bitw}>N$, the source row values will be zero-padded:
          as an example, for $N=1$ and $\texttt{lut\_bitw}=8$, the 1-bit values \texttt{\{0,1\}} would be zero-padded to a width of 8 bits (i.e., \texttt{\{00000000,00000001\}}) and used to query a 2-entry LUT whose elements may be any 8-bit value (e.g., \texttt{\{00000000,11111111\}}).}} i.e., the \emph{width} of the match logic's comparators for this \texttt{pluto\_op}.}
  {
    A \plutoop instruction always operates at the granularity of {a DRAM row}; as a result, operating on %
    $S$ input bytes requires $\ceil{\frac{S}{DRAM_{row~size}}}$ \plutoop instructions.
  }

\head{Bit Manipulation}
{{\mechanism requires} bit manipulation operations {proposed by prior works~\cite{seshadri2017ambit, li2017drisa, chang2016lisa}, as shown in \Cref{tab:isa_extensions}.
      We use these operations} to align source row values
  (bit shifting {{using} \texttt{pluto\_bit\_*} and \texttt{pluto\_byte\_*}}),
  merge operands between source rows (bitwise {\texttt{OR}} {{using} \texttt{pluto\_or}}),
  apply bit masks to input and output rows (bitwise {\texttt{AND}} {{using} \texttt{pluto\_and}}),
  and copy rows in-memory (row buffer to row buffer data copy {{using} \texttt{pluto\_move}}).}

\subsection{The \mechanism Library}
\label{sec:mech_library}

{The \mechanism library encompasses
  1) {\emph{computation} routines}, which {the programmer} may {conveniently} use to express operations at a high level of {abstraction,} and
  2) {a routine for \emph{memory allocation}} (\texttt{pluto\_alloc}), which {the programmer may use} to instantiate the data structures involved in \mechanism's operation (i.e., the source and destination rows and the LUT-holding subarrays).}

\head{{Computation}}
{Examples of \mechanism Library computation routines include common operations (e.g., \texttt{{api\_}pluto\_add} {and} \texttt{{api\_}pluto\_mul} express addition and multiplication).}
{\Cref{fig:pluto_framework}~\circledletter{b} contains {a code example} with \mechanism library calls (\texttt{{api\_}pluto\_add}, \texttt{{api\_}pluto\_mul}) in place of the addition and multiplication operations in the reference code (\Cref{fig:pluto_framework}~\circledletter{a}).
  Each of the routines in the \mechanism library {{translates} into} a {predetermined, constant} sequence of \mechanism ISA {instructions}.
  For example, the 4-bit addition operation}

\noindent
{
  \centering
  \setlength{\fboxrule}{0pt}
  \fbox{\small\texttt{api\_pluto\_add(in1,in2,out,bitwidth=4)}}
  \par
}

\noindent
{\emph{always} corresponds to the following sequence {of} \mechanism ISA {instructions}:}

\noindent
{
  \setlength{\fboxrule}{0pt}
  \resizebox{!}{0.7\baselineskip}{\fbox{\texttt{pluto\_or temp,in1,in2}}}
  \par
}

\noindent
{
  \setlength{\fboxrule}{0pt}
  \resizebox{!}{0.7\baselineskip}{\fbox{\texttt{pluto\_op dst,temp,add4\_lut,lut\_size=256,lut\_bitw=8}}}
  \par
}

\noindent
{{Here,}
    {\texttt{dst} denotes the destination row to which the result will be stored,
      \texttt{temp} {holds} the result of the bitwise {\texttt{OR}} operation that combines the {two source} {rows},}
  \texttt{add4\_lut} denotes {the subarray} that holds {the LUT} with the results for the 4-bit addition, and
  \texttt{lut\_size} and \texttt{lut\_bitw} are uniquely determined by the bit width of this operation {(256 and 8, respectively, in this example)}.
  \mechanism library routines always assume a specific data alignment (e.g., the \texttt{pluto\_add} operation assumes that the left and right operands are concatenated before performing the LUT query).
  However, these routines \emph{do not explicitly guarantee this alignment;} instead, the responsibility of {ensuring correct} input operand alignment is assumed by the {\mechcompiler}, as explained in \Cref{sec:mech_compiler}.
}

\head{Memory Allocation}
{To abstract the low-level memory allocation instructions defined in  \Cref{sec:allocation_routines}, the \mechanism Library implements the {\texttt{pluto\_malloc} routine, defined as}}
\mbox{\texttt{pluto\_malloc(size,bitwidth)}}.
{Here,} \texttt{size} is the number of bits to be allocated, and
  {\texttt{bitwidth} {is} the bit width of each element (either an input/output value, or a LUT element).
    Based on the dependences between the arguments of this function, the \mechanism Compiler (\Cref{sec:mech_compiler}) is able to infer a sequence of \mechanism ISA {instructions} (i.e., \texttt{pluto\_row\_alloc}, \texttt{pluto\_subarray\_alloc}{)} that are equivalent to it.
    \circledletter{b} and \circledletter{c} in \Cref{fig:pluto_framework} show an example of this compilation process.}

\subsection{The {\mechcompiler}}
\label{sec:mech_compiler}

{
  The role of the {\mechcompiler} is to %
    {identify} the {dependences} between operands used by \mechanism library routines to ensure the correct {allocation (i.e., \texttt{pluto\_alloc\_*})} and alignment of these operands.
  The compiler may achieve operand alignment by inserting additional \mechanism ISA {instructions} to perform bit shifting {(\texttt{pluto\_\{bit,byte\}\_*})}, bit masking {(\texttt{pluto\_and})}, and row merging {(\texttt{pluto\_or})} operations as needed, \emph{in addition to the \mechanism ISA {instructions} {specified} by each \mechanism Library routine}.
  As an example of the role of the compiler, consider the multiplication between the 2-bit elements of arrays \texttt{A} and \texttt{B} {using the \texttt{api\_pluto\_mul} instruction, as} shown in~\Cref{fig:pluto_framework}{~\circledletter{c}}.
  The translation from the {\texttt{api\_pluto\_mul} call} to \mechanism ISA {instructions} {yields only a \texttt{pluto\_or} and a \plutoop,} and therefore does {\emph{not}} guarantee that each value from \texttt{A} is combined with its counterpart from \texttt{B} to create the value to be queried in the LUT.
  In this example, the compiler guarantees correct operand alignment by performing the operations shown in \circledletter{d}, namely
  1) shifting the contents of input row \texttt{A} to the left by two bits, and
  2) merging the result of operation 1) with input row \texttt{B} using a bitwise {\texttt{OR}}.
}

\subsection{The {\mechcontroller}}
\label{sec:mech_controller}

{The {\mechcontroller}, which extends the DRAM controller, executes the \mechanism ISA {instructions} that {are} either
1) specified by the programmer using \mechanism Library routines, or
2) inserted by the {\mechcompiler} to ensure correct operand alignment.
{Each of these ISA {instructions} translates into {either}
  {i)~}a {predefined} sequence of {\cmdact} and {\cmdprech} commands, {which} the {\mechcontroller} {stores} in an internal ROM {(for the execution of bitwise logic operations, bit- and byte-level shifting, and in-DRAM data movement operations); or
    ii)~a \mechanism Row Sweep command}.}

{The \mechcontroller's hardware consists of:
1)~a small internal ROM that maps {each \mechanism ISA {instruction}} {to} appropriate DRAM commands;
2)~a small register file that holds \mechrowregisters; and
3)~a finite state machine that decodes {\mechanism} ISA {instructions}, gathers the physical addresses of all operands of a {\mechanism} ISA {instruction} and controls the execution flow of \mechanism's in-memory operations.
The {hardware and operation of the \mechcontroller resemble those of SIMDRAM's Control Unit~\cite{hajinazar2020simdram}, and thus incur negligible area overhead on the host CPU die ($<$ 0.08\%).}}}

\subsection{Loading LUT Data}
\label{sec:create_load_store_luts}
{To load the LUTs required by \lutqueries, it is necessary to
  1) \emph{allocate LUT subarrays} (using the \texttt{pluto\_subarray\_alloc} operation) that are {adjacent or in close physical} proximity to the source and destination rows, and
  2) \emph{load the LUTs} {into} these subarrays.
  The loading of LUTs may take place in one of three ways, which we quantitatively compare in \Cref{sec:loading_luts}:
}

\head{{1. First-Time} Generation}
{
The first time a LUT is required, its elements {must} be computed from {scratch.}
Optionally, these values may then be saved {in memory} for later reuse.
This {procedure} is {similar} to operand memoization~\cite{suresh2017compile} and presents an opportunity for potential further optimizations, as described in prior works~{\cite{suresh2017compile,wilcox2011mesa,besnard2019framework,pinto2021methodology,oliveira2018employing}}.
We {leave} the exploration of more complex memoization strategies for future work.
}

\head{{2.} Loading From Memory}
If a LUT already exists in memory, the most efficient way to reuse it is by copying it to the designated \enabledmechanism subarray using LISA-RBM~\cite{chang2016lisa} {(if the source and destination subarrays are {in close physical proximity}) or a CPU-mediated copy operation.}

\head{{3.} Loading From Secondary Storage}
{If a LUT {is stored in} secondary storage (e.g., at compile time, or following First-Time Generation), it may be loaded into the main memory at runtime using a direct memory access (DMA) operation.}

\subsection{Limitations of the System Integration Stack}
\label{sec:mech_stack_limitations}

\head{Address Translation}
Ensuring the physical proximity {between the source row, the {LUT-holding subarray}, and
      {the} destination row} requires knowledge of {the physical address mapping of the involved DRAM subarrays, banks, and} {ranks.
    Two} {{possible approaches to achieve this are}
    1) {via} the help of a memory controller {that} can provide this {information,} and
    2) {via} an \textit{a priori} reverse-engineering effort that allows the memory mapping scheme to be recovered~\cite{barenghi2018software}.}

\head{Coherence}
\mechanism does \emph{not} provide means to enforce coherence between the data stored in \mechanism subarrays and the data stored in other locations in the system {(e.g., CPU caches).}
  {For this reason, programmers are responsible for preventing data decoherence stemming from modifications by simultaneous CPU and \mechanism operations {(e.g., using instructions to flush cache lines belonging to memory addresses that \mechanism will operate on).
        \mechanism can leverage coherence optimizations tailored to PiM to improve overall performance~\cite{boroumand2019conda,boroumand2017lazypim,amiraliphd}.}}

\section{Methodology}
\label{sec:evaluation_methodology}

We evaluate the {three {proposed} \mechanism designs}{: \mechanismA, \mechanismB, and \mechanismC}.
{Unless stated otherwise, our} implementations assume the parallel operation of 16 subarrays with \SI{8}{\kilo\byte} row buffers for DDR4 {memory}{~\cite{kim2012case},} and 512 subarrays with \SI{256}{\byte} row buffers for {3D-stacked (3DS)~\cite{pawlowski2011hybrid} memory}.
{These two design points are comparable since
  {the} volume of data processed \emph{per operation} is identical in both cases: $16 \times 8\ kB = 512 \times 256\ B = 128\ kB$.}
We compare %
  {each} \mechanism design %
against four baselines:
1) a state-of-the-art {CPU,}
2) a state-of-the-art {GPU,}
3) a {simulated} {Processing-near-Memory (PnM) }accelerator, and
4) a {simulated} FPGA. {Table~\ref{table:evalparam} shows the main {parameters we} {use} in our evaluations.}

\begin{table}[ht]
  \setlength\tabcolsep{1.5pt} %
  \footnotesize
  \centering
  \caption{Configuration of the simulated system.}
  \resizebox{0.95\columnwidth}{!}{%
    \begin{tabular}{m{2cm}m{6.2cm}}
      \toprule
      \textbf{Parameter} & \textbf{Configuration}                                                                                                                                                                                    \\\toprule
      Main Memory        & DDR4 \SI{2400}{\mega\hertz}, \SI{8}{\giga\byte}, 1-channel, 1-rank, 4-bank groups, 4-banks per bank group, 512 rows per subarray, \SI{8}{\kilo\byte} per row; timings 17-17-17 (\SI{14.16}{\nano\second}) \\\hline
      {PnM}              & HMC Model~\cite{hmc_spec} with support for bitwise operations~\cite{seshadri2017ambit} and bit shifting~\cite{li2017drisa}; on-die core with \SI{1.25}{\giga\hertz} clock, \SI{10}{\watt} TDP             \\\hline
      FPGA               & Zynq® UltraScale+ MPSoC ZCU102~\cite{xilinxzcu102}                                                                                                                                                        \\\hline
      \mechanism         & {16-subarray parallelism~\cite{kim2012case} unless stated otherwise; \endgraf unthrottled rate of row activations ($\tfaw=0s$)}                                                                           \\\toprule
    \end{tabular}
  }
  \label{table:evalparam}
\end{table}

\subsection{Evaluation Frameworks}
\label{sec:evaluation_frameworks}

\head{Baselines}
{We evaluated the CPU and GPU baselines on a real system equipped with an Intel® Xeon Gold 5118~\cite{intel_gold_cpu} CPU and {an NVIDIA® GeForce RTX 3080 Ti~\cite{nvidia_3080ti} GPU}}.
The CPU versions of our evaluated workloads employ SSE2 and SSE4 Intel® Streaming SIMD Extensions.
  {
    We evaluate the FPGA baseline using high-level synthesis (HLS) implementations created with Vitis 2020.1~\cite{xilinx2020vitis} and Vivado 2020.1~\cite{xilinx2020vivado}{, and perform} post-synthesis simulation for a %
    Xilinx Zynq UltraScale ZCU102 FPGA~\cite{xilinxzcu102}.
  }
For the evaluation of the {PnM} baseline, we simulate an HMC-based system~\cite{hmc_spec} with support for bulk bitwise operations as described in {Ambit}~\cite{seshadri2017ambit} and {bit} shifting as described in {DRISA}~\cite{li2017drisa}.
We simulate various configurations of \mechanism on DDR4~\cite{jedec2017sdram} and HMC~\cite{hmc_spec} memory models using a custom-built simulator, which we have made publicly available at~\cite{plutorepo,plutorepozenodo} under an open-source license{.}%

\head{\mechanism}
Our simulator estimates the performance of \mechanism operations by parsing the sequence of memory commands required to perform them and enforcing the memory's timing parameters.
The simulator then outputs the total time elapsed {and energy consumed to complete
    the operations.}

\head{Energy and Area}
We evaluate the energy consumption and area overhead of \mechanism configurations using CACTI~7~\cite{balasubramonian2017cacti} DDR4 and HMC models.
These models supply the energy consumption of each memory command and the area of each memory component.
Using these values, we extrapolate \mechanism's energy consumption and area overhead by considering the transistor count associated with the logic required to implement its functionality, including 1) the addition of the match logic, 2) modifications to the subarray architecture and memory controller, and 3) the addition of the \mechanism controller.

\subsection{Workloads}

\Cref{table:workloads} shows the names and characteristics of the workloads we {evaluate.
    We select these workloads because
    1) they exemplify general-purpose, real-world functions that cannot be efficiently executed by previous Processing-using-Memory architectures~\cite{seshadri2017ambit,hajinazar2020simdram} (e.g., substitution tables~\cite{bernstein2005salsa20,zoltak2004vmpc}, polynomial division~\cite{warren2013hacker}),
    2) they include segments that are well{-}suited for LUT-based computation, and
    3) their typical working sets are much larger than the cache size of modern systems.
  }

\begin{table}[h]
  \setlength\tabcolsep{1.5pt} %
  \footnotesize
  \centering
  \caption{Evaluated workloads.} %
  \resizebox{0.95\columnwidth}{!}{%
    \begin{tabular}{m{4.8cm}m{3.7cm}}
      \toprule
      \textbf{Name} & \textbf{Parameters}                                      \\\toprule

      Vector Addition, LUT-based~\cite{wang2013augem}
                    & Element width: 4 bits                                    \\ \hline

      Vector Point-Wise Multiplication~\cite{wang2013augem}
                    & Q Format: Q1.7, Q1.15                                    \\ \hline

      {Row-Level Bitwise Logic Operations}~\cite{seshadri2017ambit}
                    & \# LUT entries: 4                                        \\ \hline

      Bit Counting~\cite{warren2013hacker}
                    & BC-4: 4 bits, 16-entry LUT; BC-8: 8 bits, 256-entry LUT  \\ \hline

      CRC-8/16/32~\cite{warren2013hacker}
                    & Packet size: \SI{128}{\byte}                             \\ \hline

      Salsa20~\cite{bernstein2005salsa20}, VMPC~\cite{zoltak2004vmpc}
                    & Packet size: \SI{512}{\byte}                             \\ \hline

      Image Binarization {(ImgBin)~\cite{imgbin2022mathworks}}
                    & 3-channel 8-bit image, 936000 pixels; threshold: 50\%    \\ \hline

      Color Grading {(ColorGrade)~\cite{colorgrade2022apple}}
                    & One 3-channel 8-bit image, 936000 pixels; 8-bit to 8-bit \\\toprule
    \end{tabular}
  }
  \label{table:workloads}
\end{table}

\section{Evaluation}
\label{sec:evaluation}

In this section, we evaluate \mechanism's {reliable and correct} operation~(\Cref{sec:spice}), performance~(\Cref{sec:performance}), energy {consumption}~(\Cref{sec:energy_analysis_of_workloads}), and area overhead~(\Cref{sec:area_overhead}).
  {
    We also carry out performance sensitivity analyses to assess the cost of loading LUTs~(\Cref{sec:loading_luts}), the scalability of \mechanism~(\Cref{sec:scalability_analysis}), \tfaw's impact on performance~(\Cref{sec:tfaw}), and the effect of varying degrees of subarray-level parallelism~(\Cref{sec:analysis_of_subarray_parallelism}).
  }
Finally, we discuss how \mechanism compares to {various prior approaches}~(\Cref{sec:comp_prior_works}).

\subsection{{Reliability and Correctness}}
\label{sec:spice}
\begin{figure}[ht]
  \centering
  \includegraphics[width=0.94\linewidth]{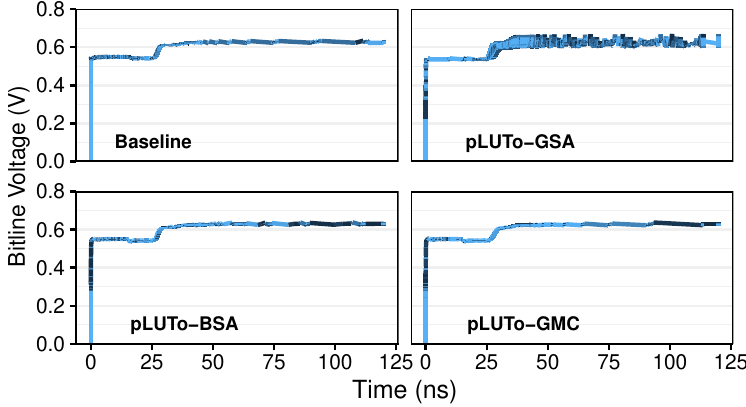}
  \vspace{-3pt}
  \caption{{Bitline voltage level over time in response to wordline activation at $time = 0$.} %
      {Shades of blue {indicate} different runs of our Monte Carlo simulation.} %
  }
  \label{fig:spice_sims}
  \vspace{-5pt}
\end{figure}

We {perform} circuit-level SPICE simulations to verify that the modifications required by each of the three \mechanism designs do not compromise the correct and reliable operation of DRAM.
We model the {effect of activating} a DRAM row {in} unmodified DRAM and {in} each of the three designs {of \mechanism}.
We model DRAM cells based on Low-Power 22nm Metal Gate PTM transistors~\cite{ptm2012transistors}, and conduct Monte Carlo simulations of 100 runs, where the process variation is assumed to be 5\%.
\Cref{fig:spice_sims} shows the {results of these simulations.}
Our results show that the proposed changes in each of the three \mechanism designs \emph{do not introduce} errors in DRAM operation.
The observed {disturbances in the final bitline voltage} {following a row activation}
correspond to {only} 0.9\% of the reference voltage value.

{
We make three key observations.
First, the correctness of the row activation is not affected in any of the proposed designs, since the bitline voltage reaches the value required to trigger the activation in all cases.
Second, {in all \mechanism} designs, the activation time is \emph{not} affected by the introduced DRAM modifications.
Third, the activation procedure is {the} noisiest for \mechanismA.
This is expected due to the {operational} principles of \mechanismA (\Cref{sec:pluto_a}), {whereby} the {contents of DRAM cells} in consecutive rows are shared with the bitline {\emph{without} {precharging the array after each activation} %
    until the end of the \rowsweep.}
However, we observe correct row activation behavior even in this case.
}

\subsection{Performance}
\label{sec:performance}

\begin{figure}[ht]
  \centering
  \includegraphics[width=\linewidth]{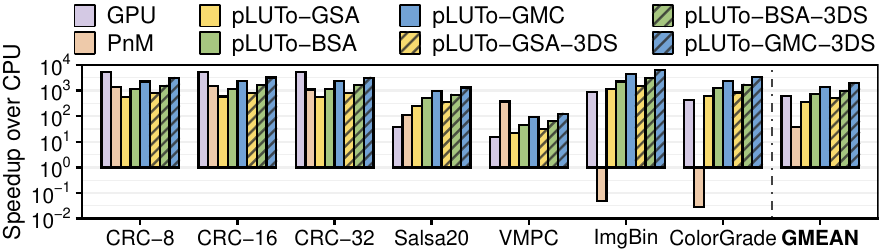}
  \caption{
    Speedup of GPU, {PnM}, and \mechanism relative to the baseline CPU.
    The y-axis uses a logarithmic scale; higher is better.
  }
  \label{fig:normperfA}

\end{figure}

\Cref{fig:normperfA} {shows the performance of GPU, PnM, and \mechanism systems, normalized to the baseline CPU.}
{
  For the DDR4 (3DS) implementation, {the performance of} \mechanismA, \mechanismB, and \mechanismC {mechanisms, on average across all workloads, is}
  {\speedupavgmechA/\speedupavgmechAovergpu/\speedupavgmechAoverpnm (\speedupavgmechATDS/\speedupavgmechATDSovergpu/\speedupavgmechATDSoverpnm), \speedupavgmechB/\speedupavgmechBovergpu/\speedupavgmechBoverpnm (\speedupavgmechBTDS/\speedupavgmechBTDSovergpu/\speedupavgmechBTDSoverpnm), and \speedupavgmechC/\speedupavgmechCovergpu/\speedupavgmechCoverpnm (\speedupavgmechCTDS/\speedupavgmechCTDSovergpu/\speedupavgmechCTDSoverpnm)}
  {that of the CPU/GPU/PnM baselines,} respectively.
  All \mechanism designs achieve performance comparable to {or higher} {than} that of the GPU, and consistently outperform the {PnM} baseline.
  We make two key observations.
  First, {the} 3DS-based \mechanism designs consistently outperform their DDR4 {counterparts} by {38\% on average across the three \mechanism designs. T}his is due to HMC's faster row activations, which lead to faster \lutqueries.
  Second, {the} CRC workloads show the smallest overall benefit from execution in \mechanism.
  The speedup in these workloads is bottlenecked by a serial reduction step, which must be performed in the CPU (\mechanism-DDR4) or in the logic layer {of HMC} (\mechanism-3DS).
  Nevertheless, the acceleration of the parallel portion of the CRC workloads still allows {most} \mechanism designs to achieve performance comparable to {or higher than} that of the GPU.
  We conclude that \mechanism significantly improves the performance of a variety of workloads compared to both processor-centric and PnM architectures.
}

\subsubsection{Performance per Area}
\label{sec:normperfarea}

\begin{figure}[ht]
  \centering
  \includegraphics[width=\columnwidth]{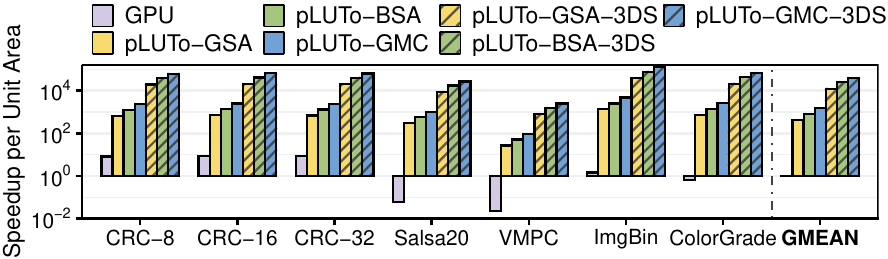}
  \caption{
    {Speedup of GPU and \mechanism relative to CPU, normalized to area.}
    The y-axis uses a logarithmic scale; higher is better.
  }
  \label{fig:normperfArea}
  \vspace{-5pt}
\end{figure}

{
\Cref{fig:normperfArea} shows the speedup per unit area of GPU and \mechanism systems, normalized to the baseline CPU.
The area overhead of \mechanismTDS designs is calculated assuming an area overhead of $4.4~mm^2$ per vault~\cite{boroumand2018google,boroumand2019conda,hmc_spec}.
{For the DDR4 (3DS) implementation, the performance per unit area of \mechanismA, \mechanismB, and \mechanismC mechanisms, on average across all workloads, is
426$\times$/441$\times$ (12405$\times$/12856$\times$),
801$\times$/830$\times$ (24747$\times$/25646$\times$), and
1504$\times$/1558$\times$ (39245$\times$/40670$\times$) that of the CPU/GPU baselines, respectively.}
We make three key observations.
  {First, all \mechanism designs provide substantially higher performance per unit area {than} both the CPU and the GPU (4283$\times$ and 2577$\times$, respectively, on average across all \mechanism designs), and consistently outperform both baselines by a wide margin.}
This improvement is considerably {greater} than the one observed {when considering} {performance} {in isolation (\Cref{sec:performance})} and highlights the potential benefits of scaling \mechanism's design further {with larger DRAM systems.} %
Second, \mechanismTDS designs are more area-efficient than their DDR4 counterparts across all workloads.
This is a consequence of the {large available area and 3D density in the} HMC substrate.
Third, we observe \mechanism's greatest improvements in Salsa20 and VMPC {({respectively} 2970$\times$/50591$\times$ and 273$\times$/106151$\times$ the performance per unit area of the CPU/GPU, on average across all \mechanism designs)}.
These workloads are very memory-intensive, and are therefore well-suited for in-memory execution. %
In contrast, {for the Salsa20 and VMPC workloads,} the GPU baseline provides performance per {unit} area results that fall below even those of the baseline CPU{, which highlights the negative impact of {data movement bottlenecks in the execution of memory-intensive workloads under a processor-centric computing paradigm}}.
}

\subsubsection{Comparison with FPGA}
\label{sec:comparison_fpga}
\begin{figure}[ht]
  \centering
  \includegraphics[width=\columnwidth]{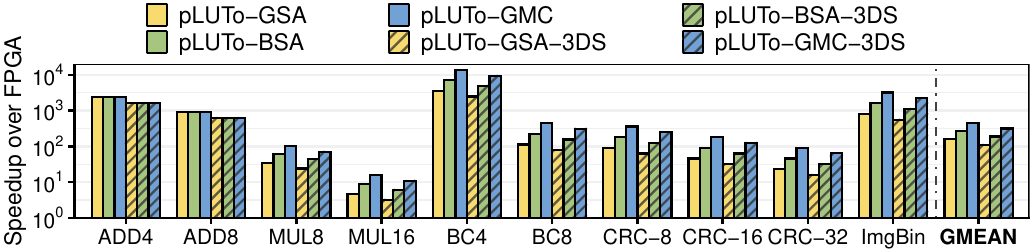}
  \caption{\mechanism speedup relative {to the baseline FPGA.} {The y-axis uses a logarithmic scale; higher is better.}}
  \label{fig:comparison_fpga}
  \vspace{-5pt}
\end{figure}

{
  Figure~\ref{fig:comparison_fpga} shows the performance of the evaluated \mechanism systems normalized to the baseline FPGA.
  We observe that \mechanism outperforms {the FPGA baseline {across all workloads we evaluate}.} {For the DDR4 (3DS) implementation, the performance of \mechanismA, \mechanismB, and \mechanismC mechanisms, on average across all workloads, is
      160$\times$ (111$\times$),
      274$\times$ (190$\times$), and
      459$\times$ (318$\times$) that of the FPGA baseline, respectively.}
  The most significant gains are associated with workloads that rely on smaller LUTs (e.g., BC4, ImgBin), and the smallest gains correspond to operations with large input bit widths (e.g., MUL16).
    {We conclude that, a}lthough both the FPGA and \mechanism rely on LUT-based computation, the former's access to data in memory is still limited {by} {{main} memory bandwidth.
        {In contrast, \mechanism exploits much higher {main memory} bandwidth via the {pLUTo LUT Query, leading to overall higher performance}}.}}

\subsection{Energy {Consumption}}
\label{sec:energy_analysis_of_workloads}

\begin{figure}[ht]
  \centering
  \includegraphics[width=\columnwidth]{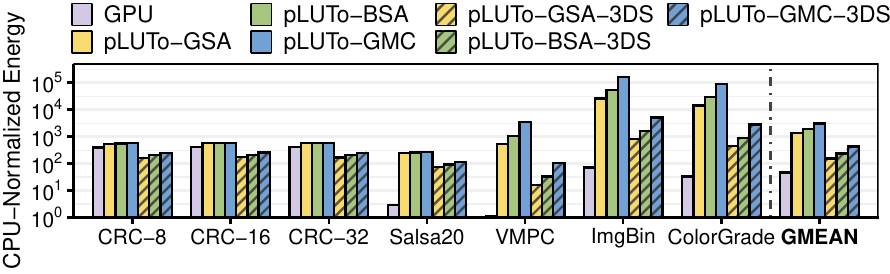}
  \caption{
    Energy {consumption} of GPU and \mechanism compared to the CPU.
    The y-axis uses a logarithmic scale; higher is better.
  }
  \label{fig:normEnergy}
  \vspace{-10pt}
\end{figure}

Figure~\ref{fig:normEnergy} shows {the energy consumed by the GPU and \mechanism systems when executing the evaluated workloads, normalized to the baseline CPU.}
  {{\mechanism's energy consumption depends on the} total number of DRAM operations required by the executed \mechanism ISA instructions (\Cref{tab:isa_extensions}), and therefore \emph{does not vary with different degrees of subarray-level parallelism.}}
  {For the DDR4 (3DS) implementation, \mechanismA, \mechanismB, and \mechanismC systems, on average across all workloads, consume}
  {\energyavgmechA/\energyavgmechAovergpu (\energyavgmechATDS/\energyavgmechATDSovergpu)
    \energyavgmechB/\energyavgmechBovergpu (\energyavgmechBTDS/\energyavgmechBTDSovergpu)
    \energyavgmechC/\energyavgmechCovergpu (\energyavgmechCTDS/\energyavgmechCTDSovergpu)
    less energy than the CPU/GPU baselines, respectively.}

  {We} make two {key} observations.
  {First, {the} energy savings enabled by \mechanism are considerably greater in workloads that are especially memory-intensive (e.g., VMPC) or require simple operations (e.g., ImgBin), but becomes {lower} as workload complexity increases (e.g., CRC-8/16/32).
    This trend is consistent with our observations from \Cref{sec:normperfarea}.
    Second, in some of the workloads ({e.g.,} CRC-8/16/32, Salsa20), the energy consumption {values} of {all} three \mechanism designs {are similar to each other}.
    This is due to the relatively small number of \rowsweep operations required to {execute} these workloads, {which are {\emph{not}} enough to highlight the differences in energy consumption of each \mechanism design}.
    As a result, the overall impact of the improved efficiency of the \rowsweep in \mechanismB and \mechanismC relative to \mechanismA becomes less pronounced.}
  {We conclude that \mechanism significantly reduces energy consumption compared to processor-centric architectures for various workloads.}

\subsection{Area Overhead}
\label{sec:area_overhead}

\Cref{tab:area_breakdown} shows {the estimated area of the baseline DRAM and three \mechanism designs, broken down by DRAM component.}
These estimates are derived from transistor {count estimates and rely on the DRAM area models provided by} CACTI 7~\cite{balasubramonian2017cacti}.

\head{\mechanismA}
{
  The estimated area overhead of the {matchline-controlled switch} (shown in \Cref{fig:pluto_designs_combined}{b}) is 20\% of the area of a sense amplifier per bitline.
  The total area overhead of \mechanismA is \areaoverheadA of the DRAM chip area.
}

\head{\mechanismB} {The estimated area overhead of the} {matchline-controlled switch} and {the} FF (shown in \Cref{fig:pluto_designs_combined}{a}) {is 60\% of the area taken up by sense amplifiers in the base DRAM chip}.
The total {area} overhead of \mechanismB is \areaoverheadB of the DRAM chip area.

\head{\mechanismC} The estimated area overhead {of the matchline-controlled switch} per 2T1C DRAM cell (shown in \Cref{fig:pluto_designs_combined}{c}) is 25\%.
The total area overhead of \mechanismC is \areaoverheadC of the DRAM chip area.

\begin{table}[h]
  \centering
  \caption{{Area breakdown for DRAM and the three \mechanism designs.}}
  \renewcommand{\arraystretch}{1.2}
  \resizebox{0.9\columnwidth}{!}{%
    {
        \begin{tabular}{lc|c|c|c|c|}
          \cline{3-6}
                                                                                                        &                                                 & \textbf{Base DRAM} & \textbf{\mechanismA}                                              & \textbf{\mechanismB}                                              & \textbf{\mechanismC}                                              \\ \hhline{--|=|=|=|=|}
          \multicolumn{1}{|l|}{\multirow{9}{*}{{\rotatebox[origin=c]{90}{\textbf{Area $(mm^{2})$  }}}}} & \multicolumn{1}{|c||}{\textbf{DRAM Cell}}       & 45.23              & 45.23                                                             & 45.23                                                             & \textbf{56.53}                                                    \\ \cline{2-6}
          \multicolumn{1}{|l|}{}                                                                        & \multicolumn{1}{|c||}{\textbf{Local WL driver}} & 12.45              & 12.45                                                             & 12.45                                                             & 12.45                                                             \\ \cline{2-6}
          \multicolumn{1}{|l|}{}                                                                        & \multicolumn{1}{|c||}{\textbf{Match Logic}}     & -                  & \textbf{4.61}                                                     & \textbf{4.61}                                                     & \textbf{4.61}                                                     \\ \cline{2-6}
          \multicolumn{1}{|l|}{}                                                                        & \multicolumn{1}{|c||}{\textbf{Match Lines}}     & -                  & \textbf{0.02}                                                     & \textbf{0.02}                                                     & \textbf{0.02}                                                     \\ \cline{2-6}
          \multicolumn{1}{|l|}{}                                                                        & \multicolumn{1}{|c||}{\textbf{Sense Amp}}       & 11.40              & \textbf{13.67}                                                    & \textbf{18.23}                                                    & 11.40                                                             \\ \cline{2-6}
          \multicolumn{1}{|l|}{}                                                                        & \multicolumn{1}{|c||}{\textbf{Row Decoder}}     & 0.16               & \textbf{0.47}                                                     & \textbf{0.47}                                                     & \textbf{0.47}                                                     \\ \cline{2-6}
          \multicolumn{1}{|l|}{}                                                                        & \multicolumn{1}{|c||}{\textbf{Column Decoder}}  & 0.01               & 0.01                                                              & 0.01                                                              & 0.01                                                              \\ \cline{2-6}
          \multicolumn{1}{|l|}{}                                                                        & \multicolumn{1}{|c||}{\textbf{Other}}           & 0.99               & 0.99                                                              & 0.99                                                              & 0.99                                                              \\ \hhline{|~|=::=|=|=|=|} %
          \multicolumn{1}{|l|}{}                                                                        & \multicolumn{1}{|c||}{\textbf{Total}}           & 70.23              & \begin{tabular}[c]{@{}c@{}}77.44\\ (+\areaoverheadA)\end{tabular} & \begin{tabular}[c]{@{}c@{}}82.00\\ (+\areaoverheadB)\end{tabular} & \begin{tabular}[c]{@{}c@{}}86.47\\ (+\areaoverheadC)\end{tabular} \\ \hline
        \end{tabular}
      }
  }
  \label{tab:area_breakdown}
\end{table}

The area overheads of the match logic and match lines described in \Cref{sec:match_logic}, which are {identical for} all three designs, are shown separately {in \Cref{tab:area_breakdown}}.
The Row Decoder overhead includes that of the logic required for the \rowsweep {operation}.
The only \mechanism design that requires modifications to the DRAM cell design is \mechanismC.
Its {per-cell} overhead is indicated in the DRAM Cell row of~\Cref{tab:area_breakdown}.
In the baseline system, {{DRAM} cell {access} transistors take up approximately \SI{15.1}{\milli\meter\squared}.
    The overhead associated with these transistors doubles} {in \mechanismC's 2T1C cell.}

\vspace{-1mm}
\subsection{LUT Loading Overhead}
\label{sec:loading_luts}
\begin{figure}[ht]
  \centering
  \includegraphics[width=0.9\columnwidth]{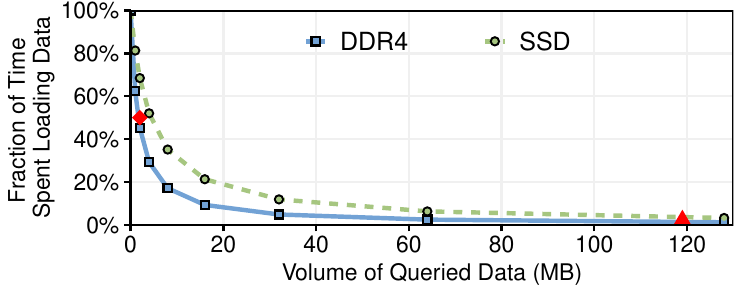}
  \caption{
    \vspace{-0.5mm}Fraction of {time spent loading} {LUTs}
      {(from DDR4 DRAM~\cite{crucialCT8G4SFS8} and M.2 SSD~\cite{pnyCS3140})} {versus} {the volume of LUT input data.}} %
  \label{fig:fraction_loading}
  \vspace{-10pt}
\end{figure}

{
  \Cref{fig:fraction_loading} shows the fraction of total computation time spent loading LUT data (y-axis), {as a function of} the total volume of data queried (x-axis).
  We evaluate two scenarios for the loading of LUTs, as described in \Cref{sec:create_load_store_luts}: \emph{Loading from Memory} and \emph{Loading from Secondary Storage}.
  For the first scenario, we assume a DDR4 memory bandwidth of \SI{19.2}{\giga\byte\per\second}~\cite{crucialCT8G4SFS8}; for the second scenario, we assume an M.2 SSD bandwidth of \SI{7500}{\mega\byte\per\second}~\cite{pnyCS3140}.
  As an example of how to interpret this plot, if data is \emph{loaded from memory} {(i.e., from DRAM)}, when querying around \SI{20}{\mega\byte} of data (x-axis), {approximately} 10\% of the {LUT query} execution time {(y-axis)} {is} spent loading the {LUT's} data into {DRAM} (accordingly, 90\% of the execution time {is} spent performing \lutqueries).
}

{We make three {key} observations.
  First{, it} is sufficient to process \SI{1.9}{\mega\byte} of data in the DDR4-based scenario (\textcolor{red}{\ding{117}} in \Cref{fig:fraction_loading}) for the LUT loading time to equal the LUT query time.
  This observation illustrates that the cost of loading LUTs into {DRAM} can be quickly amortized, even for small volumes of input data.
  Second{, as} the volume of data to be processed increases, the fraction of time spent loading the LUTs into memory quickly decreases, both for the DDR4- and the SSD-based scenarios.
  For example, for an input of \SI{120}{\mega\byte} {of data} (\textcolor{red}{\ding{115}} in \Cref{fig:fraction_loading}), the fraction of time spent {loading LUTs} is {only} about $2\%$ in the case of DDR4.
  Third{, although} {loading LUTs from the SSD takes longer than doing so from DRAM,} %
  this difference {does \emph{not} significantly impact the volume data that needs to be queried to amortize the LUT loading time.} %
    {W}e conclude {that loading LUTs {both from DRAM and} from a secondary storage device (i.e., M.2 SSD) are well-suited and practical approaches to support {\mechanism's operation}.}
}

\begin{figure}[b]
  \centering
  \begin{subfigure}{0.95\columnwidth}
    \centering
    \includegraphics[width=\columnwidth]{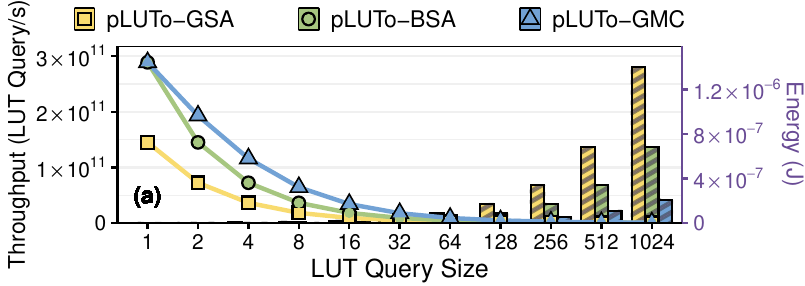}
    \label{fig:first}
  \end{subfigure}
  \hfill
  \begin{subfigure}{0.95\columnwidth}
    \centering
    \hspace{-9.5mm}
    \includegraphics[width=0.815\columnwidth]{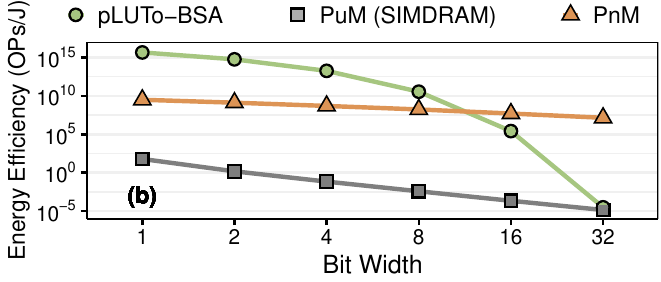}
    \label{fig:third}
  \end{subfigure}
  \caption{
    Scalability analysis for \mechanism's LUT query {operation}:
    (a) {shows LUT} query throughput {(lines in the primary/left y-axis) and energy consumption (bars in the secondary/right y-axis)} for the three \mechanism designs while varying {LUT} query size;
    (b) compares the energy efficiency (in {OPs}/J) of \mechanismB against a prior PuM mechanism (SIMDRAM~\cite{hajinazar2020simdram}) and the {PnM} baseline.
  }
  \label{fig:scal}
\end{figure}

\vspace{-0.5mm}
\subsection{Scalability Analysis}
\label{sec:scalability_analysis}

This section analyzes the scalability of \mechanism's LUT query operation.
Our goals are 1) to fundamentally understand the performance {limits} of the \lutquery, and 2) to study the suitability of different PiM architectures {for multiplication, a commonly used operation.}
First, in \Cref{fig:scal}a, {we show an analysis of {the throughput} and energy consumption scaling of the three proposed \mechanism designs, as determined by the equations and timing parameters {in \Cref{tab:pluto-modes} and by the equations derived in \Cref{sec:bsa_analysis,sec:gsa_analysis,sec:gmc_analysis}}.}
Second, in \Cref{fig:scal}b, {we compare the energy efficiency {(in $\texttt{\#\ Multiplications}/J$)} of three systems:}
1) \mechanismB,
2) a bit-serial PuM mechanism (SIMDRAM~\cite{hajinazar2020simdram}), and
3) our baseline {PnM} device, while varying the {bit width} of the operands involved in the multiplication.
  {We note that the {multiplication operation}
    is especially costly for SIMDRAM to execute, as discussed in~\Cref{sec:comp_prior_works}.}

We make two key observations from the figures.
First, all three \mechanism designs provide high throughput and low energy consumption for small LUT query sizes ($N \le 8$).
  {This happens because, as \Cref{tab:pluto-modes} shows, the latency and energy consumption of a \rowsweep increase linearly with the LUT query size (i.e., consequently, the throughput of a \rowsweep \emph{decreases} linearly with the LUT query size). Therefore, \mechanism achieves its highest throughput and lowest energy consumption for small LUT query sizes.}
Second, {\mechanism} provides higher energy efficiency {than {the} alternative} SIMDRAM and {PnM} {Processing-in-Memory architectures} for low-precision {multiplication} ({bit width} $\le 8$ bits).
Executing {multiplication} in \mechanism leads to better energy efficiency than in SIMDRAM for all evaluated {bit widths}.
This happens because executing bit-serial multiplications {(as SIMDRAM does)} incurs a quadratic number of DRAM activations~\cite{hajinazar2020simdram}.
We conclude that \mechanism is well-suited to perform low-bit-width LUT queries, which can be adopted alongside alternative solutions (e.g., {SIMDRAM and} {PnM}) to take full advantage of the underlying DRAM substrate.

\vspace{-1mm}
\subsection{Impact of \tfaw on Performance}
\label{sec:tfaw}
\begin{figure}[ht]
  \centering
  \includegraphics[width=0.87\columnwidth]{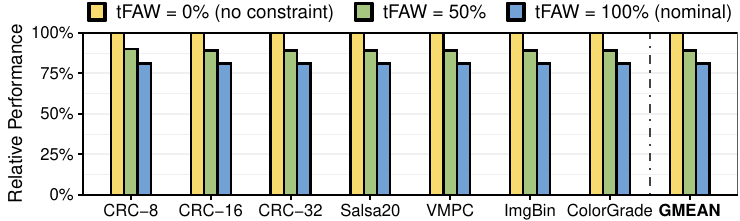}
  \caption{{The impact of tFAW on the performance of \mechanism.}}
  \label{fig:tfaw_perf}
  \vspace{-5pt}
\end{figure}

The \tfaw timing parameter {limits} the activation rate in a DRAM chip to meet power {and reliability} constraints.
Since the activation operation is central to \mechanism, it is important to evaluate the impact of this parameter on performance.
Figure~\ref{fig:tfaw_perf} shows the {effect} {on the performance of a single \lutquery} of varying \tfaw between 0\% and 100\% of its {\emph{nominal value} {(}i.e., the {{actual} \tfaw value} in the DRAM chip we model, \SI{13.328}{\nano\second}){, across our examined workloads.}
    When $\tfaw=0\%$, it is possible to issue an \emph{unlimited} number of activations concurrently; when $\tfaw=50\%$, it is possible to issue \emph{twice} as many activations per unit of time as {commodity} DRAM permits; when $\tfaw=100\%$, the number of activations issued per unit of time corresponds to \emph{exactly} as many as {commodity} DRAM permits.}

We make two key observations.
First, the performance loss is {approximately} 10\% for $\tfaw=50\%$, and {approximately} 20\% for {$\tfaw=100\%$}, {compared to not having the \tfaw limitation at all}.
Even accounting for this performance penalty {of \tfaw}, \mechanism still outperforms the CPU and GPU baselines, as shown in \Cref{sec:performance}.
Second, the {performance penalty is} very similar across all of the {evaluated} workloads for the same value of \tfaw.
{We conclude that, despite \tfaw's limited impact on \mechanism, the use of more powerful charge pumps could further relax power constraints {of} \mechanism-capable {DRAMs.}}

\subsection{Effect of Subarray{-Level} Parallelism}
\label{sec:analysis_of_subarray_parallelism}

\begin{figure}[t]
  \centering
  \includegraphics[width=0.87\columnwidth]{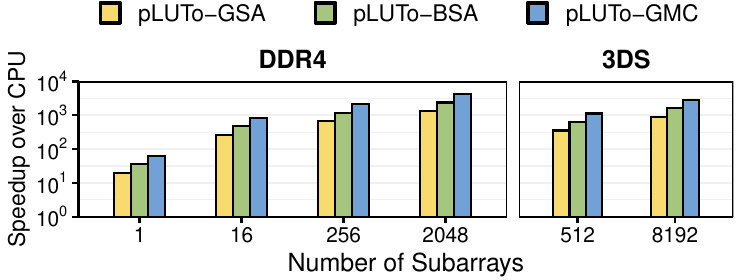}
  \vspace{-5pt}
  \caption{Geometric mean speedup of \mechanism over CPU, for varying degrees of subarray-level parallelism.
      {The y-axis uses a logarithmic scale; higher is better.}}
  \label{fig:pluto_parallelism}
  \vspace{-8pt}
\end{figure}

{We} evaluate the three \mechanism designs (\mechanismA, \mechanismB, \mechanismC) with varying degrees of subarray-level parallelism for both DDR4 and 3D-stacked memory.
Figure~\ref{fig:pluto_parallelism} plots the {speedup} (averaged across all evaluated workloads) of each configuration against the baseline CPU.
We make two key observations.
  {First, due to the different row buffer sizes of DDR4 (\SI{8}{\kilo\byte}) and 3DS (\SI{256}{\byte}) {memories} (as discussed in \Cref{sec:evaluation_methodology}), the same degree of subarray-level parallelism results in higher speedup for \mechanism-DDR4 than \mechanism-3DS.}
  {Second, performance scaling is approximately proportional to the number of subarrays operating in parallel in both cases, provided that the size of the input to be queried is sufficiently large.
    This linear relationship between subarray {count} and performance improvement shows that
    1) although the volume of internal data movement required by \mechanism's operation {increases with} the degree of subarray-level parallelism, this factor does {\emph{not}} limit \mechanism's scalability, and
    2) \mechanism should be configured to operate with as high a degree of subarray-level parallelism as the memory technology supports.}%
\footnote{{As discussed in \Cref{sec:energy_analysis_of_workloads}, energy consumption is {\emph{not}} affected by different degrees of subarray-level parallelism.}}

\subsection{Comparison With Prior {PuM Systems}}
\label{sec:comp_prior_works}

As demonstrated in~\Cref{tab:comparison_with_pum} and discussed in~\Cref{sec:motivation}, prior {DRAM-based} PuM architectures (e.g.,~\cite{seshadri2017ambit,li2017drisa,deng2019lacc}) achieve very high throughput and energy efficiency, but do so while supporting a very limited range of operations.
These works can address this limitation by exploiting alternatives to conventional bit-parallel algorithms.
For example, it is possible to efficiently realize arithmetic operations in Ambit~\cite{seshadri2017ambit} using bit-serial algorithms~\cite{hajinazar2020simdram}.
{We show} that the additional flexibility afforded by  \mechanism's native support for LUT operations allows it to outperform prior PuM {architectures.}
We substantiate this claim with \Cref{tab:comparison_with_pum}, which shows {the latency of each operation of interest under Ambit~\cite{seshadri2017ambit}, SIMDRAM~\cite{hajinazar2020simdram}, LAcc~\cite{deng2019lacc}, DRISA~\cite{li2017drisa} and \mechanism.}
  {In each case, we assume the use of ideal data {layout for each system} as defined in the original works (e.g., bit-parallel for \mechanism, bit-serial for SIMDRAM) and report the best-case \textit{achievable} performance for each design.}\footnote{{Supporting each architecture's ideal data layout requires system-level changes.}}
{As an example,} bitwise operations between input sets \texttt{A} ('$a_1a_2...$') and \texttt{B} ('$b_1b_2...$'), under {\mechanism's} LUT-based paradigm require that input operands be shuffled (\texttt{'$a_1b_1a_2b_2...$'}); in contrast, the input sets \texttt{A} and \texttt{B} are ideally stored in two separate {DRAM} rows for all prior PuM designs (Ambit~\cite{seshadri2017ambit}, SIMDRAM~\cite{hajinazar2020simdram}, LAcc~\cite{deng2019lacc}, DRISA~\cite{li2017drisa}).
To {make fair comparisons}, the memory capacity for each design is such that the area overheads for all designs remain in a narrow range {(the area values across the five PuM comparison points average \SI{62.5}{\milli\metre\squared} $\pm$ \SI{5.2}{\milli\metre\squared}), which is similar to the area of commodity DRAM devices ($\approx$\SI{70.23}{\milli\metre\squared}, see \Cref{tab:area_breakdown}).
    As a result, due to its inferior storage density, the capacity of DRISA~\cite{li2017drisa} is limited to \SI{2}{\giga\byte}.}

\begin{table}[hbt]
  \centering
  \caption{Comparison of operations supported by \mechanism vs. prior PuM.
      {All} performance per area and energy efficiency values are normalized to \mechanismB with 4-subarray parallelism.}
  \renewcommand{\arraystretch}{1.1}
  \resizebox{\columnwidth}{!}{
    \begin{tabular}{|c|c|c|c|c|c|}
      \hline
                                                             & \textbf{Ambit}~\cite{seshadri2017ambit} & \textbf{SIMDRAM}~\cite{hajinazar2020simdram} & \textbf{LAcc}~\cite{deng2019lacc} & \textbf{DRISA}~\cite{li2017drisa} & \textbf{\mechanismB}           \\ \hhline{|=|=|=|=|=|=|}
      \multicolumn{1}{|c||}{Capacity}                        & \SI{8}{\giga\byte}                      & \SI{8}{\giga\byte}                           & \SI{8}{\giga\byte}                & \SI{2}{\giga\byte}                & \SI{8}{\giga\byte}             \\ \hline
      \multicolumn{1}{|c||}{Area ($mm^2$)}                   & 61.0                                    & 61.1                                         & 54.8                              & 65.2                              & 70.5                           \\ \hline
      \multicolumn{1}{|c||}{Power ($W$)}                     & 5.3                                     & 5.3                                          & 5.3                               & 98.0                              & 11                             \\ \hline \hline
      \multicolumn{1}{|c||}{NOT ($ns$)}                      & 135.0                                   & 135.0                                        & 135.0                             & 207.6                             & 105.0                          \\ \hline
      \multicolumn{1}{|c||}{AND ($ns$)}                      & 270.0                                   & 270.0                                        & 270.0                             & 415.2                             & 165.0                          \\ \hline
      \multicolumn{1}{|c||}{OR ($ns$)}                       & 270.0                                   & 270.0                                        & 270.0                             & 415.2                             & 165.0                          \\ \hline
      \multicolumn{1}{|c||}{XOR ($ns$)}                      & 585.0                                   & 585.0                                        & 450.0                             & 691.9                             & 165.0                          \\ \hline
      \multicolumn{1}{|c||}{XNOR ($ns$)}                     & 585.0                                   & 585.0                                        & 450.0                             & 691.9                             & 165.0                          \\ \hline
      \multicolumn{1}{|c||}{\textbf{Performance Per Area}}   & \multirow{2}{*}{\textbf{0.54}}          & \multirow{2}{*}{\textbf{0.54}}               & \multirow{2}{*}{\textbf{0.67}}    & \multirow{2}{*}{\textbf{0.37}}    & \multirow{2}{*}{\textbf{1.00}} \\
      \multicolumn{1}{|c||}{\textbf{(higher is better)}}     &                                         &                                              &                                   &                                   &                                \\ \hline
      \multicolumn{1}{|c||}{\textbf{Energy Efficiency}}      & \multirow{2}{*}{\textbf{0.54}}          & \multirow{2}{*}{\textbf{0.54}}               & \multirow{2}{*}{\textbf{0.67}}    & \multirow{2}{*}{\textbf{0.02}}    & \multirow{2}{*}{\textbf{1.00}} \\
      \multicolumn{1}{|c||}{\textbf{(higher is better)}}     &                                         &                                              &                                   &                                   &                                \\ \hline \hline
      \multicolumn{1}{|c||}{4-bit Addition ($ns$)}           & 5081.0                                  & 1585.0                                       & 1142.3                            & 1756.5                            & 1920.0                         \\ \hline
      \multicolumn{1}{|c||}{4-bit Multiplication ($ns$)}     & 19065.0                                 & 7451.0                                       & 5365.4                            & 8250.1                            & 1920.0                         \\ \hline
      \multicolumn{1}{|c||}{4-bit Bit Counting ($ns$)}       & 2936.0                                  & 1156.0                                       & -                                 & 6649.9                            & 120.0                          \\ \hline
      \multicolumn{1}{|c||}{8-bit Bit Counting ($ns$)}       & 6901.0                                  & 2696.0                                       & -                                 & 13580.0                           & 1920.0                         \\ \hline
      \multicolumn{1}{|c||}{\textbf{Performance Per Area}}   & \multirow{2}{*}{\textbf{0.34}}          & \multirow{2}{*}{\textbf{0.45}}               & \multirow{2}{*}{\textbf{1.00$*$}} & \multirow{2}{*}{\textbf{0.17}}    & \multirow{2}{*}{\textbf{1.00}} \\
      \multicolumn{1}{|c||}{\textbf{(higher is better)}}     &                                         &                                              &                                   &                                   &                                \\ \hline
      \multicolumn{1}{|c||}{\textbf{Energy Efficiency}}      & \multirow{2}{*}{\textbf{0.69}}          & \multirow{2}{*}{\textbf{0.94}}               & \multirow{2}{*}{\textbf{2.00$*$}} & \multirow{2}{*}{\textbf{0.02}}    & \multirow{2}{*}{\textbf{1.00}} \\
      \multicolumn{1}{|c||}{\textbf{(higher is better)}}     &                                         &                                              &                                   &                                   &                                \\ \hline \hline
      \multicolumn{1}{|c||}{6-bit to 2-bit LUT Query ($ns$)} & -                                       & -                                            & -                                 & -                                 & 480.0                          \\ \hline
      \multicolumn{1}{|c||}{8-bit to 8-bit LUT Query ($ns$)} & -                                       & -                                            & -                                 & -                                 & 1920.0                         \\ \hline
      \multicolumn{1}{|c||}{{8-bit Binarization} ($ns$)}     & -                                       & -                                            & -                                 & -                                 & 1920.0                         \\ \hline
      \multicolumn{1}{|c||}{{8-bit Exponentiation} ($ns$)}   & -                                       & -                                            & -                                 & -                                 & 1920.0                         \\ \hline
    \end{tabular}

  }
  \label{tab:comparison_with_pum}
  \newline
  \scriptsize{$-$ indicates that the operation is \emph{not} supported by the proposed mechanism.} \\
  \scriptsize{$*$ indicates that the result was obtained from partial data.}
\end{table}

We draw three key {observations} from \Cref{tab:comparison_with_pum}.
First, {due} to their complexity, some operations (e.g., binarization, exponentiation) \emph{{cannot}} be implemented in a time-efficient manner using any prior {design}.
In \mechanism, it is possible to perform exponentiation with high efficiency when operating on small bit widths (for best results, up to 8 bits).
  {Second,} {\mechanism's throughput for} bitwise logic {operations matches or exceeds that of all} prior works.
  {Third}, \mechanism consistently outperforms all {four} other approaches for most of the considered operations in performance (absolute and normalized to area) and energy efficiency.
This improvement is not universal: for instance, \mechanism slightly lags behind all baselines {for} 4-bit addition.
  {We conclude that \mechanism achieves its main goal of extending the range and complexity of operations that the DRAM-based PuM paradigm supports.}

\section{Case Study: {Quantized Neural Networks}}
\label{sec:end_to_end}

{Building on the observation that \mechanism is especially well-suited for executing low-bit-width operations (\Cref{sec:comp_prior_works}), we {evaluate} the {benefits of} \mechanism on neural networks quantized to 1 and 4 bits, an emerging machine learning application that is especially useful for power-limited devices~\cite{hubara2017quantized,khoram2018adaptive,garofalo2020pulp}.
We evaluate a quantized version of the LeNet-5 network~\cite{lecun1998gradient} to classify digits from the MNIST dataset~\cite{lecun1998gradient} as a proof of concept.
For this evaluation, the CPU and FPGA baselines are those described in~\Cref{sec:evaluation_frameworks}; the GPU is a data-center-grade {NVIDIA P100~\cite{nvidia_p100}, which was specifically developed for machine learning applications.}
{\Cref{tab:lenet} shows the inference times for CPU, GPU, FPGA, and \mechanismB.}
\mechanismB outperforms the CPU ($10\times$, $30\times$ for 1-bit, 4-bit inference), the GPU ($2\times$, $7\times$) and the FPGA ($6\times$, $19\times$) in inference time.
\mechanism's performance improvements for these operations result from the bulk querying of input values using only short sequences of DRAM commands to perform bitwise logic operations.
Simultaneously, \mechanism provides large energy savings over both the CPU ($110\times$, $109\times$), the GPU ($80\times$, $81\times$) and the FPGA ($15\times$, $16\times$), for both 1- and 4-bit precision.
The key reason behind this increase in energy efficiency is the reduction in overall data movement (since many operations are performed {in-place}) and the energy efficiency of DRAM commands.
We conclude that \mechanism is well-suited to accelerate quantized neural network inferencing and to reduce the energy cost of this workload, especially in heavily energy-constrained devices, such as edge and IoT devices.}

\vspace{1mm}

\begin{table}[h]
  \centering
  \caption{LeNet-5 inference times
    (in \si{\micro\second}) and energy (in \si{\milli\joule}) for CPU, GPU, FPGA, and \mechanism.}
  \renewcommand{\arraystretch}{1.1}
  \resizebox{\columnwidth}{!}{%
    \begin{tabular}{|c|c|c|c|c|c|c|c|c|c|}
      \hline
      \multirow{2}{*}{\textbf{\begin{tabular}[c]{@{}c@{}}Bit\\ Width\end{tabular}}} & \multirow{2}{*}{\textbf{\shortstack{Accuracy                                                                                                                                         \\ \cite{khoram2018adaptive}}}} & \multicolumn{2}{c|}{\textbf{CPU}} & \multicolumn{2}{c|}{\textbf{GPU}} & \multicolumn{2}{c|}{\textbf{FPGA}} & \multicolumn{2}{c|}{\textbf{\mechanismB}} \\ \cline{3-10}
                                                                                    &                                              & \textbf{Time} & \textbf{Energy} & \textbf{Time} & \textbf{Energy} & \textbf{Time} & \textbf{Energy} & \textbf{Time} & \textbf{Energy} \\ \hhline{|=|=|=|=|=|=|=|=|=|=|}
      1 bit                                                                         & 97.4 \%                                      & 249           & 2.2             & 56            & 1.6             & 141           & 0.3             & 23            & 0.02            \\ \hline
      4 bits                                                                        & 99.1 \%                                      & 997           & 8.7             & 224           & 6.5             & 563           & 1.3             & 30            & 0.08            \\ \hline
    \end{tabular}
  }
  \label{tab:lenet}
\end{table}

\vspace{-3mm}

\section{Related Work}
\label{sec:related_work}

To our knowledge, \mechanism is the first work {that enables the efficient bulk querying of lookup tables (LUTs)} inside DRAM to enable {the execution} of complex operations.
In this section, we describe relevant prior works.

\head{Processing-using-Memory (PuM)}
Many prior works propose various forms of {Processing-using-Memory}~\cite{aga2017compute,akerib2012using,angizi2017design,angizi2017energy,angizi2017imc,angizi2017rimpa,angizi2018cmp,angizi2018dima,angizi2018imce,angizi2018pima,angizi2019aligns,angizi2019deep,angizi2019graphide,angizi2019graphs,angizi2019parapim,angizi2019redram,angizi2020exploring,angizi2020pimaligner,angizi2020pimassembler,besta2021sisa,chang2016lisa,chi2016prime,deng2018dracc,deng2019lacc,eckert2018neural,fan2016low,fan2017energy,fan2017leveraging,fan2017memory,flashcosmos,fujiki2019duality,gaillardon2016programmable,gao2019computedram,gu2020dlux,hajinazar2020simdram,hamdioui2015memristor,hamdioui2017memristor,he2017exploring,he2017high,he2017leveraging,he2020sparse,imani2019floatpim,kang2014energy,kvatinsky2011memristor,kvatinsky2013memristor,kvatinsky2014magic,lea2019apparatuses,levy2014logic,li2016pinatubo,li2017drisa,Li2018SCOPEAS,manning2018apparatuses,oliveira2022accelerating,parveen2017hybrid,parveen2017low,parveen2018hielm,parveen2018imcs2,rakin2018pim,ramanathan2020look,rezaei2020nom,seshadri.arxiv16,seshadri2013rowclone,seshadri2015fast,seshadri2015gather,seshadri2016processing,seshadri2017ambit,seshadri2017simple,seshadri2018rowclone,seshadri2019dram,shafiee2016isaac,song2017pipelayer,song2018graphr,tian2017approxlut,wu2022dram,xie2015fast,xin2019roc,xin2020elp2im,yang2020flexible,yu2018memristive,zawodny2018apparatuses,zha2020hyper,zhao2017apparatuses}.
All these approaches provide significant performance and energy improvements, but focus {mainly} on a reduced set of operations (e.g., data movement~\cite{chang2016lisa,seshadri2013rowclone}, bulk bitwise operations~\cite{aga2017compute,li2016pinatubo,seshadri2017ambit,xin2019roc}, or neural network \mbox{acceleration}~\cite{deng2018dracc,deng2019lacc,eckert2018neural,li2017drisa}).
By combining {the \lutquery} with {fast} and efficient bitwise logic~\cite{seshadri2017ambit} and {bit shifting} operations~\cite{li2017drisa} enabled by these prior works, \mechanism supports a much {wider} range of operations.
While pPIM~\cite{sutradhar2020ppim} and LAcc~\cite{deng2019lacc}, for example, leverage dedicated LUT hardware for neural network acceleration, the \lutquery is suitable for a greater range of operations (by supporting a broader set of input-output configurations, with greater performance and energy {efficiency).}
  {DRAF}~\cite{gao2016draf} employs a DRAM-based FPGA-like LUT-based computing paradigm {that allows it to outperform an FPGA} baseline in area and energy {efficiency; however, DRAF lags} in throughput and latency.
In contrast, \mechanism enables high-throughput LUT queries without compromising energy efficiency
  {and with {small} overhead (between \areaoverheadA and \areaoverheadC, for different versions of \mechanism) on the storage density of the DRAM array.}

\head{Processing-near-Memory (PnM)}
3D-stacked memories~{\cite{hmc_spec,lee20141,lee2016simultaneous}}
enable the stacking of memory layers {on top of} a logic layer~{\cite{loh2008isca,kim2018grim,PEI,kim2016neurocube,cali2020genasm,akin2016data,boroumand2021mitigating,boroumand2021google,fernandez2020natsa,niu2022184qps,ahn2016scalable,boroumand2018google,boroumand2016lazypim,zhang2014top,drumond2017mondrian,boroumand2019conda,hsieh2016transparent,pattnaik2016scheduling,boroumand2022polynesia,boroumand2021polynesia,amiraliphd,besta2021sisa,singh2019napel,gao2016hrl,santos2017operand,glova2019near,xie2017processing,kersey2017lightweight,guo20143d,hadidi2018performance,liu20173d,NDC_ISPASS_2014,de2018design,santos2018processing,LiM_3D_FFT_MM,Sparse_MM_LiM,smc_sim,jang2019charon,tsai:micro:2018:ams,oliveira2017generic,picorel2017near,hadidi2017cairo}}.
This technology provides higher bandwidth compared to 2D DRAM.
\mechanism is \textit{complementary} to 3D-stacked memory: the two can be combined as shown in~\Cref{sec:evaluation}.

\head{Content-Addressable Memories (CAMs)}
{
  CAM-based accelerators (e.g., ~\cite{caminal2021cape,caminal2022accelerating,yavits2021giraf,garzon2022aida,kaplan2018prins,morad2016resistive,kaplan2017resistive,zha2020hyper}) return the address of matched data given an input query and can {therefore} be used to perform LUT-based computing.
  Most CAMs are SRAM-based and provide low area density compared to DRAM-based memories.}
DRAM-based CAMs also exist~\cite{patel2006dram,makosiej2017cellule,batson2002dram},
but require a greater number of transistors per cell than \mechanismC, our most {expensive} design.

\head{Automata Processors (APs)}
{APs are specialized processing engines that support the hardware-native execution of nondeterministc-finite-automata~\cite{subramaniyan2017parallel,wang2016overview}.
  These automata may be used for pattern recognition tasks and the querying of unstructured data sources, {but {have a relatively narrow domain of applications}} and are therefore not well-suited} for the offloading of complex functions by memoization.

\section{Conclusion}
\label{sec:conclusion}

We introduced \mechanism, a {new} DRAM-based Processing-using-Memory architecture that enables the %
storage and bulk {querying} of lookup tables {completely in-DRAM}.
We build \mechanism based on the key observation that {enabling bulk lookup-table-query operations inside DRAM enables the efficient execution of complex operations with high performance and {energy} efficiency.}
We describe
  {1)} the hardware design of three different \mechanism architectures, each {providing} a different trade-off between performance, energy efficiency, and area overhead, and
  {2)} the necessary system integration support to enable the execution of in-DRAM \mechanism operations.
Our evaluations show that \mechanism significantly outperforms the baseline CPU-, GPU-, FPGA-, PnM-, and PuM-based systems {in terms of execution time, performance per area, and energy consumption}.
We {hope that future work explores new ways of taking advantage of \mechanism and in-DRAM LUT-based computing to provide even {greater} performance and energy benefits for more applications that can take advantage of the PuM paradigm.}

\vspace{-1mm}

\section*{Acknowledgments}

We thank the anonymous reviewers of HPCA (2020), ISCA (2020, 2021, 2022), ICCD (2020), MICRO (2021, 2022), and ASPLOS (2022) for their valuable comments and feedback.
We thank the SAFARI Research Group members for their valuable feedback and the stimulating intellectual environment they provide.
We acknowledge the generous gifts provided by our industrial partners: Google, Huawei, Intel, Microsoft, and VMware.
This work was supported in part by {the} Semiconductor Research Corporation (SRC) {and the ETH Future Computing Laboratory}.

This work was supported in part by the Instituto de Telecomunicações and the Fundação para a Ciência e a Tecnologia (FCT), under grant numbers UIDB/50008/2020-UIDP/50008/2020 and EXPL/EEI-HAC/1511/2021.

\setstretch{1.0}
{
  \bstctlcite{IEEEexample:BSTcontrol}
  \let\OLDthebibliography\thebibliography
  \renewcommand\thebibliography[1]{
    \OLDthebibliography{#1}
    \setlength{\parskip}{0pt}
    \setlength{\itemsep}{0pt}
  }
  \bibliographystyle{IEEEtran}
  \bibliography{references}
}
\setstretch{1.0}

\appendix
\section{Artifact Appendix}

\subsection{Abstract}

{Our artifacts span two components of the evaluation of \mechanism: 1) SPICE simulations (\Cref{sec:spice}), and 2) throughput and energy consumption estimations (\Crefrange{sec:performance}{sec:analysis_of_subarray_parallelism}).
  We evaluate the correct circuit-level functionality of \mechanism cells using DRAM cell models based on Low-Power 22nm Metal Gate PTM transistors~\cite{ptm2012transistors}.
  We evaluate the performance and energy consumption of \mechanism using a custom Python-based analytical timing and energy model.
  To aid the reproducibility of our results, we provide a Jupyter Notebook that may be used to automatically
  1) run the Python-based performance and energy model, and
  2) plot the ensuing results.
    {The {artifact repository is available}
      at
      \url{https://doi.org/10.5281/zenodo.6942058}
      and at
      \url{https://github.com/CMU-SAFARI/pLUTo}.}}

\subsection{Artifact Check-List (Meta-Information)}

{\small
  \begin{itemize}
    \item {\bf Program: LTSpice, Python}
    \item {\bf Metrics: voltage level, cycle count, energy consumption}
    \item {\bf Experiments: DRAM cell operation, evaluated workloads}
    \item {\bf Disk space requirements: 10 MB}
    \item {\bf Time required to prepare the workflow: 1 hour}
    \item {\bf Time required to complete the experiments: 1 hour}
    \item {\bf Publicly available: yes}
    \item {\bf Code license: MIT License}
    \item {\bf DOI: 10.5281/zenodo.6942058}
  \end{itemize}
}

\subsection{Description}

\subsubsection{How to Access}

The repository containing the \mbox{artifacts} may be accessed at
\url{https://doi.org/10.5281/zenodo.6942058}
or at
\mbox{\url{https://github.com/CMU-SAFARI/pLUTo}}.



\subsubsection{Software Dependencies}

{The reproduction of our \mbox{artifacts} require{s} LTSpice, Python 3, and NumPy.
  The interactive artifact generation walkthrough further requires the Jupyter, pandas, SciPy, and Matplotlib Python libraries.}

\subsection{Installation}

{No installation is required.
  The simulator may be launched either directly as a Python script or via the provided interactive Jupyter Notebook.
  Detailed step-by-step directions for achieving this are provided in the repository and in \Cref{sec:experiment_workflow}.}

\subsection{Experiment Workflow}
\label{sec:experiment_workflow}

\subsubsection{SPICE Simulation}

{Please follow the following instructions to reproduce these results:
  \begin{enumerate}
    \item Download LTSpice.
    \item Open the \texttt{.asc} files and run a Monte Carlo simulation.
    \item Probe the bitline voltage by clicking the bitline node.
          A similar (results are stochastic) plot to \Cref{fig:spice_sims} will appear.
  \end{enumerate}}

\subsubsection{Performance and Energy Model}

{
  These results can be easily reproduced by following the step-by-step instructions in the provided \texttt{sim\_walkthrough.ipynb} file.
  This Jupyter Notebook will automatically generate three output CSV files containing the performance and energy values for each of the 6 (DDR4- and 3DS-based \mechanismA, \mechanismB, and \mechanismC) configurations of \mechanism.
}

\subsection{Evaluation and Expected Results}

\subsubsection{SPICE Simulation}

{Inspection of the resulting bitline voltages should reveal {similar results} to those shown in \Cref{fig:spice_sims} {(small variations are expected due to the randomness inherent to the Monte Carlo simulation)}.}

\subsubsection{Performance and Energy Model}

{
  The execution of the provided simulator should result in the creation of files identical to the provided \texttt{pluto\_sim/pysim\_reference/\mechanism\_*.csv}.
  In addition, when executing the provided Jupyter Notebook, plots depicting the same data as shown in \Crefrange{fig:normperfA}{fig:scal} should appear.
}

\flushcolsend

\ifcameraready
\else
  \ifshowtodos
    \iftodos
    {%
      \begin{center}
        \textcolor{red}{\textbf{\totaltodos\ TODOs left}}
      \end{center}
      \listoftodos
      \vspace{2mm}
      \hrule
      \vspace{2mm}
    }
  \fi
\fi

\end{document}